\documentclass[SIG,biber,figuresright]{nowfnt}

\usepackage{textcomp}
\usepackage{tabularx}
\usepackage[utf8]{inputenc}
\usepackage{booktabs}
\usepackage{rotating}
\usepackage{amsmath}
\usepackage{eurosym}
\usepackage{multirow}
\usepackage{amssymb}
\usepackage[ruled,norelsize,linesnumbered]{algorithm2e}
\usepackage{dcolumn}
\newcolumntype{d}[1]{D{.}{.}{#1}} 

\usepackage{color, colortbl}
\usepackage[thinc]{esdiff}

\usepackage[capitalize]{cleveref}

\usepackage{color, colortbl}
\definecolor{Gray}{gray}{0.9}

\title{Min-Max Framework for Majorization-Minimization Algorithms in Signal Processing Applications: An Overview}

\maintitleauthorlist{
Astha Saini\\
Indian Institute of Technology Delhi\\
astha.saini@care.iitd.ac.in\\
\and
Petre Stoica\\
Uppsala University\\
ps@it.uu.se\\
\and
Prabhu Babu\\
Indian Institute of Technology Delhi\\
Prabhu.Babu@care.iitd.ac.in\\
\and
Aakash Arora\\
Indian Institute of Technology Delhi\\
aarora@care.iitd.ac.in
}

\issuesetup
{%
copyrightowner={A. Saini \textit{et al}.},
volume = 18,
issue = 4,
pubyear = 2024,
isbn = 978-1-63828-466-6,
eisbn = 978-1-63828-467-3,
doi = 10.1561/2000000129,
firstpage = 310,
lastpage =  389
}

\usepackage{mwe}

\author[1]{\hspace*{-3.1pt}Astha Saini}
\author[2]{Petre Stoica}
\author[3]{Prabhu Babu}
\author[4]{Aakash Arora}

\affil[1]{Indian Institute of Technology Delhi, India; astha.saini@care.iitd.ac.in}
\affil[2]{Uppsala University, Sweden; ps@it.uu.se}
\affil[3]{Indian Institute of Technology Delhi, India; Prabhu.Babu@care.iitd.ac.in}
\affil[4]{Indian Institute of Technology Delhi, India; aarora@care.iitd.ac.in}

\articledatabox{\nowfntstandardcitation}

\begin{document}

\makeabstracttitle

\begin{abstract}
This monograph presents a theoretical background and a broad introduction to the \textbf{M}in-\textbf{M}ax Framework \textbf{for} \textbf{M}a\-jori\-za\-tion-\textbf{M}inimization (MM4MM), an algorithmic methodology for solving minimization problems by formulating them as min-max problems and then employing majorization-minimization. The monograph lays out the mathematical basis of the approach used to reformulate a minimization problem as a min-max problem. With the prerequisites covered, including multiple illustrations of the formulations for convex and non-convex functions, this work serves as a guide for developing MM4MM-based algorithms for solving non-convex optimization problems in various areas of signal processing. As special cases, we discuss using the majorization-minimization technique to solve min-max problems encountered in signal processing applications and min-max problems formulated using the Lagrangian. Lastly, we present detailed examples of using MM4MM in ten signal processing applications such as phase retrieval, source localization, independent vector analysis, beamforming and optimal sensor placement in wireless sensor networks. The devised MM4MM algorithms are free of hyper-parameters and enjoy the advantages inherited from the use of the majorization-minimization technique such as monotonicity.
\end{abstract}

\begin{flushleft}
minimization; non-convex optimization.
\end{flushleft}

\chapter{Introduction}\label{sec:Intro}

{T}{he} Majorization Minimization (MM) method, with its roots in the 1970s, is a generalization of the classical Expectation-Maximization (EM) method \cite{dempster1977EM}. MM was earlier known by different names like space-alternating generalized EM (SAGE) \cite{fessler1994sage}, optimization transfer \cite{Lange1997OptTrn}, and iterative majorization \cite{heiser1995MMstats}, and it was mainly used in the field of statistics \cite{heiser1987MMstats,bijleveld1991MMstats,de1994MMstats,heiser1995MMstats,lange2000optimizationTrsfr} and image processing \cite{MMapp1_Tom,MMapp2_ImgRes,MMapp3_ImgRec,MMapp4_Tom,mm_IP1,mm_IP2,mm_IP3}. MM is a general algorithmic framework for convex and non-convex problems \cite{lange2000optimizationTrsfr,Lange_MMtut}. \cite{Lange_MMtut,lange2016mm,zhou2019mmVCM,zhou2010mmdmvd,lange2014mmGSP} triggered the attention of researchers from the statistics and signal processing fields to the MM methodology. The work \cite{Lange_MMtut}, in particular, sparked research interest in using the MM algorithmic framework for solving non-convex, non-smooth optimization problems encountered in applications such as compressive sensing \cite{blumensath2007sparse,candes2008Sparse,stoica2012sparse,Mohammadi2016sparse}, covariance estimation \cite{MMapp6_covEst,MMapp7_ScEst,MMapp8_rCovEst,MMapp9_rCovEst}, non-negative matrix factorization \cite{MMapp11_NMF,NMF1,NMF2,NMF3}, sequence design \cite{MMapp15_seqDes,wavDes,seqDes1,seqDes2,seqDes3}, localization in sensor networks \cite{mm_sp2,rLoc2011MM,loc2015MM}, and image processing \cite{MMapp12_TVImgDeconv,MMapp13_wavImgRes,MM2006IP,oliveira2009IP,MM2012IP,MM2012IPsuperRes}.

We will always abbreviate majorization-minimization as MM. However, the abbreviation MM4MM has different (but related) meanings depending on the context:
\begin{itemize}
\item In \cref{sec:minmaxMM4MM}, we use a min-max formulation to derive an MM algorithm and there MM4MM${}={}$min-max for MM.
\item In \cref{sec:maxminMM4MM}, we swap the min and max operators and therefore MM4MM = max-min for MM.
\item Finally in \cref{sec:SplCases}, we get the min-max formulation for free (without any need for a max formulation of the function to be minimized), consequently there MM4MM${}={}$MM for min-max.
\end{itemize}

\section{{MM Summary}}

Employing MM involves two basic steps: majorizing the objective function to create a surrogate function and then minimizing the surrogate function iteratively to solve the min problem. Consider the following optimization problem:
\begin{equation}
\min_{\mathbf{x} \in \mathbb{X}} f(\mathbf{x}),
\label{eqn:minxProb}
\end{equation}
where $\mathbb{X}\subseteq \mathbb{R}^n$ is a non-empty convex set and $f(\mathbf{x}){:}~\mathbb{X}\rightarrow \mathbb{R}$ is a continuous function that is bounded below. To solve \eqref{eqn:minxProb} using MM, a suitable surrogate function ${f_s}(\mathbf{x}\mid\mathbf{x}^t)$ needs to be constructed such that given a feasible point $\mathbf{x}^t\in\mathbb{X}$, the following inequality holds:
\begin{equation}
{f_s}(\mathbf{x}\mid\mathbf{x}^t)
\geq f(\mathbf{x}) + c_t\quad \forall \;\;\mathbf{x}\in\mathbb{X}.
\label{eqn:majStep_MM}
\end{equation}
The constant $c_t \in \mathbb{R}$ ensures that equality is satisfied in \eqref{eqn:majStep_MM} at $\mathbf{x} = \mathbf{x}^t$. In other words, the surrogate function ${f_s}(\mathbf{x}\mid\mathbf{x}^t)$ majorizes $f(\mathbf{x})$ and is equal to $f(\mathbf{x}^t) + c_t$ at $\mathbf{x}=\mathbf{x}^t$.

At the second step, the surrogate function is minimized yielding the next iterate point of the algorithm, that is,
\begin{equation}
\mathbf{x}^{t+1} = \arg\; \min_{\mathbf{x}\in\mathbb{X}}\; {f_s}(\mathbf{x}\mid\mathbf{x}^t).
\end{equation}
This implies,
\begin{equation}
{f_s}(\mathbf{x}^{t+1}\mid\mathbf{x}^t) \leq {f_s}(\mathbf{x}\mid\mathbf{x}^t)\quad \forall\;\mathbf{x} \in\mathbb{X}.
\label{eqn:minSurrgIneq}
\end{equation}
Using \eqref{eqn:minSurrgIneq} and \eqref{eqn:majStep_MM}, we get:
\begin{equation}
\begin{split}
f(\mathbf{x}^{t+1})&\leq{f_s}(\mathbf{x}^{t+1}\mid\mathbf{x}^t) - c_t\\
&\leq {f_s}(\mathbf{x}^t\mid\mathbf{x}^t) -c_t = f(\mathbf{x}^t).
\label{eqn:monotonicfx}
\end{split}
\end{equation}
From \eqref{eqn:monotonicfx} it follows that the sequence of function values $\{f(\mathbf{x}^t)\}$ is monotonically non-increasing.
    What plays a crucial role in developing a computationally efficient MM algorithm for an optimization problem is the construction of a suitable surrogate function. The surrogate function needs to be such that it follows the shape of the objective function as closely as possible and the computational cost of minimizing it is low. To construct a surrogate function, different techniques can be used such as Jensen's inequality, arithmetic-geometric mean inequality, and linearizing a concave function using a first-order Taylor expansion. Many of these techniques are discussed in~\cite{mm}, which also provides a general overview of the MM algorithmic framework along with applications of MM in signal processing, communication and machine learning. The MM technique has been used to solve a wide variety of optimization problems from areas such as signal processing \cite{Emilie_adaptiveSP2017_MM,parEst_MM2019St,abdallah2020sigSubMM,Koyama2021audio_MM}, communication \cite{chnEst_MIMO_mm,Gong_MIMOcomm2020MM,Arora_2020comm_MM,IRS_MIMO_mm,Kundu_RISchEst_MMcnn}, radar and sonar \cite{Ben2019SARIP,Fan2019radarWD,sankuru2020MM_radar,Babu_WavDesMIMO_mm}, machine learning and computer vision \cite{Lin2018rMatFact_MM,hu2020mvSC_mmML,VPoor2020MM_EdgeML,zheng2021MM_l2svr,chapelEcole2021MM_OT,Sandeep_2021StiefelMM_ML, geiping2019MM_EminMethod}, image recovery \cite{RED_mmIP,Marnissi2020HastingAlgoMM_IP,chalvidal2023blockMMIP}, intelligent transportation systems \cite{cachSelfDrCar_MMedgeCom,Huang2022MM_ITS,HCSo2024MM_ITS}, graph learning \cite{Kumar2019_GL_MM,kumar2019sGLlap_MM,kumar2020MM_GL,fatima2022GL_MM,lin2022regGraphMM,Rey2022_GL_MM,Rey2023_GLmotifsMM}, biomedical signal processing \cite{Muduli2017BioMed_MM,deng2018heartSoundSig_MM,mourad2019ecgMM,chatterjee2022spECGbiomed}, and neuroimaging applications \cite{gorriz2019NeuroIm_MM,hashemi2021NeuroIm_MM,Lange2024MM_NeuroimageReg}.

\section{{Need for MM4MM}}

There exist minimization problems for which surrogate functions are difficult to derive or, even if such functions can be found, they are not convenient to deal with from a computational standpoint. Indeed for functions in $\mathbf{x}\in\mathbb{R}^n$ like $\mathbf{x}^{\mathsf{T}}\mathbf{A}\mathbf{x}\log(\mathbf{x}^{\mathsf{T}}\mathbf{A}\mathbf{x})$ where $\mathbf{A}\succ \mathbf{0}$, and $(\log\|\mathbf{x}-\mathbf{a}\|)^2$ where $\mathbf{a}\in \mathbb{R}^n$ finding a suitable surrogate is difficult. On the other hand, for $\log\lvert\mathbf{X}^{\mathsf{T}}\boldsymbol{\Sigma}^{-1}\mathbf{X}\rvert$ (where $\mathbf{X}\in\mathbb{R}^{n \times m}$ and $\boldsymbol{\Sigma} \in\mathbb{S}_{++}^{n}$) a surrogate can be found using a first-order Taylor expansion (below $\mathbf{Y} = \mathbf{X}^{\mathsf{T}}\boldsymbol{\Sigma}^{-1}\mathbf{X}$):
\begin{equation}
    \log\lvert\mathbf{Y}\rvert \leq \log\lvert\mathbf{Y}_t\rvert + \textrm{Tr}(\mathbf{Y}_t^{-1}(\mathbf{Y}-\mathbf{Y}_t)).
\end{equation}
   However, the cost of computing $\mathbf{Y}_t^{-1} = (\mathbf{X}_t^{\mathsf{T}}\boldsymbol{\Sigma}^{-1}\mathbf{X}_t)^{-1}$ at each iteration is usually high.

   To deal with cases such as those above, the proposed MM4MM framework expresses the function $f(\mathbf{x})$ as maximum of an augmented function $g(\mathbf{x},\mathbf{z})$, where $\mathbf{z}$ is an auxiliary variable:
\begin{equation}
    f(\mathbf{x}) = \max_{\mathbf{z}}~g(\mathbf{x},\mathbf{z}).
\label{eqn:fAsMax}
\end{equation}
We call this representation of $f(\mathbf{x})$ the \textit{max formulation}. Finding the function $g(\mathbf{x},\mathbf{z})$ requires ingenuity. This function should be such that suitable surrogate function(s) for the non-convex term(s) of $g(\mathbf{x},\mathbf{z})$ are easy to obtain, making it possible to come up with an efficient MM algorithm. The problem of minimizing $f(\mathbf{x})$ is reformulated as a min-max problem with $g(\mathbf{x},\mathbf{z})$ as the objective function:
\begin{equation}
\min_{\mathbf{x}}~\max_{\mathbf{z}}~g(\mathbf{x},\mathbf{z}).
\label{eqn:minMaxl}
\end{equation}
The problem in \eqref{eqn:minMaxl} can be solved using MM as we explain in \cref{sec:MM4MM}. We show how the proposed MM4MM algorithmic framework can be used to find suitable surrogate functions for the examples mentioned in the previous paragraph in \cref{app:orpr} (for $\mathbf{x}^{\mathsf{T}}\mathbf{A}\mathbf{x}\log(\mathbf{x}^{\mathsf{T}}\mathbf{A}\mathbf{x})$), \cref{app:rss} (for $(\log\|\mathbf{x}-\mathbf{a}\|)^2$), and \cref{app:osp} (for $\log\lvert\mathbf{X}^{\mathsf{T}}\boldsymbol{\Sigma}^{-1}\mathbf{X}\rvert$).

One of the first algorithms using the basic ideas of the MM4MM framework is the PDMM (Primal-Dual Majorization Minimization) \cite{GFatimaPDMM} algorithm that solves the inverse problem of phase retrieval for Poisson noise. \cite{NSahuEoptimal} proposes a special case of MM4MM algorithm using a Lagrangian min-max formulation of the problem of E-optimal experiment design (see \cref{sec:LagSpCase}). Also, \cite{SelesnickTVF} uses the max formulation of the objective function and then employs MM for the problem of total variation filtering. However, this problem is convex, and thus \cite{SelesnickTVF} uses the max formulation only to deal with non-differentiability of $\ell_1$ norm penalty. More details on the applications in \cite{SelesnickTVF,PBabuFPCA,KPanwarHybrid,GFatimaPDMM,AsainiRKLDMM,NSahuEoptimal,rss24ASaini,dfrc24Tang} and on the way in which these works use the MM4MM framework can be found in \cref{sec:applcs}.

\section{Organization}

Before describing the Min-Max framework for MM, we present some preliminary results in Sections~\ref{sec:dualForm} and~\ref{sec:minMaxTh}. In \cref{sec:dualForm}, we describe the max formulation for convex and non-convex functions, along with ten illustrative examples for functions which are found in various signal processing problems. The minimax theorem is discussed in \cref{sec:minMaxTh}. These sections cover the prerequisites for the MM4MM framework, which is described in \cref{sec:MM4MMdesc}. \cref{sec:SplCases} describes two special cases of the MM4MM framework. The bulk of the monograph is \cref{sec:applcs} in which we present detailed derivations of the MM4MM algorithms for ten signal processing applications.

\section{Notation}
Italic letters $(x)$, lower case bold letters $(\mathbf{x})$ and upper case bold letters     $(\mathbf{X})$ denote scalars, vectors and matrices respectively. $|x|$ is the absolute value of $x$ and $|\mathbf{x}|$ is the element-wise absolute value of vector $\mathbf{x}$. The $i$th element of a vector $\mathbf{x}$ is denoted $x_i$. $\mathbf{x}_i$ denotes the $i$th column and $X_{ij}$ the $(i,j)$th element of the matrix $\mathbf{X}$. $\mathbf{I}$ is the identity matrix and $\mathbf{0}$ the zero matrix. $\textrm{Tr}(\mathbf{X})$ and $\lvert\mathbf{X}\rvert$ denote the trace and determinant of the matrix $\mathbf{X}$. $\mathbf{X} \succ \mathbf{0}$ ($\mathbf{X} \succeq \mathbf{0}$) denotes the positive definiteness (positive semi-definiteness) of the matrix $\mathbf{X}$. $\mathrm{vec}(\mathbf{X})$ is the vectorization operator reshaping a matrix $\mathbf{X}$ of size $m \times n$ in a vector of size of $mn \times 1$.

The sets of real numbers and complex numbers are denoted $\mathbb{R}$ and $\mathbb{C}$. $\mathbb{R}_+$ is the set of non-negative real numbers. $\mathbb{S}^n_{+}$ and $\mathbb{S}^n_{++}$ represent the sets of positive semi-definite matrices and positive definite matrices of size $n \times n$. A subspace is represented by $\mathcal{X}$. The indicator function $I_{\mathbb{X}}(\mathbf{x})$ is zero $0$ for all $\mathbf{x}$ lying in the set $\mathbb{X}$ and $\infty$ otherwise. $\mathcal{N}(\mu,\sigma^2)$ denotes Gaussian distribution with mean $\mu$ and variance $\sigma^2$. The notation $\mathbf{x}\sim \mathrm{Poisson}(\lambda)$ means that $\mathbf{x}$ follows a Poisson distribution with mean and variance $\lambda$. The natural logarithm is denoted $\log(\cdot)$ and the logarithm to the base $10$ is denoted $\log_{10}(\cdot)$. $f'(x)$ is the derivative of the function $f(x)$. The symbols $\odot$ and $\oslash$ denote the element-wise multiplication and division of two vectors. $\otimes$ denotes the Kronecker product of two matrices.
   Infimal convolution between two functions $f_1(\mathbf{x})$ and $f_2(\mathbf{y})$ is defined as:
\begin{equation}
    (f_1\square f_2)(\mathbf{x}) = \min_{\mathbf{y}} \;f_1(\mathbf{x}-\mathbf{y}) + f_2(\mathbf{y}).
\end{equation}
    The proximal operator of a scaled function $\textbf{prox}_{\alpha f}(\mathbf{z})$ is defined as follows:
\begin{equation}
    \textbf{prox}_{\alpha f}(\mathbf{z}) = \arg~\min_{\mathbf{x}}~\left(f(\mathbf{x}) + \frac{1}{2\alpha}\|\mathbf{x} - \mathbf{z}\|_2^2\right).
\end{equation}

Superscripts $(\cdot)^{\mathsf{T}}$, $(\cdot)^{\mathsf{H}}$ and $(\cdot)^{-1}$ denote the transpose, conjugate transpose and inverse operations respectively. $(\mathbf{X})^{\frac{1}{2}}$ or $\sqrt{\mathbf{X}}$ refers to the matrix square root of the positive semi-definite matrix $\mathbf{X}\in\mathbb{S}_{+}^{n}$ such that $(\mathbf{X})^{\frac{1}{2}}(\mathbf{X})^{\frac{1}{2}} = \mathbf{X}$. The symbol $\|\mathbf{x}\|_p$ denotes the $\ell_p$ norm of vector $\mathbf{x}\in\mathbb{R}^n$ defined as:
\begin{equation}
    \|\mathbf{x}\|_p = (|x_1|^p + |x_2|^p + \cdots + |x_n|^p)^{1/p},
\end{equation}
which has the following special cases:
\begin{align}
    \|\mathbf{x}\|_1 &= |x_1| +\cdots + |x_n|,\\
    \|\mathbf{x}\|_{\infty} &= \max\{|x_1|, \ldots, |x_n|\}.
\end{align}
The dual norm of any norm $\|\mathbf{x}\|$ is represented by $\|\mathbf{z}\|_*$ and is defined as:
\begin{equation}
    \|\mathbf{z}\|_* = \max\{\mathbf{z}^{\mathsf{T}}\mathbf{x}\mid \|\mathbf{x}\|\leq 1\}.
\end{equation}
Huber norm $\|\mathbf{x}\|_{\text{H}}$  of vector $\mathbf{x}$ is given as follows:
\begin{equation}
\|\mathbf{x}\|_{\text{H}} = \sum_{i=1}^n f_{\alpha}(x_i);\quad f_{\alpha}({x_i}) =  \begin{cases}
    \displaystyle\frac{|{x_i}|^2}{2\alpha} & \text{if}\; |{x_i}|\le \alpha\\[5pt]
    |{x_i}|-\displaystyle\frac{\alpha}{2} & \text{if} \;|{x_i}| > \alpha.
\end{cases}
\end{equation}
Unless otherwise stated, $\|\mathbf{x}\|$ will be used to denote the $\ell_2$ norm of the vector $\mathbf{x}$, i.e., $\|\mathbf{x}\|_2$. For matrix $\mathbf{X}\in\mathbb{R}^{m \times n}$, the $\|\mathbf{X}\|_1$ norm is given by:
\begin{equation}
    \|\mathbf{X}\|_1 = \max_{j=1,\ldots,n}\sum_{i=1}^m |X_{ij}|.
\end{equation}
The nuclear norm, defined as sum of the singular values and denoted $\|\mathbf{X}\|_{2*}$, is given by:
\begin{equation}
\|\mathbf{X}\|_{2*} = \sigma_1(\mathbf{X}) + \cdots + \sigma_r(\mathbf{X}) = \textrm{Tr}(\mathbf{X}^{\mathsf{T}}\mathbf{X})^{\frac{1}{2}},
\end{equation}
where $\{\sigma_i\}$ are the singular values and $r$ is the rank of the matrix $\mathbf{X}$. The spectral norm, defined as the maximum singular value and denoted $\|\mathbf{X}\|_2$, is given by:
\begin{equation}
    \|\mathbf{X}\|_2 = \sigma_{\text{max}}(\mathbf{X}).
\end{equation}

\chapter{Max Formulation}\label{sec:dualForm}

A key step of the MM4MM framework is reformulating a minimization problem as a min-max problem, which necessitates the max formulation of the objective function. Consider the objective function $f{:}~\mathbb{X}\rightarrow\mathbb{R}$ of the minimization problem in \eqref{eqn:minxProb}, and decompose it as follows:
\begin{equation}
    f(\mathbf{x}) = f_{\textrm{c}}(\mathbf{x}) + f_{\textrm{o}}(\mathbf{x}),
\label{eqn:objfDec}
\end{equation}
where $f_{\textrm{o}}(\mathbf{x})$ corresponds to the non-convex term(s) of $f(\mathbf{x})$, for which a max formulation is to be derived, and $f_{\textrm{c}}(\mathbf{x})$ corresponds to the remaining term(s) of $f(\mathbf{x})$ that are either convex, differentiable functions of $\mathbf{x}$ or functions of $\mathbf{x}$ for which a suitable MM surrogate function is not hard to find. The max formulation of $f_{\textrm{o}}(\mathbf{x})$ requires introducing an auxiliary variable $\mathbf{z}$ and expressing $f_{\textrm{o}}(\mathbf{x})$ as maximum, with respect to $\mathbf{z}$, of an augmented function in $(\mathbf{x},\mathbf{z})$:
\begin{equation}
f_{\textrm{o}}(\mathbf{x}) = \max_{\mathbf{z}\in\mathbb{Z}}~g_{\textrm{o}}(\mathbf{x}, \mathbf{z}),
\label{eqn:foMaxRep}
\end{equation}
where $\mathbb{Z} \subseteq \mathbb{R}^n$ and ${g_{\textrm{o}}}{:}~\mathbb{X}\times\mathbb{Z}\rightarrow\mathbb{R}$ is the augmented function. Using \eqref{eqn:objfDec} and \eqref{eqn:foMaxRep}, one can represent  the function $f(\mathbf{x})$ as follows:
\begin{align}
f(\mathbf{x}) &=  f_{\textrm{c}}(\mathbf{x}) + \max_{\mathbf{z} \in \mathbb{Z}}~  g_{\textrm{o}}(\mathbf{x}, \mathbf{z}),\\
    \implies f(\mathbf{x})  &= \max_{\mathbf{z} \in \mathbb{Z}}~ g(\mathbf{x}, \mathbf{z}),
\label{eqn:fMaxRep}
\end{align}
   where
\begin{equation}
    g(\mathbf{x}, \mathbf{z}) = f_{\textrm{c}}(\mathbf{x}) + g_{\textrm{o}}(\mathbf{x}, \mathbf{z}).
\end{equation}

The representation of $f_{\textrm{o}}(\mathbf{x})$ as given in \eqref{eqn:foMaxRep} is facilitated by the theory of conjugate functions in the case of unconstrained problems and the Lagrangian formulation for constrained optimization problems. In some cases, the derivation of the augmented function may involve introducing auxiliary variables using both the conjugate function as well as the Lagrangian formulation. The MM4MM algorithm devised for the dual function beamforming design problem in \cref{DFRC} is one such example. Using the Lagrangian for constrained problems and then employing MM is treated as a special case of the MM4MM in \cref{sec:LagSpCase}.

In this section, we discuss the max formulation using conjugate functions. To lay the ground for the max formulations of non-convex functions, we first discuss the max formulation for convex functions. In the following we also present ten examples of max formulations for non-convex functions encountered in signal processing optimization problems.

\section{Conjugate Function}\label{sec:FenchelDual}

Consider a function $f{:}~\mathbb{X} \rightarrow \mathbb{R}$. The conjugate function of $f(\mathbf{x})$ is defined as:
    \begin{equation}
        f^*(\mathbf{z}) = \max_{\mathbf{x}\in\mathbb{X}} \;\; \mathbf{x}^\mathsf{T}\mathbf{z} - f(\mathbf{x}).
    \label{eqn:FenchelConj}
    \end{equation}
    The function $f^*{:}~\mathbb{Z}\rightarrow\mathbb{R}$ is convex and has a domain $\mathbb{Z}$ consisting of all $\mathbf{z} \in \mathbb{R}^n$ such that $\mathbf{x}^\mathsf{T}\mathbf{z} - f(\mathbf{x}) < \infty$ \cite[Ch. 3]{SBoydCvxBook}.
\begin{lemma}
\label{lem:FenchConjcvx}
The conjugate function $f^*(\mathbf{z})$ is a closed convex function in $\mathbf{z}$ irrespective of whether $f(\mathbf{x})$ is convex or not.
\end{lemma}

    The conjugate of the conjugate function $f^*(\mathbf{z})$ is:
    \begin{equation}
        f^{**}(\mathbf{x}) = \max_{\mathbf{z}\in\mathbb{Z}} \;\; \mathbf{x}^\mathsf{T}\mathbf{z} - f^*(\mathbf{z}).
    \label{eqn:cnjcnjf}
    \end{equation}
    The function $f^{**}(\mathbf{x})$ is identical to $f(\mathbf{x})$ if $f(\mathbf{x})$ is a proper, closed and convex function \cite{RockafellarCvxBook,bertsekas2003convex}.
\begin{lemma}\label{lem:cnjcnjorf}
    If $f(\mathbf{x})$ is proper, closed and convex, then the function $f^{**}(\mathbf{x})$ is equal to $f(\mathbf{x})$.
\end{lemma}
    If $f(\mathbf{x})$ is a closed convex function, then by Lemma~\ref{lem:cnjcnjorf} $f(\mathbf{x})$ can be represented as follows:
\begin{equation}
    f(\mathbf{x}) = \max_{\mathbf{z}\in\mathbb{Z}}\; \{{g}(\mathbf{x},\mathbf{z}) = \mathbf{x}^\mathsf{T}\mathbf{z} - f^*(\mathbf{z})\},
\label{eqn:dualRep_cvx}
\end{equation}
   where $f^*(\mathbf{z})$ is the conjugate function of $f(\mathbf{x})$.
    The max formulation in \eqref{eqn:dualRep_cvx} has been used to solve convex optimization problems in e.g., \cite{SelesnickTVF,DTianPDImageP,AChambolleFenchImaging}. In the following sub-section we make use of \eqref{eqn:dualRep_cvx} to derive the max formulations for some non-convex functions.

\section{Max Formulation for Certain Non-Convex Functions}
\label{sec:FenDualComp}

Let $f{:}~\mathbb{X}\rightarrow\mathbb{R}$ be a given non-convex function that can be written as the composition of two functions $\Hat{\mathbf{f}}{:}~\mathbb{X}\rightarrow\mathbb{Y}$ and $\Tilde{f}{:}~\mathbb{Y}\rightarrow\mathbb{R}$, as follows:
\begin{equation}
\begin{split}
    f(\mathbf{x}) &= \Tilde{f}(\Hat{\mathbf{f}}(\mathbf{x})),\\
                &= \Tilde{f}(\mathbf{y}),
\end{split}
\label{eqn:fxasComp}
\end{equation}
   where $\mathbb{Y}\subseteq\mathbb{R}^n$ and
\begin{equation}
     \mathbf{y} = \Hat{\mathbf{f}}(\mathbf{x}).
\label{eqn:y}
\end{equation}

Assume that $\Tilde{f}(\mathbf{y})$ is a closed convex function and let $\Tilde{f}^*{:}~\mathbb{Z}\rightarrow\mathbb{R}$ be its conjugate function. Then using \eqref{eqn:dualRep_cvx}, $\Tilde{f}(\mathbf{y})$ can be represented as:
\begin{equation}
   \Tilde{f}(\mathbf{y}) = \max_{\mathbf{z}\in\mathbb{Z}}\; \mathbf{y}^\mathsf{T}\mathbf{z} - \Tilde{f}^*(\mathbf{z}).
\label{eqn:dualftily}
\end{equation}
    Because the maximization is with respect to $\mathbf{z}$, we can substitute $\mathbf{y}$ from \eqref{eqn:y} in \eqref{eqn:dualftily} to write:
\begin{equation}
    \Tilde{f}(\Hat{\mathbf{f}}(\mathbf{x})) = \max_{\mathbf{z}\in\mathbb{Z}}~ (\Hat{\mathbf{f}}(\mathbf{x}))^\mathsf{T}\mathbf{z} - \Tilde{f}^*(\mathbf{z}).
\label{eqn:dualftilfhatx}
\end{equation}
Using \eqref{eqn:fxasComp} the following max formulation of $f(\mathbf{x})$ is obtained from \eqref{eqn:dualftilfhatx}:
\begin{equation}
    f(\mathbf{x}) = \max_{\mathbf{z}\in\mathbb{Z}} ~\{{g}(\mathbf{x},\mathbf{z}) = (\Hat{\mathbf{f}}(\mathbf{x}))^\mathsf{T}\mathbf{z} - \Tilde{f}^*(\mathbf{z})\}.
\label{eqn:FenchDualInGnrl}
\end{equation}
In the particular case that $f(\mathbf{x})$ is a convex function, we let $\Hat{\mathbf{f}}(\mathbf{x}) = \mathbf{x}$ so $\Tilde{f}^*(\mathbf{z})$ is simply $f^*(\mathbf{z})$ and thus the representation in \eqref{eqn:FenchDualInGnrl} becomes identical to \eqref{eqn:dualRep_cvx}.
\renewcommand\qedsymbol{$\blacksquare$}
\begin{lemma}
\label{lem:ccvgtilz}
    The function ${g}(\mathbf{x},\mathbf{z})$ in the max formulation of $f(\mathbf{x})$ in \eqref{eqn:FenchDualInGnrl} is concave in $\mathbf{z}$.
\end{lemma}
\begin{proof}
    In \eqref{eqn:FenchDualInGnrl} $-\Tilde{f}^*(\mathbf{z})$ is concave (see Lemma~\ref{lem:FenchConjcvx}).
    Hence, ${g}(\mathbf{x},\mathbf{z})$ is the sum of a concave function and an affine function of $\mathbf{z}$ and thus it is concave in $\mathbf{z}$.
\end{proof}
    The concavity of ${g}(\mathbf{x},\mathbf{z})$ in $\mathbf{z}$ will turn out to be useful later when introducing the Min-Max framework for MM in \cref{sec:MM4MM}.

\section{Illustrative Examples of Max Formulations}\label{subsec:IllEx}

In the following, we present the max formulations for some common convex functions, which will be used to derive max formulations for certain non-convex functions (as explained in \cref{sec:FenDualComp}). The examples of max formulations that follow are directly related to some signal processing applications. The representations of non-convex functions in these examples need interesting re-parameterizations, a fact that will become even more clear in \cref{sec:applcs}.

\subsection{$\ell_1$-Norm}

For a generic norm function $f(\mathbf{x}) = \|\mathbf{x}\|{:}~\mathbb{R}^n \rightarrow \mathbb{R}_+$, the conjugate function $f^*(\mathbf{z})$ is an indicator function of the dual norm unit ball $(\|\mathbf{z}\|_*\le1)$ \cite{SBoydCvxBook}. The max formulation for the norm function $f(\mathbf{x})$ can therefore be written as follows:
\begin{align}
    &\|\mathbf{x}\| = \max_{\mathbf{z}}\; \mathbf{z}^\mathsf{T}\mathbf{x} - f^*(\mathbf{z}),\\
\implies &\|\mathbf{x}\| = \max_{\|\mathbf{z}\|_* \le 1}\; \mathbf{z}^\mathsf{T}\mathbf{x},
\label{eqn:normDualrep}
\end{align}
where $f^*(\mathbf{z}) = \begin{cases}
    0 & \|\mathbf{z}\|_* \le 1\\
    \infty & \text{otherwise}\\
\end{cases}$.
    In particular, the $\ell_1$ norm can be written as:
\begin{equation}
\label{eq:dualL1norm}
    \|\mathbf{x}\|_1 = \max_{|\mathbf{z}|\le \mathbf{1}} \;\mathbf{z}^\mathsf{T}\mathbf{x}.
\end{equation}

This representation of the convex $\ell_1$ norm is mainly used in image processing problems to deal with non-differentiable $\ell_1$ norm regularization terms. Examples of the use of the representation in \eqref{eq:dualL1norm} include total variation based $1-$D signal filtering \cite{SelesnickTVF}, fractional-order ROF (Rudin Osher Fatemi) model based image denoising \cite{DTianPDImageP}, as well as image inpainting and motion estimation \cite{AChambolleFenchImaging}.

\subsection{Nuclear Norm}

Nuclear norm minimization is used in applications like matrix completion and subspace segmentation \cite{NVaswani2018NuclearNorm,RVidal2011NucNorm}. The max formulation for the nuclear norm of the matrix $\mathbf{X} \in \mathbb{R}^{m \times n}$ is:
\begin{equation}
    \|\mathbf{X}\|_{2*} = \max_{\|\mathbf{Z}\|_{2} \le 1}\; \textrm{Tr}(\mathbf{Z}^\mathsf{T}\mathbf{X}),
\end{equation}
where the dual norm of the nuclear norm, $\|\mathbf{Z}\|_{2} = \sigma_1(\mathbf{Z})$, is the spectral norm or the $\ell_2$ norm of the matrix \cite{SBoydCvxBook}. This follows from the max formulation of the norm function given in \eqref{eqn:normDualrep}.

\subsection{Squared Norm}\label{sec:normSq}

The max formulation of the squared $\ell_p$ norm is given by,
\begin{equation}
\label{eqn:dualNormSq}
    \|\mathbf{x}\|_{p}^2 = \max_{\mathbf{z}}\; 2\mathbf{z}^\mathsf{T}\mathbf{x} - \|\mathbf{z}\|_{q}^2,
\end{equation}
where $p^{-1} + q^{-1} = 1$. Here, $\|\mathbf{z}\|_{q}$ is the dual norm of $\|\mathbf{x}\|_{p}$ \cite{SBoydCvxBook} and the conjugate function of $\|\mathbf{x}\|_p^2/2$ is $\|\mathbf{z}\|_q^2/2$.
    The conjugate of the scaled squared norm $\displaystyle\frac{1}{2\alpha}\|\mathbf{x}\|_2^2$ is $\displaystyle\frac{\alpha}{2}\|\mathbf{z}\|_2^2$, where $\alpha$ is a scalar \cite{SBoydCvxBook}.
\begin{example}
\label{egLogSq}
    Consider the non-convex function,
\begin{equation}
    f(\mathbf{r}) = \displaystyle\sum_{i=1}^n (\log\|\mathbf{r}-\mathbf{s}_i\|)^2.
\end{equation}
    The function $f(\mathbf{r})$ may be represented as follows:
    \begin{equation}
    \label{eqn:dualSumLog2}
        \sum_{i=1}^n (\log\|\mathbf{r}-\mathbf{s}_i\|)^2  = \max_{\mathbf{z}}
        ~2\sum_{i=1}^nz_i\log\|\mathbf{r}-\mathbf{s}_i\| - \|\mathbf{z}\|^2.
    \end{equation}
\end{example}
\begin{proof}
    The function in the right hand side of \eqref{eqn:dualSumLog2} is concave in $\mathbf{z}$, and its maximizer with respect to $\mathbf{z}$ is:
\begin{equation}
    z_i^* = \log\|\mathbf{r}-\mathbf{s}_i\| \quad\forall~i=1,\ldots,n.
\end{equation}
    Inserting $z_i^*$ in \eqref{eqn:dualSumLog2}, the right hand side becomes:
\begin{equation}
    \sum_{i=1}^n [ 2z_i^*\log\|\mathbf{r}-\mathbf{s}_i\| - ({z}^*_i)^2 ] = \sum_{i=1}^n (\log\|\mathbf{r}-\mathbf{s}_i\|)^2,
\end{equation}
    which is equal to the left hand side of \eqref{eqn:dualSumLog2}.
\end{proof}

The max formulation of $f(\mathbf{r})$ in \eqref{eqn:dualSumLog2} can be alternatively derived using the conjugate function of the squared $\ell_2$ norm. Defining $\mathbf{x}(\mathbf{r}) = [\log\|\mathbf{r}-\mathbf{s}_1\|, \ldots, \log\|\mathbf{r}-\mathbf{s}_n\|]$, $f(\mathbf{r})$ becomes $\|\mathbf{x}(\mathbf{r})\|^2$. Using the max formulation in \eqref{eqn:dualNormSq} (with $p=2$), we obtain the representation of $f(\mathbf{r})$ in \eqref{eqn:dualSumLog2}.

\subsection{Huber-Norm}

Huber norm or Huber loss function $f_{\alpha}({x})$ is a differentiable, convex function that is quadratic (square term) for values of $\lvert x\rvert$ less than a scalar $\alpha$ and linear (absolute term) for values greater than $\alpha$:
\begin{equation}
f_{\alpha}({x}) =  \begin{cases}
    \displaystyle\frac{|{x}|^2}{2\alpha} & \text{if}\; |{x}|\le \alpha\\[10pt]
    |{x}|-\displaystyle\frac{\alpha}{2} & \text{if} \;|{x}| > \alpha.
\end{cases}
\label{eqn:HuberExp}
\end{equation}
The parameter $\alpha$ controls the transitioning point from quadratic to linear.
    For vector $\mathbf{x}$, the Huber norm $\|\mathbf{x}\|_{\text{H}} = \displaystyle\sum_{i=1}^n f_{\alpha}(x_i)$. The max formulation of Huber norm is given by:
\begin{equation}
\label{eqn:DualRepHuber}
    \|\mathbf{x}\|_{\text{H}} = \max_{|\mathbf{z}|\le \mathbf{1}} \mathbf{x}^\mathsf{T}\mathbf{z} -\frac{\alpha}{2}\|\mathbf{z}\|_2^2,
\end{equation}
\renewcommand\qedsymbol{$\blacksquare$}
\begin{proof}
    The Huber norm can be written as the infimal convolution of $f_1(\mathbf{x}) = \frac{1}{2\alpha}\|\mathbf{x}\|_2^2$ ($\ell_2$ norm) and $f_2(\mathbf{x}) = \|\mathbf{x}\|_1$ ($\ell_1$ norm), i.e.,
    \begin{equation}
    \begin{split}
                \|\mathbf{x}\|_{\text{H}} &=  \;(f_1\square f_2)(\mathbf{x})\\
                   &= \min_{\mathbf{y}} \;f_1(\mathbf{x}-\mathbf{y}) + f_2(\mathbf{y}),\\
                   &= \min_{\mathbf{y}} \;\frac{1}{2\alpha}\|\mathbf{x} - \mathbf{y}\|_2^2 + \|\mathbf{y}\|_1
    \end{split}
    \label{eqn:HuberasInfConv}
    \end{equation}
    The  minimizer of \eqref{eqn:HuberasInfConv} is $\mathbf{y}^* = \textbf{prox}_{\alpha f_2}(\mathbf{x})$ (the proximal operator of $\ell_1$ norm). Substituting $\mathbf{y}^*$ in \eqref{eqn:HuberasInfConv} yields the expression of the Huber norm.

    The conjugate of the infimal convolution of convex functions \cite{RockafellarCvxBook} is given by:
    \begin{equation*}
        (f_1\square f_2)^*(\mathbf{z}) = f_1^*(\mathbf{z}) + f_2^*(\mathbf{z}).
    \end{equation*}
    Hence, the conjugate function of the Huber norm $\|\mathbf{x}\|_{\text{H}}$ is ${\alpha}\|\mathbf{z}\|_2^2/2 + I_{\mathbb{L}_n}(\mathbf{z})$, where $\mathbb{L}_n = \{\mathbf{z}\mid\|\mathbf{z}\|_{\infty}\le 1\}$, and $I_{\mathbb{L}_n}(\mathbf{z})$ is the indicator function of the $\ell_{\infty}$ norm unit ball. Using the conjugate function of $\|\mathbf{x}\|_{\text{H}}$ in the representation of the convex function given in \eqref{eqn:dualRep_cvx}, we get the representation of the Huber norm in \eqref{eqn:DualRepHuber}.
\end{proof}

Regularization using Huber norm is popular in Machine Learning applications such as robust regression, Huber principal component analysis, and point forecasting. For the use of the dual representation for the convex optimization problem of total variation based image denoising see, e.g., \cite{AChambolleFenchImaging}.

\subsection{Log-Determinant}

Consider the function $\Tilde{f}(\mathbf{X})=\log|\mathbf{X}^{-1}|$, where $\mathbf{X} \in \mathbb{S}^n_{++}$. $\Tilde{f}(\mathbf{X})$ is a convex function, whose convex conjugate function is $\Tilde{f}^*(\mathbf{Y})  = \log|-\mathbf{Y}^{-1}| -n$ with $\mathbf{Y}\prec \mathbf{0}$ \cite{SBoydCvxBook}. Consequently, the max formulation of $\Tilde{f}(\mathbf{X})=\log|\mathbf{X}^{-1}|$ is:
\begin{equation}
    \log|\mathbf{X}^{-1}| = \max_{\mathbf{Y}\prec \mathbf{0}}\; \textrm{Tr}(\mathbf{X}\mathbf{Y}) - \log|-\mathbf{Y}^{-1}| + n,
\end{equation}
or equivalently,
\begin{equation}
\label{eqn:logdetInZ}
     \log|\mathbf{X}^{-1}| = \max_{\mathbf{Z}\succ \mathbf{0}} -\textrm{Tr}(\mathbf{X}\mathbf{Z}) - \log|\mathbf{Z}^{-1}| + n,
\end{equation}
    where $\mathbf{Z} = -\mathbf{Y}$.
    We can verify \eqref{eqn:logdetInZ} solving the above maximization problem. The maximizer is $\mathbf{Z}^* = \mathbf{X}^{-1}$. Inserting $\mathbf{Z}^*$ in \eqref{eqn:logdetInZ}, we obtain:
\begin{equation}
\begin{split}
    (-\textrm{Tr}(\mathbf{X}\mathbf{Z}) + \log|\mathbf{Z}| + n)\rvert_{\mathbf{Z} = \mathbf{Z}^*} &= -\textrm{Tr}(\mathbf{X}\mathbf{X}^{-1}) + \log|\mathbf{X}^{-1}| + n,\\
    &= -n - \log|\mathbf{X}| +n,\\
    &= -\log|\mathbf{X}|.
\end{split}
\end{equation}
\begin{example}\label{eg:logWAW}
    Using \eqref{eqn:logdetInZ} the non-convex function $f(\mathbf{W}) =\break -\log|\mathbf{W}^\mathsf{T}\mathbf{A}\mathbf{W}|$, where $\mathbf{W} \in \mathbb{R}^{m \times n}$ $(m \geq n)$ and $\mathbf{A}\succ \mathbf{0}$, can be represented as:
    \begin{equation}
        -\log|\mathbf{W}^\mathsf{T}\mathbf{A}\mathbf{W}| = \max_{\mathbf{Z}\succ \mathbf{0}} \;-\textup{Tr}(\mathbf{Z}\mathbf{W}^\mathsf{T}\mathbf{A}\mathbf{W}) + \log|\mathbf{Z}| + n.
    \label{eqn:dualReplogdetWAW}
    \end{equation}
\end{example}
    \begin{example}\label{eglogWWHiva}
        The non-convex function $f(\mathbf{W}) = -2\log|\mathbf{W}|_{+}$, where $\mathbf{W}\in\mathbb{C}^{n \times n}$ is a non-singular matrix and $\lvert\mathbf{W}\rvert_{+}$ denotes the absolute value of its determinant, can be represented as follows:
    \begin{equation}
        -2\log|\mathbf{W}|_{+} = \max_{\mathbf{Z}\succ\mathbf{0}}~-\textrm{Tr}(\mathbf{W}\mathbf{W}^{\mathsf{H}}\mathbf{Z}) + \log|\mathbf{Z}| + n.
    \label{eqn:dualReplogWWH}
    \end{equation}
    \end{example}
\begin{proof}
    This result follows from the analog of \eqref{eqn:dualReplogdetWAW} for complex-valued matrices.
    Indeed using \eqref{eqn:logdetInZ} and the fact that $f(\mathbf{W}) = -2\log\lvert\mathbf{W}\rvert_{+} = $ $-\log\lvert\mathbf{W}\mathbf{W}^{\mathsf{H}}\rvert$
we obtain the formulation given in \eqref{eqn:dualReplogWWH}.
\end{proof}

\subsection{Negative Logarithm}\label{sec:negLog}

The convex conjugate of $\Tilde{f}(x) = -\log(x)$ for $x >0$ is $f^*(y) = -\log(-y)\break -1$ for $y<0$ \cite{SBoydCvxBook}. Thus the max formulation for the convex function $\Tilde{f}(x)$ is:
\begin{align}
    -\log(x) &= \max_{y <0} \; xy + \log(-y) + 1.\\
\implies -\log(x) &= \max_{z >0} \;\log(z) - zx + 1.
\label{eqn:nlogx}
\end{align}
\begin{example}\label{egRSSnLog}
    The non-convex function $f(\mathbf{x}) = -\log(\Vert \mathbf{x}-\mathbf{s}\Vert^{2} )$ can be expressed as:
\begin{equation}
    -\log(\Vert \mathbf{x}-\mathbf{s}\Vert^{2} )=\underset{z>0}{\textrm{max}}\;\log(z)-z(\Vert \mathbf{x}-\mathbf{s}\Vert^{2} )+1.
\label{eq:8}
\end{equation}
\end{example}
\begin{proof}
    The maximizer $z^*$ of \eqref{eq:8} is given by:
\begin{equation}
z^* = \frac{1}{\|\mathbf{x}-\mathbf{s}\|^2}.
\end{equation}
    Substituting $z^*$ in the right hand side of \eqref{eq:8}, we get the function $f(\mathbf{x})$.
\end{proof}

Alternatively, because $-\log(y)$, where $y = \|\mathbf{x}-\mathbf{s}\|^2$, is convex in $y$, we can use the representation of $-\log(y)$ in \eqref{eqn:nlogx} with $y = \|\mathbf{x}-\mathbf{s}\|^2$ to obtain the max formulation in \eqref{eq:8}. Doing so verifies \eqref{eq:8} using the theory of conjugate functions.

\begin{example}\label{egPDMMnlog}
    The non-convex function $f(\mathbf{w}) =  -\log(\mathbf{w}^\mathsf{T}\mathbf{A}\mathbf{w})$, where $\mathbf{A} \in \mathbb{S}_{++}^{n}$ and $\mathbf{w}\in \mathbb{R}^n$, has the following representation:
    \begin{equation}
        -\log(\mathbf{w}^\mathsf{T}\mathbf{A}\mathbf{w}) = \max_{z> 0} \; \log(z) - z(\mathbf{w}^\mathsf{T}\mathbf{A}\mathbf{w}) + 1.
    \label{eqn:dualRepNlog}
    \end{equation}
\end{example}
\begin{proof}
    The maximizer $z^*$ of \eqref{eqn:dualRepNlog} is given by:
\begin{equation}
    z^* = \frac{1}{\mathbf{w}^\mathsf{T}\mathbf{A}\mathbf{w}}.
\end{equation}
    Inserting ${z}= {z}^*$ in \eqref{eqn:dualRepNlog} proves the result.
\end{proof}

For an alternative derivation, note that the function $f(\mathbf{w})$ can be written as $f(\mathbf{w}) = \Tilde{f}(\Hat{f}(\mathbf{w}))$, where $\Hat{f}(\mathbf{w}) = \mathbf{w}^\mathsf{T}\mathbf{A}\mathbf{w} = x$. Using the representation of the convex $\Tilde{f}(x) = -\log(x)$ in \eqref{eqn:nlogx} with $x = \mathbf{w}^\mathsf{T}\mathbf{A}\mathbf{w}$, we get the max formulation of $f(\mathbf{w})$ in \eqref{eqn:dualRepNlog}.

\subsection{Negative Entropy}

The negative entropy function $\Tilde{f}(x) = x\log x$ for $x>0$ ($\Tilde{f}(x) = 0$ for $x=0$) has the following representation:
\begin{equation}
\label{eqn:dualxlogx}
    x\log(x) = \max_{z} \;xz - e^{(z-1)},
\end{equation}
where $e^{(z-1)}$ is the conjugate function of $x\log x$ \cite{SBoydCvxBook}.

\begin{example}\label{egRKLDxlogx}
    Consider the non-convex function $f(\mathbf{x}) = \mathbf{x}^\mathsf{T}\mathbf{A}\mathbf{x} \cdot\break\log(\mathbf{x}^\mathsf{T}\mathbf{A}\mathbf{x})$, where $\mathbf{A}\in \mathbb{S}_{++}^n$. The function $f(\mathbf{x})$ can be represented as:
    \begin{equation}
    \label{eqn:dualRepNent}
        \mathbf{x}^\mathsf{T}\mathbf{A}\mathbf{x} \log(\mathbf{x}^\mathsf{T}\mathbf{A}\mathbf{x}) = \max_{z} \; z\mathbf{x}^\mathsf{T}\mathbf{A}\mathbf{x} - e^{z-1}.
    \end{equation}
\end{example}
\begin{proof}
    The maximizer $z^*$ of \eqref{eqn:dualRepNent} is:
\begin{equation}
    z^* = \log(\mathbf{x}^{\mathsf{T}}\mathbf{A}\mathbf{x}) + 1,
\end{equation}
    and the corresponding maximum value is equal to $f(\mathbf{x})$.
\end{proof}

The above max formulation of $f(\mathbf{x})$ can also be obtained using \eqref{eqn:dualxlogx}. The function $f(\mathbf{x})$ can be written as the composition of the convex function $\Tilde{f}(x) = x\log(x)$ and the function $\Hat{f}(\mathbf{x}) = \mathbf{x}^{\mathsf{T}}\mathbf{A}\mathbf{x}$. Replacing $x$ in the max formulation of $x\log(x)$ in \eqref{eqn:dualxlogx} with $\mathbf{x}^{\mathsf{T}}\mathbf{A}\mathbf{x}$ gives \eqref{eqn:dualRepNent}.

\subsection{Inverse Function}

The max formulation for the inverse function $\Tilde{f}(x) = 1/x$, where $x>0$ is:
\begin{align}
    \frac{1}{x} &= \max_{y<0} \;xy + 2\sqrt{-y}.\\
\implies    \frac{1}{x} &= \max_{z\geq 0} \;-xz + 2\sqrt{z}.
\label{eqn:invxz}
\end{align}
Here, $z = -y$ and $-2\sqrt{z}$ is the conjugate of $\Tilde{f}(x)$ \cite{SBoydCvxBook}.
\begin{example}
\label{egDFRCmm}
    The max formulation of $f(\mathbf{w}) = 1/(\mathbf{w}^\mathsf{T}\mathbf{A}\mathbf{w})$, where $\mathbf{A}\in \mathbb{S}_{++}^n$, is:
    \begin{equation}
    \label{eqn:InxAx}
        \frac{1}{\mathbf{w}^\mathsf{T}\mathbf{A}\mathbf{w}} = \max_{z\geq 0} ~ -z(\mathbf{w}^\mathsf{T}\mathbf{A}\mathbf{w}) + 2\sqrt{z}.
    \end{equation}
\end{example}
\begin{proof}
    The maximizer $z^*$ of \eqref{eqn:InxAx} is:
\begin{equation}
    z^* = \frac{1}{(\mathbf{w}^\mathsf{T}\mathbf{A}\mathbf{w})^2},
\end{equation}
    and the corresponding maximum value of \eqref{eqn:InxAx} is $f(\mathbf{w})$.
\end{proof}

The representation in \eqref{eqn:InxAx} can also be verified using the conjugate function of $\Tilde{f}(x) = 1/x$. With $f(\mathbf{w}) = \Tilde{f}(\Hat{f}(\mathbf{w}))$ replace $x$ with $\mathbf{w}^\mathsf{T}\mathbf{A}\mathbf{w}$ in the max formulation for the inverse function in \eqref{eqn:invxz} to obtain \eqref{eqn:InxAx}.

\subsection{Scaled Inverse Function}

For a scaled inverse function $f(x) = vf_1(x) = v/x$, where $v$ is a positive scalar and $f_1(x) = 1/x$, the conjugate function is $f^*(z) = vf_1^*(z/v)$ \cite{SBoydCvxBook}, which gives $f^*(z) = -2\sqrt{vz}$ with $z>0$. Therefore, the function $f(x)$ has the following representation:
\begin{equation}
    \frac{v}{x} = \max_{z\geq 0}\; -xz + 2\sqrt{vz}.
\label{eqn:vbyx}
\end{equation}
This representation can be used to derive max formulations for fractional functions like $({\mathbf{x}^\mathsf{T}\mathbf{B}\mathbf{x}})/({\mathbf{x}^\mathsf{T}\mathbf{A}\mathbf{x}})$ as shown in the following example.

\begin{example}
The fractional function $f(\mathbf{x}) = \displaystyle\frac{\mathbf{x}^\mathsf{T}\mathbf{B}\mathbf{x}}{\mathbf{x}^\mathsf{T}\mathbf{A}\mathbf{x}}$, where $\mathbf{A},\mathbf{B} \in \mathbb{S}_{++}^n$, can be represented as follows:
\begin{equation}
\label{eqn:dualFracInv}
\frac{\mathbf{x}^\mathsf{T}\mathbf{B}\mathbf{x}}{\mathbf{x}^\mathsf{T}\mathbf{A}\mathbf{x}} = \max_{z\geq 0}~ -z\mathbf{x}^\mathsf{T}\mathbf{A}\mathbf{x} + 2\;\|\mathbf{B}^{\frac{1}{2}}\mathbf{x}\|\sqrt{z}.
\end{equation}
\end{example}

\begin{proof}
The maximizer $z^*$ of \eqref{eqn:dualFracInv} is given by:
\begin{equation}
    z^* = {\mathbf{x}^\mathsf{T}\mathbf{B}\mathbf{x}}/{(\mathbf{x}^\mathsf{T}\mathbf{A}\mathbf{x})^2},
\end{equation}
    Substituting $z^*$ in \eqref{eqn:dualFracInv} gives $\displaystyle\frac{\mathbf{x^\mathsf{T}\mathbf{B}\mathbf{x}}}{\mathbf{x}^\mathsf{T}\mathbf{A}\mathbf{x}}$.
\end{proof}

Alternatively replacing $x$ with $\mathbf{x}^\mathsf{T}\mathbf{A}\mathbf{x}$ and $v$ with $\mathbf{x}^\mathsf{T}\mathbf{B}\mathbf{x}$ in the max formulation in \eqref{eqn:vbyx} yields the max formulation in \eqref{eqn:dualFracInv}.

\subsection{Trace $(\mathbf{X}^{-1})$}

The max formulation for the convex function $f(\mathbf{X})= \textrm{Tr}(\mathbf{X}^{-1})$, where $\mathbf{X} \in \mathbb{S}_{++}^{n}$, is as follows:
\begin{equation}
    \textrm{Tr}(\mathbf{X}^{-1}) = \max_{\mathbf{Z}\succ \mathbf{0}}\; -\textrm{Tr}(\mathbf{X}\mathbf{Z}) +2\textrm{Tr}(\sqrt{\mathbf{Z}}).
\label{eqn:dualTrXI}
\end{equation}

\begin{proof}
    The conjugate of $f(\mathbf{X})$ is given by:
    \begin{align}
         f^*(\mathbf{Z})&= \max_{\mathbf{X} \succ \mathbf{0}}\; -\textrm{Tr}(\mathbf{X}\mathbf{Z}) -\textrm{Tr}(\mathbf{X}^{-1})\\
         &= - \min_{\mathbf{X} \succ \mathbf{0} }~\textrm{Tr}[(\mathbf{X}^{-1} - \sqrt{\mathbf{Z}})\mathbf{X}(\mathbf{X}^{-1} - \sqrt{\mathbf{Z}}) + 2\sqrt{\mathbf{Z}}] \\
         &= -2\;\textrm{Tr}(\sqrt{\mathbf{Z}}),
    \end{align}
    where $\mathbf{Z}\succ\mathbf{0}$. Using this conjugate function, $f(\mathbf{X})$ can be represented as:
    \begin{equation}
        \textrm{Tr}(\mathbf{X}^{-1}) = \max_{\mathbf{Z}\succ \mathbf{0}}\; -\textrm{Tr}(\mathbf{X}\mathbf{Z}) +2\textrm{Tr}(\sqrt{\mathbf{Z}}).
    \end{equation}
\end{proof}
\begin{example}\label{egTrHybrid}
    For the function $f(\mathbf{W}) = \textup{Tr}[(\mathbf{W}^\mathsf{T}\mathbf{A}\mathbf{W})^{-1}]$, the max formulation is as follows:
\begin{equation}
    \textup{Tr}[(\mathbf{W}^\mathsf{T}\mathbf{A}\mathbf{W})^{-1}] = \max_{\mathbf{Z}\succ \mathbf{0}} -\textup{Tr}(\mathbf{Z}\mathbf{W}^\mathsf{T}\mathbf{A}\mathbf{W}) + 2\textup{Tr}(\sqrt{\mathbf{Z}}).
\label{eqn:dualTrYIn}
\end{equation}
\end{example}
\begin{proof}
    The maximizer $\mathbf{Z}^*$ of \eqref{eqn:dualTrYIn} is:
\begin{equation}
    \mathbf{Z}^* = (\mathbf{W}^\mathsf{T}\mathbf{A}\mathbf{W})^{-2},
\end{equation}
    and the corresponding maximum is $f(\mathbf{W}) = \textup{Tr}[(\mathbf{W}^\mathsf{T}\mathbf{A}\mathbf{W})^{-1}]$.
\end{proof}

Alternatively replacing $\mathbf{X}$ with $\mathbf{W}^\mathsf{T}\mathbf{A}\mathbf{W}$ in \eqref{eqn:dualTrXI} yields the representation of $f(\mathbf{W})$ in \eqref{eqn:dualTrYIn}.

\subsection{Maximum Element of a Vector}

The conjugate function of an indicator function $I_{\mathbb{X}}(\mathbf{x})$, where $\mathbb{X}$ is a non-empty convex set, is
    the support function of the set $\mathbb{X}$ \cite{SBoydCvxBook}:
\begin{equation}
    S_{\mathbb{X}}(\mathbf{z}) = \displaystyle\max_{\mathbf{x} \in \mathbb{X}} ~\mathbf{x}^\mathsf{T}\mathbf{z}
\end{equation}
    The conjugacy of the support and indicator functions of a convex set can be used for the representation of many functions. The maximum element of a vector can be represented using the indicator function of the $(n-1)$-dimensional probability simplex $\mathbb{Z} = \{\mathbf{z}\mid z_i \ge 0, \mathbf{z}^\mathsf{T}\mathbf{1}=\mathbf{1}\}$ as follows:
    \begin{align}
    \label{eqn:dualMaxEl}
        \max_{i}\; x_i = &\max_{\mathbf{z}}\; \mathbf{x}^\mathsf{T}\mathbf{z}\\
         &\;\;\mathrm{s.t.}\; \{z_i \ge 0\}, \mathbf{z}^\mathsf{T}\mathbf{1} = 1.\nonumber
    \end{align}
\begin{proof}
    The support function of the convex set $\mathbb{Z} = \{\mathbf{z}\mid z_i \ge 0, \mathbf{z}^\mathsf{T}\mathbf{1}=\mathbf{1}\}$ is $S_{\mathbb{Z}} = \mathrm{max}\{x_i\mid i = 1,\ldots,n\}$ \cite{RockafellarCvxBook}. Since the set $\mathbb{Z}$ is a convex set, the conjugate function $S_{\mathbb{Z}}^*$ of the support function $S_{\mathbb{Z}}$ is the indicator function of the set ${\mathbb{Z}}$. Hence, the support function $S_{\mathbb{Z}}$ has the following representation:
    \begin{equation}
        S_{\mathbb{Z}} = \max_{\mathbf{z}} \mathbf{x}^\mathsf{T}\mathbf{z} - I_{\mathbb{Z}},
    \end{equation}
    which gives \eqref{eqn:dualMaxEl}.
\end{proof}
\begin{example}
\label{egMaxElt}
    The maximum of $n$ functional values $\{f_i(\mathbf{w})\}$ has the following max formulation:
 \begin{align} \label{eqn:selmaxf}
\begin{array}{r@{\quad}l}
\mathop{\max}\limits_{i}~f_i(\mathbf{w}) = \mathop{\max}\limits_{\mathbf{z}} & \displaystyle \sum_{i=1}^n z_if_i(\mathbf{w})\\[10pt]
\hspace*{-1pc}\mathrm{s.t.} &  z_i\ge0, \;\mathbf{z}^\mathsf{T}\mathbf{1} = 1.
\end{array}
\end{align}
\end{example}
\begin{proof}
The maximizer $\mathbf{z}^*$ of the right hand side of \eqref{eqn:selmaxf} is the vector whose $i$th element corresponding to the index of the maximum value of $\{f_i(\mathbf{w})\}$ is $1$ and the rest of the elements are zero, thus \eqref{eqn:selmaxf} follows.
\end{proof}

\begin{sidewaystable}
\caption{Representations of ten non-convex functions and the related signal processing applications}\label{tab:IllEx}
{\fontsize{7}{9}\selectfont\tabcolsep=0pt\begin{tabular*}{\columnwidth}{@{\extracolsep{\fill}}lll}%
  \toprule
  \textbf{Non-Convex Function} & \multicolumn{1}{c}{\textbf{Max Formulation}} & \multicolumn{1}{c}{\textbf{Application}}\\
\midrule
      $\displaystyle\sum_{i=1}^n (\log\|\mathbf{r}-\mathbf{s}_i\|)^2 $  & $\displaystyle\max_{\mathbf{z}}~2\displaystyle\sum_{i=1}^nz_i\log\|\mathbf{r}-\mathbf{s}_i\| - \|\mathbf{z}\|^2$ & Received Signal Strength based Source Localization \cite{rss24ASaini}\\
      \midrule
      $-\log(\mathbf{w}^\mathsf{T}\mathbf{A}\mathbf{w})$ & $\displaystyle\max_{z> 0} \; \log(z) - z\mathbf{w}^\mathsf{T}\mathbf{A}\mathbf{w} + 1$ & Phase Retrieval for Poisson Noise \cite{GFatimaPDMM}\\
    \midrule
    $-2\log|\mathbf{W}|_{+}$ & $\displaystyle\max_{\mathbf{Z}\succ\mathbf{0}}~-\textrm{Tr}(\mathbf{W}\mathbf{W}^{\mathsf{H}}\mathbf{Z}) +\log|\mathbf{Z}| + n$ & Independent Vector Analysis \cite{ono2012auxIVA}\\
    \midrule
    $-\log(\Vert \mathbf{x}-\mathbf{s}\Vert^{2} )$& $\displaystyle\max_{z>0}\;\log(z)-z(\Vert \mathbf{x}-\mathbf{s}\Vert^{2} )+1$ & Received Signal Strength based Source Localization \cite{rss24ASaini}\\
    \midrule
    $-\log|\mathbf{W}^\mathsf{T}\mathbf{A}\mathbf{W}|$ &
    $\displaystyle\max_{\mathbf{Z}\succ \mathbf{0}} \;-\textrm{Tr}(\mathbf{Z}\mathbf{W}^\mathsf{T}\mathbf{A}\mathbf{W}) + \log|\mathbf{Z}| +n$ & Optimal Sensor Placement (D-Optimal Design)  \cite{KPanwarHybrid}\\
    \midrule
     $\mathbf{x}^\mathsf{T}\mathbf{A}\mathbf{x} \log(\mathbf{x}^\mathsf{T}\mathbf{A}\mathbf{x})$ & $\displaystyle\max_{z}~ z\mathbf{x}^\mathsf{T}\mathbf{A}\mathbf{x} - e^{z-1}$ & Outlier Robust Phase Retrieval \cite{AsainiRKLDMM}\\
     \midrule
     $ {1}/{\mathbf{w}^\mathsf{T}\mathbf{A}\mathbf{w}}$ & $ \displaystyle\max_{z>0} ~ -z\mathbf{w}^\mathsf{T}\mathbf{A}\mathbf{w} + 2\sqrt{z}$ & Dual-Function Radar Communication Beamforming Design \cite{dfrc24Tang}\\
     \midrule     ${\mathbf{x^\mathsf{T}\mathbf{B}\mathbf{x}}}/{\mathbf{x}^\mathsf{T}\mathbf{A}\mathbf{x}}$ & $\displaystyle\max_{z>0}~ -z\mathbf{x}^\mathsf{T}\mathbf{A}\mathbf{x} + 2\;\|\mathbf{B}^{\frac{1}{2}}\mathbf{x}\|\sqrt{z}$ & Beamforming \cite{Wu2018BFAAF}\\
     \midrule
     $\textup{Tr}[(\mathbf{W}^\mathsf{T}\mathbf{A}\mathbf{W})^{-1}]$ & $\displaystyle\max_{\mathbf{Z}\succ \mathbf{0}} -\textup{Tr}(\mathbf{Z}\mathbf{W}^\mathsf{T}\mathbf{A}\mathbf{W}) + 2\textup{Tr}(\sqrt{\mathbf{Z}})$ & Optimal Sensor Placement (A-optimal Design) \cite{KPanwarHybrid}\\
     \midrule
     $\displaystyle\max_{i}~f_i(\mathbf{w})$ &  $\displaystyle\max_{\substack{z_i \ge 0,\\ \mathbf{z}^\mathsf{T}\mathbf{1}=\mathbf{1}}}~ \sum_{i=1}^n z_if_i(\mathbf{w})$ & Fair Principal Component Analysis \cite{PBabuFPCA}\\
    \bottomrule
    \end{tabular*}}{}
\end{sidewaystable}

Alternatively we can use \eqref{eqn:dualMaxEl} to obtain the representation in \eqref{eqn:selmaxf}.

\cref{tab:IllEx} summarizes the max formulations for the non-convex functions in the previous examples, each along with a related signal processing application. A max formulation converts a minimization problem to a min-max problem. However, the so-obtained min-max problem might not be convex-concave. In such cases, the principle of MM comes to the rescue$\colon$ the problematic terms of $g(\mathbf{x},\mathbf{z})$ are suitably majorized to get a convex-concave function. The next section shows how the MM can be employed to solve the min-max problems obtained using the max formulations in \cref{tab:IllEx}.

\chapter{Min-Max Framework for Majorization Minimization}\label{sec:MM4MM}

Before discussing the MM4MM algorithmic framework, we present the minimax theorem.

\section{Minimax Theorem}\label{sec:minMaxTh}

In general terms the minimax theorem \cite{MSionMinimax,ha1981noncompact} specifies the requirements for the equivalence of the min-max and max-min problems. We present the minimax theorem for a function $g(\mathbf{x},\mathbf{z})$ which is convex in the variable $\mathbf{x}$ and concave in the auxiliary variable $\mathbf{z}$.


\begin{theorem}\label{MinimaxTh}
    If $g(\mathbf{x},\mathbf{z})$ defined on $ \mathbb{X} \times \mathbb{Z}$ is convex in $\mathbf{x}$ and concave in $\mathbf{z}$ and $\mathbb{X}$ and $\mathbb{Z}$ are convex sets, then the max and min operators in the min-max problem can be swapped:
    \begin{equation}
        \min_{\mathbf{x} \in \mathbb{X}} \; \; \max_{\mathbf{z} \in \mathbb{Z}} \;\;g(\mathbf{x},\mathbf{z}) = \max_{\mathbf{z} \in \mathbb{Z}} \;\;\min_{\mathbf{x}\in \mathbb{X}}\;\; g(\mathbf{x},\mathbf{z}),
    \label{eqn:minimaxTh}
    \end{equation}
\end{theorem}
See, e.g., \cite{MSionMinimax,ha1981noncompact}. For a simple proof of the equality in \eqref{eqn:minimaxTh} under additional differentiability conditions, see the Appendix.

In the following subsection, we discuss how the max representations of the previous section, the minimax theorem and MM can be used to develop an MM4MM algorithm.

\section{MM4MM}\label{sec:MM4MMdesc}

Consider the optimization problem in \eqref{eqn:minxProb} where $f(\mathbf{x}){:}~\mathbb{X} \rightarrow \mathbb{R}$ and $\mathbb{X}$ is a convex set. Assume that the function $f(\mathbf{x})$ can be expressed as maximum of a function $g(\mathbf{x},\mathbf{z}){:}~\mathbb{X}\times\mathbb{Z}\rightarrow\mathbb{R}$, where $\mathbb{Z}$ is a convex set and $\mathbf{z}$ is an auxiliary variable:
\begin{equation}
    f(\mathbf{x}) = \max_{\mathbf{z} \in \mathbb{Z}}\; {g}(\mathbf{x}, \mathbf{z}).
\label{eqn:fasmaxg}
\end{equation}
    This max formulation of the objective function $f(\mathbf{x})$ was discussed in \cref{sec:dualForm}. Another way in which we can write $f(\mathbf{x})$ as in \eqref{eqn:fasmaxg} is using the Lagrangian for a constrained problem (see \cref{sec:LagSpCase}).

The set $\mathbb{Z}$ must be such that ${g}(\mathbf{x},\mathbf{z})<\infty$ for all $\mathbf{z} \in \mathbb{Z}$. If ${g}(\mathbf{x},\mathbf{z})$ is unbounded above for some $\mathbf{z} \in \mathbb{Z}$, then \eqref{eqn:fasmaxg} does not hold. Moreover, the function ${g}(\mathbf{x},\mathbf{z})$ in \eqref{eqn:fasmaxg} is chosen such that it is concave in $\mathbf{z}$. For the case of the max formulation using the conjugate function, this property follows from the concavity of the conjugate function (see Lemma~\ref{lem:ccvgtilz} in \cref{sec:dualForm}). For the case of the max formulation of $f(\mathbf{x})$ using the Lagrangian, ${g}(\mathbf{x},\mathbf{z})$ is affine in $\mathbf{z}$ and thus is a concave function of $\mathbf{z}$.

Using \eqref{eqn:fasmaxg} the non-convex optimization problem in \eqref{eqn:minxProb} is re-formulated as a min-max problem:
\begin{equation}
    \min_{\mathbf{x}\in\mathbb{X}}\; \max_{\mathbf{z}\in \mathbb{Z}} \; {g}(\mathbf{x},\mathbf{z}).
\label{eqn:minmaxProb}
\end{equation}
As indicated above, in many cases the function ${g}(\mathbf{x},\mathbf{z})$ can be chosen such that it is concave in $\mathbf{z}$, but it might not be convex in $\mathbf{x}$. To deal with the non-convexity of ${g}(\mathbf{x},\mathbf{z})$ in $\mathbf{x}$, we utilize the MM principle. Majorizing ${g}(\mathbf{x},\mathbf{z})$ in $\mathbf{x}$ at a point $\mathbf{x}^t \in \mathbb{X}$, a surrogate function $g_s(\mathbf{x},\mathbf{z}\mid\mathbf{x}^t)$ is constructed such that:
\begin{equation}
    g_s(\mathbf{x},\mathbf{z}\mid\mathbf{x}^{t})\geq {g}(\mathbf{x},\mathbf{z}).
\label{eqn:gasmajx}
\end{equation}
    The equality in \eqref{eqn:gasmajx} must hold for $\mathbf{x} = \mathbf{x}^t$:
\begin{equation}
     g_s(\mathbf{x}^t,\mathbf{z}\mid\mathbf{x}^{t})={g}(\mathbf{x}^t,\mathbf{z}).
\label{eqn:gasEqxt}
\end{equation}
We refer to \cite{mm} for guidelines on constructing surrogate functions. Construction of an appropriate surrogate function requires ingenuity and should be such that the function ${g_s}(\mathbf{x},\mathbf{z})$ is convex in $\mathbf{x}$.
    The min-max problem for the surrogate objective function $g_s(\mathbf{x},\mathbf{z}\mid\mathbf{x}^t)$ (convex in $\mathbf{x}$ and concave in $\mathbf{z}$) is as follows:
\begin{equation}
    \min_{\mathbf{x}\in \mathbb{X}}\; \max_{\mathbf{z}\in \mathbb{Z}} \; g_s(\mathbf{x},\mathbf{z}\mid\mathbf{x}^t).
\label{eqn:minimaxProbxt}
\end{equation}
We can solve the above min-max problem using one of the two approaches described in the following subsections. The selection of the approach for the algorithm development depends on factors such as existence of closed-form solutions, computational requirements, and algorithm performance. Both approaches follow the principle of MM but they differ in the order of min and max operations.

\subsection{Min-Max for MM}\label{sec:minmaxMM4MM}
For the min-max problem in \eqref{eqn:minimaxProbxt}, the inner maximization problem is:
\begin{equation}
    \max_{\mathbf{z}\in \mathbb{Z}}\; g_s(\mathbf{x},\mathbf{z}\mid\mathbf{x}^t),
\end{equation}
and its solution (assumed to have a closed form) is denoted:
\begin{equation}
    \Tilde{\mathbf{z}}(\mathbf{x}\mid\mathbf{x}^t) = \arg\;\max_{\mathbf{z}\in\mathbb{Z}}\; g_s(\mathbf{x},\mathbf{z}\mid\mathbf{x}^t).
\label{eqn:zTilxt}
\end{equation}
Substituting the maximizer \eqref{eqn:zTilxt} in the min-max problem \eqref{eqn:minimaxProbxt}, we get the following problem:
\begin{equation}
\begin{split}
    \min_{\mathbf{x}\in\mathbb{{X}}}\; &\{d(\mathbf{x}\mid\mathbf{x}^t)= g_s(\mathbf{x},\Tilde{\mathbf{z}}(\mathbf{x}\mid\mathbf{x}^t)\mid\mathbf{x}^t)\}.
\end{split}
\label{eqn:mindx}
\end{equation}
     The function $d(\mathbf{x}\mid\mathbf{x}^t)$ is convex, see e.g., Lemma \ref{fact_dcvx} in the Appendix, and it majorizes $f(\mathbf{x})$ at $\mathbf{x}^t$ as shown in the following lemma.
\begin{lemma}\label{upperboundMM4MM}
$d(\mathbf{x}\mid\mathbf{x}^t)$ majorizes $f(\mathbf{x})$ at $\mathbf{x} = \mathbf{x}^{t}$.
\end{lemma}
\begin{proof}
     The surrogate function majorizes the augmented function for all $\mathbf{x}\in\mathbb{X}$ and $\mathbf{z}\in\mathbb{Z}$ (see \eqref{eqn:gasmajx}):
\begin{equation}
    g_s(\mathbf{x},\mathbf{z}\mid\mathbf{x}^{t}) \geq {g}(\mathbf{x},\mathbf{z}).
\end{equation}
Therefore
\begin{equation}
     d(\mathbf{x}\mid\mathbf{x}^t) = g_s(\mathbf{x},\Tilde{\mathbf{z}}(\mathbf{x}\mid\mathbf{x}^t)\mid\mathbf{x}^{t}) \geq \max_{\mathbf{z}\in\mathbb{Z}}~{g}(\mathbf{x},\mathbf{z}) = f(\mathbf{x}).
\end{equation}
    Moreover the equality holds at $\mathbf{x}=\mathbf{x}^t$. Indeed we have from \eqref{eqn:gasEqxt} that:
\begin{align}
    d(\mathbf{x}^t\mid\mathbf{x}^t)  = g_s(\mathbf{x}^t,\Tilde{\mathbf{z}}(\mathbf{x}^t\mid\mathbf{x}^t)\mid\mathbf{x}^{t})=\max_{\mathbf{z}\in\mathbb{Z}}~{g}(\mathbf{x}^t,\mathbf{z}) =f(\mathbf{x}^t), ~\mathbf{x}^t\in \mathbb{X},
\label{eqn:dUbfEq}
\end{align}
    and the proof is concluded.
\end{proof}

The MM update $\mathbf{x}^{t+1}$ is obtained as the solution of the convex problem in \eqref{eqn:mindx}:
\begin{equation}
    \mathbf{x}^{t+1} = \arg \; \min_{\mathbf{x}\in\mathbb{{X}}}\; d(\mathbf{x}\mid\mathbf{x}^t).
\label{eqn:xiter_minmax}
\end{equation}
    The main steps of the MM4MM algorithm using \eqref{eqn:xiter_minmax} are summarized in \cref{alg:AlgoMinMax4MM}.
\begin{algorithm}[!htb]
\caption{Min-Max for MM}
\label{alg:AlgoMinMax4MM}
    \KwIn{$\mathbf{x}^0 \in \mathbb{X} (t=0),\;\eta$}
    \Repeat{\emph{Convergence criterion:} $(f(\mathbf{x}^{t-1}) - f(\mathbf{x}^t))/f(\mathbf{x}^t)\leq \eta$}{
        Find the surrogate function $d(\mathbf{x}\mid\mathbf{x}^t)$\;
        Obtain $\mathbf{x}^{t+1}$ using \eqref{eqn:xiter_minmax}\;
        $t+1\gets t$\;
    }
    \KwOut{$\hat{\mathbf{x}}$}
\end{algorithm}

\par

\subsection{Max-Min for MM}\label{sec:maxminMM4MM}

Consider again the min-max problem in \eqref{eqn:minimaxProbxt}. Since $g_s(\mathbf{x},\mathbf{z}\mid\mathbf{x}^t)$ is a convex-concave function, the min and max operators can be swapped using the minimax theorem (see \cref{MinimaxTh} in \cref{sec:minMaxTh}) which results in the following max-min problem:
\begin{equation}
    \max_{\mathbf{z}\in \mathbb{Z}}\; \min_{\mathbf{x}\in\mathbb{X}}\; g_s(\mathbf{x},\mathbf{z}\mid\mathbf{x}^t).
\label{eqn:maxminProbxt}
\end{equation}
    We consider \eqref{eqn:maxminProbxt} in lieu of \eqref{eqn:minimaxProbxt} whenever the inner minimization in \eqref{eqn:maxminProbxt} has a closed-form solution but such a solution for the inner maximization of \eqref{eqn:minimaxProbxt} is difficult to find.
    We assume that the inner minimization problem has a closed-form solution:
\begin{equation}
    \Tilde{\mathbf{x}}(\mathbf{z}\mid\mathbf{x}^t) = \arg \;\min_{\mathbf{x} \in \mathbb{X}}\; g_s(\mathbf{x},\mathbf{z}\mid\mathbf{x}^t).
\label{eqn:xTil_ming}
\end{equation}
    The corresponding minimum function is given by:
\begin{equation}
    h(\mathbf{z}\mid\mathbf{x}^t) = g_s(\Tilde{\mathbf{x}}(\mathbf{z}\mid\mathbf{x}^t),\mathbf{z}\mid\mathbf{x}^t),
\label{eqn:hzdefn}
\end{equation}
        which is concave in $\mathbf{z}$ as shown in the following lemma.
\begin{lemma}
    The function $h(\mathbf{z}\mid\mathbf{x}^t)$ is concave in $\mathbf{z}$.
\end{lemma}
\begin{proof}
    Using \eqref{eqn:xTil_ming} and \eqref{eqn:hzdefn}, we have:
\begin{equation}
    h(\mathbf{z}\mid\mathbf{x}^t) = \min_{\mathbf{x} \in \mathbb{X}}~ g_s(\mathbf{x},\mathbf{z}\mid\mathbf{x}^t),
\end{equation}
    which can be re-written as:
\begin{equation}
    h(\mathbf{z}\mid\mathbf{x}^t) = -\Big[\max_{\mathbf{x} \in \mathbb{X}}~ (-g_s(\mathbf{x},\mathbf{z}\mid\mathbf{x}^t))\Big].
\end{equation}
    The fact that the above function is concave in $\mathbf{z}$ follows from Lemma \ref{fact_dcvx} in the Appendix.
\end{proof}

\begin{algorithm}[!b]
\caption{Max-Min for MM}
\label{alg:AlgoMaxMin4MM}
    \KwIn{$\mathbf{x}^0 \in \mathbb{X} (t=0),\;\eta$}
    \Repeat{\emph{Convergence criterion:} $(f(\mathbf{x}^{t-1}) - f(\mathbf{x}^t))/f(\mathbf{x}^t)\leq \eta$}{

        Obtain $\mathbf{z}^{t+1}$ using \eqref{eqn:zopt_maxh}\;
        Obtain $\mathbf{x}^{t+1}$ using \eqref{eqn:xopt_mingzopt}\;
        $t+1\gets t$\;

    }
    \KwOut{$\hat{\mathbf{x}}$}
\end{algorithm}

Using $h(\mathbf{z}\mid\mathbf{x}^t)$ in \eqref{eqn:maxminProbxt} leads to the following maximization problem:
\begin{equation}
    \max_{\mathbf{z}\in \mathbb{Z}} \; h(\mathbf{z}\mid\mathbf{x}^t),
\label{eqn:maxhz}
\end{equation}
    which is convex. Let $\mathbf{z}^{t+1}$ be the solution of \eqref{eqn:maxhz}:
\begin{equation}
    \mathbf{z}^{t+1} = \arg \;\max_{\mathbf{z}\in\mathbb{Z}} \; h(\mathbf{z}\mid\mathbf{x}^t).
\label{eqn:zopt_maxh}
\end{equation}
    Then the solution $\mathbf{x}$ to \eqref{eqn:maxminProbxt} is given by (see \eqref{eqn:xTil_ming}):
\begin{equation}
    \mathbf{x}^{t+1} = \Tilde{\mathbf{x}}(\mathbf{z}^{t+1}).
\label{eqn:xopt_mingzopt}
\end{equation}

    The main steps of the MM4MM algorithm based on \eqref{eqn:zopt_maxh} and \eqref{eqn:xopt_mingzopt} are summarized in \cref{alg:AlgoMaxMin4MM}.

    The above two MM4MM algorithms are based on the MM principle and they both monotonically decrease $f(\mathbf{x})$ at each iteration. Therefore their convergence follows from the general properties of the MM algorithm \cite{jacobson2007TheorConvMM,razaviyayn2013CongMM}.

\chapter{Special Cases}\label{sec:SplCases}

In this section we discuss two cases in which we get the max formulation for free.

\section{Constrained Minimization Problem}\label{sec:LagSpCase}

A constrained minimization problem can be formulated as an unconstrained min-max problem using the Lagrangian. To be specific, consider the following problem:
\begin{equation}
\begin{split}
    \min_{\mathbf{x}}~&f_0(\mathbf{x})\\
    \mathrm{s.t.}~&\Tilde{f}_i(\mathbf{x}) \leq 0, \quad i=1,\ldots,m\\
                 &\Bar{f}_j(\mathbf{x})= 0, \quad j=1,\ldots,p.
\end{split}
\label{eqn:constMinProb}
\end{equation}
    Using the vectors $\mathbf{\Tilde{f}} = [\Tilde{f}_1(\mathbf{x}),\ldots,\Tilde{f}_m(\mathbf{x})]^{\mathsf{T}}$ and $\mathbf{\Bar{f}} = [\Bar{f}_{1}(\mathbf{x}),\ldots,\Bar{f}_{p}(\mathbf{x})]^{\mathsf{T}}$, the min-max formulation of \eqref{eqn:constMinProb} using the Lagrangian is given by:
\begin{equation}
    \min_{\mathbf{x}}~\max_{\{z_i\}\geq0,\;\mathbf{q}}~\{g(\mathbf{x},\mathbf{z},\mathbf{q}) = f_0(\mathbf{x}) + \mathbf{z}^{\mathsf{T}}\mathbf{\Tilde{f}} + \mathbf{q}^{\mathsf{T}}\mathbf{\Bar{f}}\},
\label{eqn:minmaxLag}
\end{equation}
    where $\mathbf{z}$ and $\mathbf{q}$ are the Lagrange multiplier vectors associated with the inequality and equality constraints. The function $g(\mathbf{x},\mathbf{z},\mathbf{q})$ is concave in $\mathbf{z}$ and $\mathbf{q}$ but possibly non-convex in $\mathbf{x}$. If a suitable majorizer for $f_0(\mathbf{x})$ is found, then the min-max problem \eqref{eqn:minmaxLag} can be solved using the MM4MM procedure.

\section{Worst-Case Minimization Problem}\label{sec:MMmin-max}
In some applications, the optimization problem consists of minimizing the worst-case objective function:
\begin{equation}
    \min_{\mathbf{x}\in\mathbb{X}}~\{f(\mathbf{x}) = \max_{k\in [1,\ldots,K]}~f_k(\mathbf{x})\},
\label{eqn:worstCaseObj}
\end{equation}
    where the maximum of the $\{f_k(\mathbf{x})\}$ represents the worst case.
    The problem \eqref{eqn:worstCaseObj} can be equivalently written as (see \eqref{eqn:selmaxf}):
\begin{equation}
    \min_{\mathbf{x}\in\mathbb{X}}~\left\{\max_{\mathbf{z}\in \mathbb{Z}}~g(\mathbf{x},\mathbf{z}) = \sum_{k=1}^K z_kf_k(\mathbf{x})\right\},
\label{eqn:minmaxObj}
\end{equation}
    where $\mathbb{Z} = \{z_k\mid z_k\geq 0, \mathbf{1}^{\mathsf{T}}\mathbf{z} = 1\}$.
    The above problem can be solved using the MM4MM algorithm.

\chapter{Signal Processing Applications of MM4MM}\label{sec:applcs}
    In this section, we show how the MM4MM algorithmic framework can be employed to solve ten signal processing optimization problems. Each subsection covers an MM4MM algorithm developed for a particular signal processing application.
    Most of these algorithms have been published in different forms and we cast them here in the unified framework of MM4MM. For performance studies of these algorithms we refer to the works that are cited in the discussion on the corresponding application.

\section{Total Variation Filtering}\label{applTVF}

Total variation (TV) based filtering is used for image denoising and image reconstruction \cite{rudin1992TVF,chambolle2005total}. The idea of using TV for the removal of noise from images was introduced in \cite{rudin1992TVF}. The authors of \cite{SelesnickTVF} proposed a method similar to the MM4MM algorithm for the problem of TV filtering of one-dimensional signals. When the signal $\mathbf{x}\in\mathbb{R}^n$ is contaminated by additive white Gaussian noise $\boldsymbol{\eta}\in\mathbb{R}^n$, the observed signal $\mathbf{y}$ is given by:
\begin{equation}
    \mathbf{y} = \mathbf{x} + \boldsymbol{\eta},
\end{equation}
    and the TV filtering problem is:
\begin{equation}
    \min_{\mathbf{x}} \{f(\mathbf{x}) = \|\mathbf{y}-\mathbf{x}\|^2 + \lambda\|\mathbf{D}\mathbf{x}\|_1\},
\label{eqn:minProb_TVF}
\end{equation}
where $\lambda$ is a regularization parameter and $\mathbf{D}$ is the following finite difference matrix of size $(n-1)\times n$:
\begin{equation}
    \mathbf{D} = \begin{bmatrix}
        -1 & 1 & & &\multicolumn{2}{c}{\multirow{2}{*}{\scalebox{1.5}{$0$}}}\\
         & -1 & 1 & & \\
        \multicolumn{2}{c}{\multirow{2}{*}{\scalebox{1.5}{$0$}}}& & & \ddots &\\
        & & & & -1 & 1
    \end{bmatrix}.
\end{equation}
    The objective function $f(\mathbf{x})$ is convex but not differentiable. \cite{SelesnickTVF} made use of the representation of the $\ell_1$ norm in \eqref{eq:dualL1norm} to deal with the non-differentiable term in \eqref{eqn:minProb_TVF}. Using this representation, the max formulation for $f(\mathbf{x})$ is given by:
\begin{equation}
    f(\mathbf{x}) = \max_{\lvert\mathbf{z}\rvert\leq 1} \{{g}(\mathbf{x},\mathbf{z}) =  \|\mathbf{y}-\mathbf{x}\|^2 + \lambda\mathbf{z}^{\mathsf{T}}\mathbf{D}\mathbf{x}\},
\end{equation}
    and therefore the min-max problem is:
\begin{equation}
    \min_{\mathbf{x}}\;\max_{\lvert\mathbf{z}\rvert\leq 1}\;\|\mathbf{y}-\mathbf{x}\|^2 + \lambda\mathbf{z}^{\mathsf{T}}\mathbf{D}\mathbf{x}.
\label{eqn:minmaxTVF}
\end{equation}
    Because ${g}(\mathbf{x},\mathbf{z})$ is convex-concave, invoking the minimax theorem leads to the following max-min problem:
\begin{equation}
    \max_{\lvert\mathbf{z}\rvert\leq 1}\;\min_{\mathbf{x}}\;\|\mathbf{y}-\mathbf{x}\|^2 + \lambda\mathbf{z}^{\mathsf{T}}\mathbf{D}\mathbf{x}.
\label{eqn:maxminTVF}
\end{equation}
    The minimizer $\mathbf{\Tilde{x}}(\mathbf{z})$ is given by:
\begin{equation}
    \mathbf{\Tilde{x}}(\mathbf{z}) = \mathbf{y} - \frac{\lambda}{2}\mathbf{D}^{\mathsf{T}}\mathbf{z}
\label{eqn:xTil_TVF}
\end{equation}
    The problem that remains to be solved after substituting \eqref{eqn:xTil_TVF} in \eqref{eqn:maxminTVF} is:
\begin{equation}
    \min_{\lvert\mathbf{z}\rvert\leq 1}\;\left\{h(\mathbf{z}) = \mathbf{z}^{\mathsf{T}}\mathbf{D}\mathbf{D}^{\mathsf{T}}\mathbf{z} - \frac{4}{\lambda}\mathbf{z}^{\mathsf{T}}\mathbf{D}\mathbf{y}\right\},
\label{eqn:dualProbTVF}
\end{equation}
While this problem is convex, for computational convenience, \cite{SelesnickTVF} used the MM, in lieu of an off-the-shelf convex solver, to tackle \eqref{eqn:dualProbTVF}. Let $\alpha$ be the maximum eigenvalue of $\mathbf{D}\mathbf{D}^{\mathsf{T}}$ and $\mathbf{A} = \mathbf{D}\mathbf{D}^{\mathsf{T}} - \alpha \mathbf{I}$ (a negative semi-definite matrix). Then adding and subtracting $\alpha\mathbf{I}$ from $\mathbf{D}\mathbf{D}^{\mathsf{T}}$ in $h(\mathbf{z})$, we get:
\begin{equation}
    \min_{\lvert\mathbf{z}\rvert\leq 1} \; \mathbf{z}^{\mathsf{T}}\mathbf{A}\mathbf{z} + \alpha\mathbf{z}^{\mathsf{T}}\mathbf{z} - \frac{4}{\lambda}\mathbf{z}^{\mathsf{T}}\mathbf{D}\mathbf{y}.
\end{equation}
    Since the first term in the objective function is concave, majorizing the objective function at $\mathbf{z}^k$ gives the following problem:
\begin{equation}
    \min_{\lvert\mathbf{z}\rvert\leq 1}\; 2(\mathbf{z}^k)^{\mathsf{T}}\mathbf{A}\mathbf{z} + \alpha \mathbf{z}^{\mathsf{T}}\mathbf{z} -\frac{4}{\lambda}\mathbf{z}^{\mathsf{T}}\mathbf{D}\mathbf{y}.
\end{equation}
    The next iterate $\mathbf{z}^{k+1}$ is given by:
\begin{equation}
\label{eqn:zUpd_TVF}
    {z}_i^{k+1} = \begin{cases}
        {b}_i^k \; & \lvert{b}_i^k\rvert \leq 1\\
        \mathrm{sign}({b}_i^k) \;& \lvert{b}_i^k\rvert \geq 1,
    \end{cases}
\quad \forall~i=1,\ldots,n-1.
\end{equation}
    where $z_i^{k+1}$ and $b_i^k$ are the $i$th elements of the vectors $\mathbf{z}^{k+1}$ and $\mathbf{b}^k$ respectively, and
\begin{equation}
    \mathbf{b}^k = \frac{2}{\alpha\lambda}\mathbf{D}\mathbf{y} - \frac{1}{\alpha}\mathbf{A}\mathbf{z}^k.
\end{equation}
    The update of the variable $\mathbf{x}$ (see \eqref{eqn:xTil_TVF}) is obtained using $\mathbf{z}^k$:
\begin{equation}
\label{eqn:xUpd_TVF}
    \mathbf{{x}}^{k+1} = \mathbf{y} - \frac{\lambda}{2}\mathbf{D}^{\mathsf{T}}\mathbf{z}^k.
\end{equation}
    A main point of this example is that the MM4MM algorithm can be a useful option even if the objective function is convex (but not differentiable).


\section{Phase Retrieval for Poisson Noise}\label{pdmm}

Phase retrieval problem is an inverse problem generally encountered in such applications as optical imaging, X-ray crystallography, speech processing and electron microscopy. The problem is to recover an unknown complex signal $\mathbf{x}\in\mathbb{C}^n$ from magnitude or intensity measurements:
\begin{equation}
    y_i = \lvert\mathbf{a}_i^\mathsf{H}\mathbf{x}\rvert^2 + b_i + \nu_i\quad \forall\;i=1,\ldots,M;
\label{eqn:phsRetModel}
\end{equation}
    where $\mathbf{a}_i\in\mathbb{C}^n$ is the known sampling vector, $b_i\in\mathbb{R}_+$ is the mean background signal and $\nu_i$ is the noise. In the Gaussian noise case, there exist several algorithms for obtaining the maximum likelihood (ML) estimate of $\mathbf{x}$.
    However, in some applications, a more appropriate assumption is to consider Poisson noise in the measurements \cite{bian2016ptychoPoissonpr}. \cite{GFatimaPDMM} has proposed an algorithm for solving the ML phase retrieval problem in the case of Poisson noise.
    It is this algorithm that we will cast in the MM4MM framework when:
\begin{equation}
    y_i \sim \mathrm{Poisson}(\lvert\mathbf{a}_i^\mathsf{H}\mathbf{x}\rvert^2+b_i).
\end{equation}
    The corresponding ML problem is:
\begin{equation}
    \min_{\mathbf{x}}\bigg\{ f(\mathbf{x})=\sum_{i=1}^M [\lvert\mathbf{a}_i^\mathsf{H}\mathbf{x}\rvert^2 + b_i - y_i\log( \lvert\mathbf{a}_i^\mathsf{H}\mathbf{x}\rvert^2 + b_i ) + y_i\log(y_i) -2y_i]\bigg\}.
\label{eqn:pphsRet}
\end{equation}
    The max formulation of the negative logarithmic term of $f(\mathbf{x})$ is obtained using the representation in {Example} \ref{egPDMMnlog}. For $i=1,\ldots,M$, we have:
\begin{equation}
\begin{split}
    -y_i&\log(\lvert\mathbf{a}_i^\mathsf{H}\mathbf{x}\rvert^2+b_i) + y_i\log(y_i) -2y_i \\&= \displaystyle\max_{{\Bar{z}_i}>{0}} y_i[\log({\Bar{z}}_i) - \Bar{z}_i(\lvert\mathbf{a}_i^\mathsf{H}\mathbf{x}\rvert^2+b_i) + 1] + y_i\log(y_i) -2y_i,
    \\&= \displaystyle\max_{{\Bar{z}_i}>{0}} \;y_i\log({y_i\Bar{z}}_i) - y_i\Bar{z}_i(\lvert\mathbf{a}_i^\mathsf{H}\mathbf{x}\rvert^2+b_i)- y_i.
\end{split}
\end{equation}
    Defining a new variable $\mathbf{z}$, whose $i$th element is $z_i = y_i\Bar{z}_i \geq 0$ (as $y_i\geq0$ by assumption), the max formulation for $f(\mathbf{x})$ becomes:
\begin{equation}
\begin{split}
    f(\mathbf{x}) = \max_{\mathbf{z}\geq\mathbf{0}}\Bigg\{{g}(\mathbf{x},\mathbf{z}) = &\sum_{i=1}^M [\lvert\mathbf{a}_i^\mathsf{H}\mathbf{x}\rvert^2 + b_i + y_i(\log(z_i)-1)\\& - z_i( \lvert\mathbf{a}_i^\mathsf{H}\mathbf{x}\rvert^2 + b_i )]\Bigg\}.
\end{split}
\end{equation}
    The term $-\lvert\mathbf{a}_i^{\mathsf{H}}\mathbf{x}\rvert^2$ is concave in $\mathbf{x}$, and can be majorized at $\mathbf{x} = \mathbf{x}^t$ in the following way:
\begin{equation}
    -\lvert\mathbf{a}_i^\mathsf{H}\mathbf{x}\rvert^2\leq \lvert\mathbf{a}_i^\mathsf{H}\mathbf{x}^t\rvert^2 -2\mathrm{Re}((\mathbf{x}^t)^\mathsf{H}\mathbf{a}_i\mathbf{a}_i^\mathsf{H}\mathbf{x}),
\label{eqn:mod2Ineq}
\end{equation}
    where the equality holds for $\mathbf{x} = \mathbf{x}^t$. Using \eqref{eqn:mod2Ineq}, the min-max surrogate problem (after ignoring some constants) is as follows:
\begin{equation}
\begin{split}
    \min_{\mathbf{x}}\; \max_{\mathbf{z}\geq\mathbf{0}}\; \Bigg\{g_s(\mathbf{x},\mathbf{z}\mid\mathbf{x}^t) = \sum_{i=1}^M [\lvert\mathbf{a}_i^\mathsf{H}\mathbf{x}\rvert^2 + y_i \log(z_i) \\+ z_i\lvert\mathbf{a}_i^\mathsf{H}\mathbf{x}^t\rvert^2 -2\mathrm{Re}(z_i(\mathbf{x}^t)^\mathsf{H}\mathbf{a}_i\mathbf{a}_i^\mathsf{H}\mathbf{x}) -z_ib_i]\Bigg\}
\end{split}
\label{eqn:minmaxpdmm}
\end{equation}
Invoking the minimax theorem we swap the min and max operators and solve the inner minimization problem to obtain:
\begin{equation}
\label{eqn:xtilz_pr}
    \mathbf{\Tilde{x}}(\mathbf{z}\mid\mathbf{x}^t)= (\mathbf{A}^{\mathsf{H}}\mathbf{A})^{-1}\mathbf{A}^{\mathsf{H}}\mathbf{D}\mathbf{z},
\end{equation}
where $\mathbf{A} = [\mathbf{a}_1, \ldots, \mathbf{a}_M]^{\mathsf{H}}$, and $\mathbf{D} = \mathrm{diag}(\mathbf{A}\mathbf{x}^t)$.
    Inserting $\mathbf{x} = \Tilde{\mathbf{x}}(\mathbf{z}\mid\mathbf{x}^t)$ in $g_s(\mathbf{x},\mathbf{z}\mid\mathbf{x}^t)$ yields the following maximization problem with respect to $\mathbf{z}$:
\begin{equation}
\begin{split}
    \max_{\mathbf{z}\geq\mathbf{0}}\Big\{h(\mathbf{z}\mid\mathbf{x}^t) = -\mathbf{z}^{\mathsf{H}}\mathbf{D}^\mathsf{H}\mathbf{A}(\mathbf{A}^\mathsf{H}\mathbf{A})^{-1}\mathbf{A}^\mathsf{H}\mathbf{D}\mathbf{z} + \mathbf{y}^\mathsf{T}\log(\mathbf{z}) - \mathbf{z}^\mathsf{T}\mathbf{b} + \mathbf{z}^\mathsf{T}\mathbf{\Bar{d}}
    \Big\},
\end{split}
\label{eqn:maxqpdmm}
\end{equation}
    where $\mathbf{b} = [b_1,\ldots,b_M],$ and $ \mathbf{\Bar{d}} = [\lvert\mathbf{a}_1^\mathsf{H}\mathbf{x}^t\rvert^2, \ldots, \lvert\mathbf{a}_M^{\mathsf{H}}\mathbf{x}^t\rvert^2]^\mathsf{T}$.

    Solving the convex problem in \eqref{eqn:maxqpdmm} (for instance using CVX \cite{refcvx}) gives the maximizer $\mathbf{z}^{t+1}$. Using $\mathbf{z}^{t+1}$, the update of $\mathbf{x}$ is obtained as follows (see \eqref{eqn:xtilz_pr}):
\begin{align}
\label{eqn:updxpdmm}
    \mathbf{x}^{t+1} &= \mathbf{\Tilde{x}}(\mathbf{z}^{t+1}\mid\mathbf{x}^t).
\end{align}
    Because convex solvers are slow for large values of $M$, as an alternative we can use MM also to solve the problem in \eqref{eqn:maxqpdmm}. To do so we re-write \eqref{eqn:maxqpdmm} as a minimization problem:
\begin{equation}
    \min_{\mathbf{z}\geq\mathbf{0}}\; \mathbf{z}^{\mathsf{H}}\mathbf{D}^\mathsf{H}(\mathbf{P} - \mathbf{I}_M)\mathbf{D}\mathbf{z} + \mathbf{z}^{\mathsf{H}}\mathbf{D}^{\mathsf{H}}\mathbf{D}\mathbf{z} - \mathbf{y}^\mathsf{T}\log(\mathbf{z}) + \mathbf{z}^\mathsf{T}\mathbf{b} - \mathbf{z}^\mathsf{T}\mathbf{\Bar{d}},
\label{eqn:minzpdmm}
\end{equation}
where $\mathbf{P}$ is the orthogonal projector onto the column space of $\mathbf{A}$. Because $\mathbf{P} - \mathbf{I}_M$ is a negative semi-definite matrix, the first term in \eqref{eqn:minzpdmm} is concave. Therefore we can majorize this term at $\mathbf{z}^k$ as follows:
\begin{equation}
    \mathbf{z}^{\mathsf{H}}\mathbf{D}^\mathsf{H}(\mathbf{P} - \mathbf{I}_M)\mathbf{D}\mathbf{z} \leq - (\mathbf{z}^k)^{\mathsf{H}}\mathbf{D}^\mathsf{H}(\mathbf{P} - \mathbf{I}_M)\mathbf{D}\mathbf{z}^k + 2\mathrm{Re}(\mathbf{z}^{\mathsf{H}}\mathbf{D}^\mathsf{H}(\mathbf{P} - \mathbf{I}_M)\mathbf{D}\mathbf{z}^k).
\end{equation}
    Then the surrogate minimization problem in the variable $\mathbf{z}$ (after ignoring some constants) is:
\begin{equation}
    \min_{\mathbf{z}\geq\mathbf{0}}\; 2\mathrm{Re}(\mathbf{z}^{\mathsf{H}}\mathbf{D}^\mathsf{H}(\mathbf{P} - \mathbf{I}_M)\mathbf{D}\mathbf{z}^k) + \mathbf{z}^{\mathsf{H}}\mathbf{D}^{\mathsf{H}}\mathbf{D}\mathbf{z} - \mathbf{y}^\mathsf{T}\log(\mathbf{z}) + \mathbf{z}^\mathsf{T}\mathbf{b} - \mathbf{z}^\mathsf{T}\mathbf{\Bar{d}},
\label{eqn:minzProb_pdmm}
\end{equation}
which is separable in $\{z_i\}$ and can be written as follows:
\begin{equation}
\label{eqn:ziprobpdmm}
    \min_{z_i\geq 0}~ c_iz_i + \Bar{d}_iz_i^2 -y_i\log(z_i) + b_iz_i -\Bar{d}_iz_i,\quad i=1,\ldots,M
\end{equation}
    where $c_i$ is the $i$th element of $\mathbf{c} = 2\mathrm{Re}(\mathbf{D}^\mathsf{H}(\mathbf{P} - \mathbf{I}_M)\mathbf{D}\mathbf{z}^k)$. The MM update $\mathbf{z}^{k+1}$, obtained by solving the problem in \eqref{eqn:ziprobpdmm}, is given by:
\begin{equation}
    z_i^{k+1} = \begin{cases}
        \displaystyle\frac{-(b_i + c_i - \Bar{d}_i) + \sqrt{(b_i + c_i - \Bar{d}_i)^2 + 8\Bar{d}_iy_i}}{4\Bar{d}_i}, & \text{if}~\Bar{d}_i \neq 0\\[10pt]
        \displaystyle\frac{y_i}{b_i + c_i}, & \text{if}~\Bar{d}_i = 0
    \end{cases}
\end{equation}
    The above update of $\mathbf{z}$ is iterated till the function $h(\mathbf{z}\mid\mathbf{x}^t)$ in \eqref{eqn:maxqpdmm} converges.

The problem \eqref{eqn:minmaxpdmm} can also be solved using the Min-Max for MM approach (see \cref{sec:minmaxMM4MM}). The solution of the inner maximization with respect to $\mathbf{z}$, is:
\begin{align}
    {\Tilde{z}_i}(\mathbf{x}\mid\mathbf{x}^t) &= \frac{y_i}{b_i - \lvert\mathbf{a}_i^{\mathsf{H}}\mathbf{x}^t\rvert^2 + 2\mathrm{Re}((\mathbf{x}^t)^\mathsf{H}\mathbf{a}_i\mathbf{a}_i^\mathsf{H}\mathbf{x})},\nonumber\\ &\hspace*{4pc}\text{for}~{b_i - \lvert\mathbf{a}_i^{\mathsf{H}}\mathbf{x}^t\rvert^2 + 2\mathrm{Re}((\mathbf{x}^t)^\mathsf{H}\mathbf{a}_i\mathbf{a}_i^\mathsf{H}\mathbf{x})}>0
\end{align}
    (for ${b_i - \lvert\mathbf{a}_i^{\mathsf{H}}\mathbf{x}^t\rvert^2 + 2\mathrm{Re}((\mathbf{x}^t)^\mathsf{H}\mathbf{a}_i\mathbf{a}_i^\mathsf{H}\mathbf{x})} \leq 0$, the function in \eqref{eqn:minmaxpdmm} does not have a maximum with respect to $\mathbf{z}$).
    Using $\mathbf{z} = \mathbf{\Tilde{z}}(\mathbf{x}\mid \mathbf{x}^t)$ in \eqref{eqn:minmaxpdmm}, we get $d(\mathbf{x}\mid\mathbf{x}^t) \triangleq g_s(\mathbf{x},\mathbf{\Tilde{z}}(\mathbf{x}\mid\mathbf{x}^t)\mid\mathbf{x}^t)$ as the surrogate function, and the following constrained minimization problem with respect to $\mathbf{x}$:
\begin{align}
\label{eqn:mindpdmm}
    \min_{\mathbf{x}}~ &~\{d(\mathbf{x}\mid\mathbf{x}^t) = \mathbf{x}^{\mathsf{H}}\mathbf{A}^{\mathsf{H}}\mathbf{A}\mathbf{x} - \mathbf{y}^{\mathsf{T}}(\log(\mathbf{b} - \mathbf{\Bar{d}} + 2\mathrm{Re}(\mathbf{D}^{\mathsf{H}}\mathbf{A}\mathbf{x})))\}
    \\\nonumber
    \textrm{s.t.}~ &~b_i - \lvert\mathbf{a}_i^{\mathsf{H}}\mathbf{x}^t\rvert^2 + 2\mathrm{Re}((\mathbf{x}^t)^\mathsf{H}\mathbf{a}_i\mathbf{a}_i^\mathsf{H}\mathbf{x}) > 0 \quad \forall\; i
\end{align}
    The convex problem in \eqref{eqn:mindpdmm} is solved to obtain $\mathbf{x}^{t+1}$, an operation that is iterated till the objective function $f(\mathbf{x})$ converges. An interesting aspect of the problem in \eqref{eqn:mindpdmm} is that its objective function $d(\mathbf{x}\mid\mathbf{x}^t)$ is not a global majorizer of $f(\mathbf{x})$ (as in Lemma~\ref{upperboundMM4MM}) but only a local majorizer over the constraint set in \eqref{eqn:mindpdmm}. Furthermore, $d(\mathbf{x}\mid\mathbf{x}^t)$ changes with the iteration. However the monotonicity of MM, which is essential for its convergence, is maintained because $\mathbf{x}^t$ belongs to the constraint set in \eqref{eqn:mindpdmm}:
\begin{equation}
    [b_i - \lvert\mathbf{a}_i^{\mathsf{H}}\mathbf{x}^t\rvert^2 + 2\mathrm{Re}((\mathbf{x}^t)^\mathsf{H}\mathbf{a}_i\mathbf{a}_i^\mathsf{H}\mathbf{x})]_{\mathbf{x} = \mathbf{x}^t} = b_i + \lvert\mathbf{a}_i^{\mathsf{H}}\mathbf{x}^t\rvert^2 > 0
\end{equation}\pagebreak

\section{Outlier Robust Phase Retrieval}\label{app:orpr}

The recorded magnitude or intensity measurements are at times contaminated by outliers, which generally occur as a result of erroneous recording, or faulty instruments. The vanilla  phase retrieval algorithms for Gaussian or Poisson noise lack robustness in the presence of outliers, and hence outlier robust phase retrieval algorithms are required. Consider the phase retrieval model in \eqref{eqn:phsRetModel} in the presence of outliers $\boldsymbol{\theta} \in \mathbb{R}^M$. The measurement vector $\mathbf{y}\in \mathbb{R}^M$ for a sampling matrix $\mathbf{A} = [\mathbf{a}_1, \ldots, \mathbf{a}_M]^{\mathsf{H}}$ can be written as:
\begin{equation}
    \mathbf{y} = \lvert\mathbf{A}\mathbf{x}\rvert^2 + \boldsymbol{\nu} + \boldsymbol{\theta}.
\end{equation}
    The additive noise $\boldsymbol{\nu}$ in the measurements may have uniform, Gaussian or Poisson distribution. The mean background signal in the phase retrieval model given in \eqref{eqn:phsRetModel} is absorbed in $\boldsymbol{\theta}$. The robust phase retrieval problem is to reconstruct the signal $\mathbf{x}$ from the outlier corrupted measurements $\mathbf{y}$.

The most common technique for handling outliers is truncation in which highly erroneous samples are discarded, based on the sample mean or median of the residuals, at each iteration of an iterative algorithm for phase retrieval \cite{chen2017TWF,Zhang2018MedianTWF,wang2017TAF}. However, the truncation procedures require tuning several parameters, which increases the computational complexity and at times results in losing large magnitude measurements that are not contaminated by outliers. In \cite{AsainiRKLDMM}, a truncation free and noise distribution independent outlier robust phase retrieval algorithm that minimizes the  Reverse Kullback-Leibler Divergence (RKLD) loss function \cite{neumann2011RKLD,chan2022greedRKLD} was proposed. Here we derive this algorithm within the MM4MM framework. The RKLD based robust phase retrieval problem is the following one (see the cited work):
\begin{equation}
\begin{split}
    \displaystyle\min_{\mathbf{x}}&\left\{f(\mathbf{x}) = \sum_{i=1}^M \left[\lvert\mathbf{a}_i^{\mathsf{H}}\mathbf{x}\rvert^2 \log\left(\frac{\lvert\mathbf{a}_i^{\mathsf{H}}\mathbf{x}\rvert^2}{y_i}\right) - (\lvert\mathbf{a}_i^{\mathsf{H}}\mathbf{x}\rvert^2 - y_i)\right]\right\}.
\end{split}
\label{eqn:minProb_RKLD}
\end{equation}
The function $f(\mathbf{x})$ is not convex. However, observe that the first term of $f(\mathbf{x})$ can be written as the negative entropy function of $u_i = \displaystyle\frac{\lvert\mathbf{a}_i^{\mathsf{H}}\mathbf{x}\rvert^2}{y_i}$:
\begin{equation}
\begin{split}
    \lvert\mathbf{a}_i^{\mathsf{H}}\mathbf{x}\rvert^2 \log\left(\frac{\lvert\mathbf{a}_i^{\mathsf{H}}\mathbf{x}\rvert^2}{y_i}\right) & = y_i\left(\frac{\lvert\mathbf{a}_i^{\mathsf{H}}\mathbf{x}\rvert^2}{y_i}\right)\log\left(\frac{\lvert\mathbf{a}_i^{\mathsf{H}}\mathbf{x}\rvert^2}{y_i}\right)\\
    & = y_i(\mathbf{x}^{\mathsf{H}}\mathbf{\Bar{A}}_i\mathbf{x})\log(\mathbf{x}^{\mathsf{H}}\mathbf{\Bar{A}}_i\mathbf{x})
\end{split}
\label{eqn:yixlogx}
\end{equation}
    where $\mathbf{\Bar{A}}_i = \mathbf{a}_i\mathbf{a}_i^{\mathsf{H}}/y_i$. Using the max representation in {Example} \ref{egRKLDxlogx} for the function in \eqref{eqn:yixlogx}, we can express $f(\mathbf{x})$ as:
\begin{equation}
\hspace*{-10pt}\begin{split}
    f(\mathbf{x}) = \max_{\mathbf{z}>\mathbf{0}}\left\{{g}(\mathbf{x},\mathbf{z})  = \sum_{i=1}^M \big[
    z_i \lvert\mathbf{a}_i^{\mathsf{H}}\mathbf{x}\rvert^2 - y_ie^{z_i -1} - \lvert\mathbf{a}_i^{\mathsf{H}}\mathbf{x}\rvert^2 + y_i
    \big]
\right\}.
\end{split}
\label{eqn:maxFrkldapp}
\end{equation}
    For $z_i > 0$, the first term in ${g}(\mathbf{x},\mathbf{z})$ is convex in $\mathbf{x}$ but the third term is not. However, the latter term is concave in $\mathbf{x}$ and thus it can be majorized at $\mathbf{x}^t$ as follows:
\begin{equation}
    -\mathbf{x}^{\mathsf{H}}\mathbf{A}^{\mathsf{H}}\mathbf{A}\mathbf{x} \leq (\mathbf{x}^t)^{\mathsf{H}}\mathbf{A}^{\mathsf{H}}\mathbf{A}\mathbf{x}^t -2\mathrm{Re}((\mathbf{x}^t)^{\mathsf{H}}\mathbf{A}^{\mathsf{H}}\mathbf{A}\mathbf{x} ).
\label{eqn:conMajRKLD}
\end{equation}
    Using \eqref{eqn:maxFrkldapp} and \eqref{eqn:conMajRKLD} we obtain the following surrogate min-max problem:
\begin{equation}
\label{eqn:minmax_RKLD}
\begin{split}
    \min_{\mathbf{x}}\;&\max_{\mathbf{z}>\mathbf{0}}\;\Bigg\{g_s(\mathbf{x},\mathbf{z}\mid\mathbf{x}^t) = \;\mathbf{x}^{\mathsf{H}}\mathbf{A}^{\mathsf{H}}\mathbf{Z}\mathbf{A}\mathbf{x} \\ &-2\mathrm{Re}((\mathbf{x}^t)^{\mathsf{H}}\mathbf{A}^{\mathsf{H}}\mathbf{A}\mathbf{x})- \sum_{i=1}^My_ie^{z_i -1}
    \Bigg\},
\end{split}
\end{equation}
    where $\mathbf{Z}$ is a diagonal matrix with $\{z_i\}$ as diagonal elements. The function $g_s(\mathbf{x},\mathbf{z}\mid\mathbf{x}^t)$ is convex-concave and hence the minimax theorem can be invoked. The resulting max-min problem is:
\begin{equation}
\begin{split}
    \max_{\mathbf{z}>\mathbf{0}}\;&\min_{\mathbf{x}}\;\Bigg\{g_s(\mathbf{x},\mathbf{z}\mid\mathbf{x}^t) = \;\mathbf{x}^{\mathsf{H}}\mathbf{A}^{\mathsf{H}}\mathbf{Z}\mathbf{A}\mathbf{x} \\ &-2\mathrm{Re}((\mathbf{x}^t)^{\mathsf{H}}\mathbf{A}^{\mathsf{H}}\mathbf{A}\mathbf{x})- \sum_{i=1}^My_ie^{z_i -1}
    \Bigg\}.
\end{split}
\label{eqn:maxminRKLD}
\end{equation}
    The inner minimization problem in $\mathbf{x}$ has the following solution:
\begin{equation}
\label{eqn:xtil_RKLD}
    \mathbf{\Tilde{x}}(\mathbf{z}\mid\mathbf{x}^t) = (\mathbf{A}^{\mathsf{H}}\mathbf{Z}\mathbf{A})^{-1}\mathbf{A}^{\mathsf{H}}\mathbf{A}\mathbf{x}^t.
\end{equation}
    Substituting \eqref{eqn:xtil_RKLD} in \eqref{eqn:maxminRKLD}, the following convex problem in $\mathbf{z}$ is obtained:
\begin{equation}
\label{eqn:dualProb_RKLD}
    \min_{\mathbf{z}>\mathbf{0}}\; \left\{h(\mathbf{z}\mid\mathbf{x}^t)  = \mathbf{b}_t^{\mathsf{H}}(\mathbf{A}^{\mathsf{H}}\mathbf{Z}\mathbf{A})^{-1}\mathbf{b}_t + \sum_{i=1}^My_ie^{z_i -1}\right\},
\end{equation}
    where
\begin{equation}
    \mathbf{b}_t = \mathbf{A}^{\mathsf{H}}\mathbf{A}\mathbf{x}^t.
\end{equation}
    The problem \eqref{eqn:dualProb_RKLD} can be solved using any convex programming solver or once again by means of MM. The minimizer of $h(\mathbf{z}\mid\mathbf{x}^t)$ is the update of $\mathbf{z}$ given $\mathbf{x}^t$:
\begin{equation}
    \mathbf{z}^{t+1} = \arg \; \min_{\mathbf{z}>\mathbf{0}} \;h(\mathbf{z}\mid\mathbf{x}^t).
\end{equation}
    Using $\mathbf{z} = \mathbf{z}^{t+1}$ in \eqref{eqn:xtil_RKLD}, we obtain the update of $\mathbf{x}$:
\begin{equation}
    \mathbf{x}^{t+1} = (\mathbf{A}^{\mathsf{H}}\mathbf{Z}^{t+1}\mathbf{A})^{-1}\mathbf{A}^{\mathsf{H}}\mathbf{A}\mathbf{x}^t,
\end{equation}
    where $\mathbf{Z}^{t+1} = \mathrm{diag}(\mathbf{z}^{t+1})$.

Finally note that the majorization step in the formulation of the min-max problem in \eqref{eqn:minmax_RKLD} does not involve $\mathbf{z}$. Therefore solving the inner maximization problem in $\mathbf{z}$ results in the same non-convex term $\lvert\mathbf{a}_i^{\mathsf{H}}\mathbf{x}\rvert^2\log\left(\displaystyle\frac{\lvert\mathbf{a}_i^{\mathsf{H}}\mathbf{x}\rvert^2}{y_i}\right)$ as in $f(\mathbf{x})$. Consequently, the min-max approach for MM is not possible for this problem.

\section{RSS Based Source Localization}\label{app:rss}

Source localization techniques are widely used in communication, radar, sonar and wireless sensor networks. Localization based on received signal strength (RSS) measurements \cite{LLS_HCSo} is widely used in indoor applications such as inventory management, and indoor air quality monitoring due to its low-cost hardware implementation. The ML estimators for estimating source location from RSS measurements are computationally complex and their performance may be sub-optimal due to convergence problems \cite{ML_Patwari_WSN}. Thus, the most commonly used methods are based on linearizing the RSS data model and using the least squares approach \cite{WLS_JRCasar,LLS_HCSo}. However, \cite{rss24ASaini} has proposed an ML estimation algorithm with good convergence properties. This algorithm is derived in the following within the MM4MM framework.

In a wireless sensor network of $M$ sensors with location vectors $\{\mathbf{s}_i\in\mathbb{R}^n\}$ (for $i=1,\ldots,M$ and $n=2 \;\textrm{or}\; 3$) the RSS, i.e., the average power in dB, measured at the $i$th sensor can be modeled as:
\begin{equation}
    p_i = p_o -10\alpha\log_{10}(\Vert \mathbf{x}-\mathbf{s}_{i}\Vert )+\nu_{i} \quad \forall\;i=1,\ldots,M,
\label{eqn:rss_dataModel}
\end{equation}
where $\mathbf{x}\in\mathbb{R}^n$ is the source position, $p_o$ is the reference power at unit distance from the source, $\alpha$ is the path loss constant and $\nu_i\sim\mathcal{N}(0,\sigma^2)$ is the noise. The ML problem for the RSS data model in \eqref{eqn:rss_dataModel} is as follows:
\begin{equation}
\label{eqn:rss_mlProb}
\begin{split}
    \min_{\mathbf{x}}\bigg\{f(\mathbf{x}) = \displaystyle\sum_{i=1}^M[y_i^2 - y_i\log(\|\mathbf{x}-\mathbf{s}_i\|^2) + (\log\|\mathbf{x}-\mathbf{s}_i\|)^2]\bigg\},
\end{split}
\end{equation}
    where $y_i = 10\alpha\displaystyle\frac{\left(p_o - p_i\right)}{\log(10)}$.

The max formulation of the second term in \eqref{eqn:rss_mlProb} is given by (see {Example} \ref{egRSSnLog}):
\begin{equation}
    -\log(\|\mathbf{x}-\mathbf{s}_{i}\|^{2})=\underset{z_{i}>0}{\textrm{max}}\;\log\left(z_{i}\right)-z_{i}\| \mathbf{x}-\mathbf{s}_{i}\|^{2} +1,
\label{eqn:maxFnegLog}
\end{equation}
    where $\mathbf{z}=[z_1,\ldots,z_M]^\mathsf{T}$ is an auxiliary variable. Also, as shown in {Example} \ref{egLogSq}, the max formulation for the third term of $f(\mathbf{x})$ is as follows:
\begin{equation}
    (\log(\| \mathbf{x}-\mathbf{s}_{i}\| ))^{2} = \underset{q_{i}>0}{\textrm{max}}\;-q_{i}^2+2{q_{i}}\log(\| \mathbf{x}-\mathbf{s}_{i}\| ),
\label{eqn:maxFlogSq}
\end{equation}
where $\mathbf{q}=[q_1,\ldots,q_M]^\mathsf{T}$ is also an auxiliary variable.
    Using \eqref{eqn:maxFnegLog} and \eqref{eqn:maxFlogSq}, we obtain the following max formulation for $f(\mathbf{x})$ (after omitting an additive constant):
\begin{equation}
\begin{split}
    f(\mathbf{x}) = \max_{\substack{\{q_i\}>0,\\\{z_i\}>0}}\Bigg\{{g}(\mathbf{x},\mathbf{q},\mathbf{z}) = &\sum_{i=1}^M\big[-q_{i}^2+q_{i}\log(\|\mathbf{x}-\mathbf{s}_{i}\|^{2})\\ &+ y_{i}\log(z_{i}) - y_{i}z_{i}\|\mathbf{x}-\mathbf{s}_{i}\|^{2}\big]\Bigg\}.
\end{split}
\label{eqn:maxFRSS}
\end{equation}
The function ${g}(\mathbf{x},\mathbf{q},\mathbf{z})$ is concave in $\mathbf{z}$ but it is not convex in $\mathbf{x}$. To obtain a convex-concave surrogate function, the second term and the fourth term are majorized at $\mathbf{x}^{t}$ as follows:
\begin{equation}
\begin{split}
    \log(\| \mathbf{x}-\mathbf{s}_{i}\|^{2})\leq \log&(\|\mathbf{x}^{t}-\mathbf{s}_{i}\|^{2})+\frac{1}{\|\mathbf{x}^{t}-\mathbf{s}_{i}\|^{2}}(\|\mathbf{x}-\mathbf{s}_{i}\|^{2}-\|\mathbf{x}^{t}-\mathbf{s}_{i}\|^{2}),
\end{split}
\label{eqn:majLogSq}
\end{equation}
\begin{equation}
-\|\mathbf{x}-\mathbf{s}_{i}\|^{2}\leq-\|\mathbf{x}^{t}-\mathbf{s}_{i}\|^{2}-2(\mathbf{x}^{t}-\mathbf{s}_{i})^{\mathsf{T}}(\mathbf{x}-\mathbf{x}^{t}).
\label{eqn:majNegNormSq}
\end{equation}
Using \eqref{eqn:majLogSq} and \eqref{eqn:majNegNormSq} in \eqref{eqn:maxFRSS} leads to the following min-max problem:
\begin{equation}
\hspace*{-10pt}\begin{split}
    \displaystyle\min_{\mathbf{x}}\displaystyle\max_{\substack{\{q_i\}>0,\\\{z_i\}>0}} \Bigg\{&g_s(\mathbf{x},\mathbf{q},\mathbf{z}\mid\mathbf{x}^t) =\displaystyle\sum_{i=1}^M\Bigg[ -q_i^2 ~+~ q_ia_i~ + q_i\frac{\|\mathbf{x}-\mathbf{s}_i\|^2}{\|\mathbf{b}_i\|^2} ~\\&+~ y_i\log(z_i)~ - y_i z_i\|\mathbf{b}_i\|^2 ~-~2y_i z_i\mathbf{b}_i^\mathsf{T}(\mathbf{x}-\mathbf{x}^t)\Bigg]\Bigg\},
\end{split}
\label{eqn:minmaxrssml}
\end{equation}
    where $\mathbf{b}_i = \mathbf{x}^t-\mathbf{s}_i$, and $a_i = \log(\|\mathbf{b}_i\|^2)-1$. We can solve the min-max problem in \eqref{eqn:minmaxrssml} or use the minimax theorem and solve the corresponding max-min problem.

    First we derive the Max-Min for MM algorithm for solving \eqref{eqn:minmaxrssml}. The solution to the inner minimization problem is:
\begin{equation}
    \mathbf{\Tilde{x}}(\mathbf{q},\mathbf{z}\mid\mathbf{x}^t) = \frac{\mathbf{C}\mathbf{q} + \mathbf{D}\mathbf{z}}{2\mathbf{q}^{\mathsf{T}}\mathbf{\Bar{b}}},
\label{eqn:xtil_rssml}
\end{equation}
    where $\mathbf{\Bar{b}} = [\Bar{b}_1,\ldots,\Bar{b}_M]^{\mathsf{T}}$ with ${\Bar{b}}_i = {1}/{\|\mathbf{b}_i\|^2}$, $\mathbf{C} = 2[\Bar{b}_1\mathbf{s}_1, \ldots, \Bar{b}_M\mathbf{s}_M]$, and $\mathbf{D} = 2\big[y_1\mathbf{b}_1, \ldots,y_M\mathbf{b}_M\big]$. Substituting $\mathbf{\Tilde{x}}(\mathbf{q},\mathbf{z}\mid\mathbf{x}^t)$ for $\mathbf{x}$ in the max-min problem, we get the following maximization problem:
\begin{equation}
\begin{split}
    \max_{\substack{\{q_i\}>0,\\\{z_i\}>0}}\; \Bigg\{&g_s(\mathbf{\Tilde{x}}(\mathbf{q},\mathbf{z}\mid\mathbf{x}^t),\mathbf{q},\mathbf{z}\mid\mathbf{x}^t) = -\mathbf{q}^\mathsf{T}\mathbf{q}+ \mathbf{q}^\mathsf{T}(\mathbf{a}+ \mathbf{u})\\
     & - \displaystyle\frac{\|\mathbf{C}\mathbf{q}+\mathbf{D}\mathbf{z}\|^2}{4\mathbf{q}^\mathsf{T}\mathbf{\mathbf{\Bar{b}}}}  +\mathbf{y}^\mathsf{T}\log(\mathbf{z}) + \mathbf{z}^\mathsf{T}\mathbf{v}\Bigg\},
\end{split}
\label{eqn:dualRSSprob}
\end{equation}
    where $\mathbf{u}$ and $\mathbf{v}$ are vectors with the $i$th element ${u}_i = \Bar{b}_i\|\mathbf{s}_i\|^2$ and ${v}_i = y_i \mathbf{b}_i^\mathsf{T}(2\mathbf{x}^t -\mathbf{b}_i)$ respectively. We solve the convex problem in \eqref{eqn:dualRSSprob} to get the updates $(\mathbf{q}^{t+1},\mathbf{z}^{t+1})$ and use $(\mathbf{q}^{t+1},\mathbf{z}^{t+1})$ in \eqref{eqn:xtil_rssml} to update $\mathbf{x}^{t}$:
\begin{equation}
    \mathbf{x}^{t+1} = \mathbf{\Tilde{x}}(\mathbf{q}^{t+1},\mathbf{z}^{t+1}\mid\mathbf{x}^t).
\end{equation}

Next we derive the Min-Max for MM algorithm for solving \eqref{eqn:minmaxrssml} without swapping the min and max operators. The solutions of the inner maximization problem in \eqref{eqn:minmaxrssml} are:
\begin{align}
     \Tilde{q}_i(\mathbf{x}\mid\mathbf{x}^t) &= \frac{1}{2}\left[\frac{\|\mathbf{x}-\mathbf{s}_i\|^2}{\|\mathbf{b}_i\|^2} + a_i\right]\\
     \Tilde{z}_i(\mathbf{x}\mid\mathbf{x}^t)\; &=\; \frac{1}{\|\mathbf{b}_i\|^2 + 2\mathbf{b}_i^\mathsf{T}(\mathbf{x}-\mathbf{x}^t)}, ~\text{for}~\|\mathbf{b}_i\|^2 + 2\mathbf{b}_i^\mathsf{T}(\mathbf{x}-\mathbf{x}^t)>0.
\end{align}
    Note that the expression of $\Tilde{q}_i(\mathbf{x}\mid\mathbf{x}^t)$ is $1/2$ times the upperbound on $\log\left(\| \mathbf{x}-\mathbf{s}_{i}\|^{2}\right)$ in \eqref{eqn:majLogSq}, which is greater than zero (under far field assumption $\|\mathbf{x}-\mathbf{s}_i\|>1$). Hence, $\Tilde{q}_i(\mathbf{x}\mid\mathbf{x}^t)>0$ as required.
    Plugging $\mathbf{q} = \mathbf{\Tilde{q}}(\mathbf{x}\mid\mathbf{x}^t)$ and $\mathbf{z} = \mathbf{\Tilde{z}}(\mathbf{x}\mid\mathbf{x}^t)$ in the objective function $g_s(\mathbf{x},\mathbf{q},\mathbf{z}\mid\mathbf{x}^t)$ of the min-max problem in \eqref{eqn:minmaxrssml}, we get the following constrained minimization problem:
\begin{align}
\label{eqn:consminxEpi}
&\mbox{\fontsize{8}{10}\selectfont$\displaystyle\min_{\mathbf{x}}~\Bigg\{d(\mathbf{x}\mid\mathbf{x}^t) = \sum_{i=1}^M \Bigg[\frac{1}{4} \displaystyle\frac{\|\mathbf{x}-\mathbf{s}_i\|^4}{\|\mathbf{b}_i\|^4} + \frac{1}{2}\displaystyle\frac{a_i\|\mathbf{x}-\mathbf{s}_i\|^2}{\|\mathbf{b}_i\|^2}- y_i\log\left(\mathbf{b}_i^\mathsf{T}(2\mathbf{x}-\mathbf{x}^t-\mathbf{s}_i)\right)\Bigg]\Bigg\}$}\nonumber\\
&\mbox{\fontsize{8}{10}\selectfont$\displaystyle   \textrm{s.t.} \quad 2\mathbf{b}_i^\mathsf{T}(\mathbf{x}-\mathbf{s}_i) > \|\mathbf{b}_i\|^2\quad \forall ~i=1,\ldots,M.$}
\end{align}
    Solving the problem \eqref{eqn:consminxEpi} gives the update $\mathbf{x}^{t+1}$. Note that $\mathbf{b}_i$ changes with $\mathbf{x}^{t}$ and thus so does the constraint set in \eqref{eqn:consminxEpi}. Therefore this is also a case in which the surrogate function in \eqref{eqn:consminxEpi} majorizes $f(\mathbf{x})$ locally (for more details on this aspect see the comment made in Section~\ref{pdmm}).

\section{Optimal Sensor Placement}\label{app:osp}

The placement of sensors in a network affects the accuracy of the source location estimate. Thus, to improve the accuracy of the estimated source location, the sensors should be optimally placed. The optimal sensor orientation can be obtained by minimizing a metric of the Cram\'{e}r Rao Bound (CRB) on the variance of the source localization error \cite{bishop2010optimality}. Specifically there are three possible optimal criteria for designing the sensor orientation: the A-optimal, D-optimal, and E-optimal criteria for which the trace, the determinant, and respectively the maximum eigenvalue of the CRB matrix is minimized.

In this section we will focus on the A-optimal design for time of arrival (TOA) measurements and show how an algorithm proposed in \cite{KPanwarHybrid} can be obtained within the MM4MM framework. For a network of $M$ sensors, the design problem is to find the optimal orientation $\mathbf{X}\in\mathbb{R}^{M \times n}$ of the $M$ sensors in an $n-$dimensional coordinate system such that the source location $\mathbf{r}\in\mathbb{R}^n$ can be estimated accurately using the measurements provided by the sensors. If the sensors are positioned at $\mathbf{s}_i\in\mathbb{R}^n$, then the TOA measurement recorded at the $i$th sensor is given by:
\begin{equation}
    t_i = \frac{\|\mathbf{r}-\mathbf{s}_i\|}{c} + \nu_i \quad  \forall\; i=1,\ldots,M,
\end{equation}
    where $c$ is the speed of light and $\nu_i$ is the noise. The corresponding distance model is:
\begin{equation}
    d_i = \|\mathbf{r}-\mathbf{s}_i\| + \eta_i,
\end{equation}
    where $d_i = ct_i$ and $\eta_i = c\nu_i$. When $\boldsymbol{\eta} = [\eta_1,\ldots,\eta_M]^{\mathsf{T}}$ follows a Gaussian distribution with mean vector $\mathbf{0}$ and covariance matrix $\boldsymbol{\Sigma}$, the probability distribution function of $\mathbf{d} = [d_1,\ldots,d_M]^{\mathsf{T}}$ is given by:
\begin{equation}
    p(\mathbf{d};\mathbf{r}) = \frac{ e^{\left[-\displaystyle\frac{1}{2}\left(\mathbf{d} - \Hat{\mathbf{f}}(\mathbf{r})\right)^{\mathsf{T}}\boldsymbol{\Sigma}^{-1}\left(\mathbf{d} - \Hat{\mathbf{f}}(\mathbf{r})\right)\right]}}{\sqrt{\left(2\pi\right)^M \lvert\boldsymbol{\Sigma}\rvert}},
\label{eqn:probdtoa}
\end{equation}
    where
\begin{equation}
    \Hat{\mathbf{f}}(\mathbf{r}) = [\|\mathbf{r}-\mathbf{s}_1\|,\ldots, \|\mathbf{r}-\mathbf{s}_M\|]^{\mathsf{T}}.
\end{equation}
    The corresponding CRB matrix for any unbiased estimate of $\mathbf{r}$ is given by:
\begin{equation}
    \mathbf{C} = \left[\mathbb{E}\left\{\left(\frac{\partial \log p(\mathbf{d};\mathbf{r})}{\partial \mathbf{r}}\right)\left(\frac{\partial \log p(\mathbf{d};\mathbf{r})}{\partial \mathbf{r}}\right)^{\mathsf{T}}\right\}\right]^{-1}.
\label{eqn:crlbtoa}
\end{equation}
    From \eqref{eqn:probdtoa} and \eqref{eqn:crlbtoa} it follows that (see e.g., Appendix B in \cite{stoica2005spectral}):
\begin{equation}
    \mathbf{C} = [\mathbf{X}^{\mathsf{T}}\boldsymbol{\Sigma}^{-1}\mathbf{X}]^{-1},
\end{equation}
    where $\mathbf{X}$ is the orientation matrix:
\begin{equation}
    \mathbf{X} = \begin{bmatrix}
        \displaystyle\frac{\left(\mathbf{r}-\mathbf{s}_1\right)^{\mathsf{T}}}{\|\mathbf{r}-\mathbf{s}_1\|}\\
        \vdots\\
        \displaystyle\frac{\left(\mathbf{r}-\mathbf{s}_M\right)^{\mathsf{T}}}{\|\mathbf{r}-\mathbf{s}_M\|}
    \end{bmatrix}.
    \end{equation}
    The optimal sensor placement is obtained by minimizing the trace of the CRB matrix:
\begin{equation}
\begin{split}
    \min_{\mathbf{X}}\; &\textrm{Tr}[\mathbf{X}^{\mathsf{T}}\boldsymbol{\Sigma}^{-1}\mathbf{X}]^{-1}\\
    \mathrm{s.t.}\;\; &\;\mathbf{x}_i^{\mathsf{T}}\mathbf{x}_i = 1 \quad \forall\; \;i=1,\ldots,M.
\end{split}
\label{eqn:TOAd-optProb}
\end{equation}
    The vector $\mathbf{x}_i$ is the $i$th column of $\mathbf{X}^{\mathsf{T}}$ and the constraint is due to the fact that $\|\mathbf{x}_i\| = 1$. Let the constraint set be denoted by $\mathbb{X}= \{\mathbf{X}\mid \mathbf{x}_i^{\mathsf{T}}\mathbf{x}_i = 1 \;\forall\;i\}$.
    Using the max representation in {Example} \ref{egTrHybrid}, we get the following min-max problem:
\begin{equation}
    \min_{\mathbf{X}\in\mathbb{X}}~\max_{\mathbf{Z}\succeq\mathbf{0}}~\{{g}(\mathbf{X},\mathbf{Z}) = -\textrm{Tr}(\mathbf{Z}\mathbf{X}^{\mathsf{T}}\boldsymbol{\Sigma}^{-1}\mathbf{X}) + 2\textrm{Tr}(\sqrt{\mathbf{Z}})\},
\end{equation}
    where $\mathbf{Z}$ is an auxiliary variable. The term $-\textrm{Tr}(\mathbf{Z}\mathbf{X}^{\mathsf{T}}\boldsymbol{\Sigma}^{-1}\mathbf{X})$ is concave in $\mathbf{X}$ and can be majorized at $\mathbf{X}^t\in\mathbb{X}$ as follows:
\begin{equation}
\hspace*{-15pt}\begin{split}
    -\textrm{Tr}(\mathbf{Z}\mathbf{X}^{\mathsf{T}}\boldsymbol{\Sigma}^{-1}\mathbf{X}) \;\leq \;-2\textrm{Tr}(\mathbf{Z}(\mathbf{X}^t)^{\mathsf{T}}\boldsymbol{\Sigma}^{-1}\mathbf{X}) + \textrm{Tr}(\mathbf{Z}(\mathbf{X}^t)^{\mathsf{T}}\boldsymbol{\Sigma}^{-1}\mathbf{X}^t)
\end{split}
\label{eqn:minTrZXAX}
\end{equation}
    Using \eqref{eqn:minTrZXAX} we get the following surrogate problem:
\begin{equation}
\begin{split}
    \min_{\mathbf{X}\in\mathbb{X}}~\max_{\mathbf{Z}\succeq\mathbf{0}}~\Big\{g_s(\mathbf{X},\mathbf{Z}\mid\mathbf{X}^t) = -2\textrm{Tr}(\mathbf{Z}(\mathbf{X}^t)^{\mathsf{T}}\boldsymbol{\Sigma}^{-1}\mathbf{X}) \\+  \textrm{Tr}(\mathbf{Z}(\mathbf{X}^t)^{\mathsf{T}}\boldsymbol{\Sigma}^{-1}\mathbf{X}^t) + 2\textrm{Tr}(\sqrt{\mathbf{Z}})
    \Big\},
\end{split}
\label{eqn:maxMinAopt}
\end{equation}
    where the function $g_s(\mathbf{X},\mathbf{Z}\mid\mathbf{X}^t)$ is convex-concave.
    Because the objective function in \eqref{eqn:maxMinAopt} is linear in $\mathbf{X}$, the constraint set $\mathbb{X}$ can be relaxed to $\mathbb{X}_r = \{\mathbf{X}\mid \mathbf{x}_i^{\mathsf{T}}\mathbf{x}_i \leq 1\; \forall\; i\}$. The constraint set $\mathbb{X}_r$ is convex and we can invoke the minimax theorem to swap the min and max operators in \eqref{eqn:maxMinAopt}:
\begin{equation}
\begin{split}
    \max_{\mathbf{Z}\succeq\mathbf{0}}~\min_{\mathbf{X}\in\mathbb{X}_r}\;\Big\{g_s(\mathbf{X},\mathbf{Z}\mid\mathbf{X}^t) = -2\textrm{Tr}(\mathbf{Z}(\mathbf{X}^t)^{\mathsf{T}}\boldsymbol{\Sigma}^{-1}\mathbf{X}) \\+ \textrm{Tr}(\mathbf{Z}(\mathbf{X}^t)^{\mathsf{T}}\boldsymbol{\Sigma}^{-1}\mathbf{X}^t) + 2\textrm{Tr}(\sqrt{\mathbf{Z}})
    \Big\}
\end{split}
\end{equation}
Solving the inner minimization problem yields,
\begin{equation}
    \mathbf{\Tilde{x}_i}(\mathbf{Z}\mid\mathbf{X}^t) = \frac{\mathbf{v}_i(\mathbf{Z})}{\|\mathbf{v}_i(\mathbf{Z})\|},
\label{eqn:xtil_Dopt}
\end{equation}
    where $\mathbf{v}_i$ is the $i$th column of the matrix $\mathbf{V} = \mathbf{Z}\left(\mathbf{X}^t\right)^{\mathsf{T}}\boldsymbol{\Sigma}^{-1}$. Substituting $\mathbf{\Tilde{x}_i}(\mathbf{Z}\mid\mathbf{X}^t)$ in $g_s(\mathbf{X},\mathbf{Z}\mid\mathbf{X}^t)$ we get the following convex problem whose solution is the update of $\mathbf{Z}$:
\begin{equation}
\begin{split}
    \mathbf{Z}^{t+1} \!= \! \arg\,\max_{\mathbf{Z}\succeq\mathbf{0}}&\Bigg\{h(\mathbf{Z}\mid\mathbf{X}^t) = -2\sum_{i=1}^M \|\mathbf{Z}\mathbf{w}_i\| + \textrm{Tr}(\mathbf{Z}\mathbf{W}\mathbf{X}^t) + 2\textrm{Tr}(\sqrt{\mathbf{Z}})
    \Bigg\},
\end{split}
\label{eqn:zupd_senPl}
\end{equation}
    where $\mathbf{w}_i$ is the $i$th column of $\mathbf{W} = \left(\mathbf{X}^t\right)^{\mathsf{T}}\boldsymbol{\Sigma}^{-1}$. Finally using $\mathbf{Z}^{t+1}$ in \eqref{eqn:xtil_Dopt} yields the update of $\mathbf{X}$:
\begin{equation}
    \mathbf{{x}_i^{t+1}} = \frac{\mathbf{v}_i(\mathbf{Z}^{t+1})}{\|\mathbf{v}_i(\mathbf{Z}^{t+1})\|}.
\end{equation}

\section{Independent Vector Analysis}

Independent vector analysis (IVA) is of interest in many applications such as multichannel audio processing, biomedical signal processing, digital communication, neural networks and machine learning \cite{comon2010handbookBSS}. IVA deals with separating sources from multiple parallel mixtures with the sources in one mixture being independent among them but possibly depending on at most one source from other mixtures. Given $D$ mixtures of $M$ sources, which are assumed to be statistically independent and following a multivariate distribution, under the linear mixing model the observation vector $\{\mathbf{y}_{dn} \in \mathbb{C}^M\}$ of the $d$th mixture is given by:
\begin{equation}
    \mathbf{y}_{dn} = \mathbf{A}_d \mathbf{s}_{dn}, \quad n = 1,\ldots,N,
\end{equation}
    where $n$ is the sample index such as time or pixel, $\mathbf{A}_d \in \mathbb{C}^{M\times M}$ is the $d$th mixing filter and $\mathbf{s}_{dn} \in \mathbb{C}^M$ is the source vector in the $d$th mixture. To separate the sources from $D$ mixtures, demixing filters $\mathbf{W}_d \in\mathbb{C}^{m \times m}$:
\begin{equation}
    \mathbf{W}_d = [\mathbf{w}_{1d} \ldots \mathbf{w}_{Md}]^{\mathsf{H}},
\end{equation}
    are used to get source estimates:
\begin{equation}
    \mathbf{\Hat{s}}_{dn} = \mathbf{W}_{d}\mathbf{y}_{dn}.
\end{equation}

    The matrices $\mathbf{W}_{d}$ can be estimated by using the ML method, which relies on a contrast function of the source component vectors ($\mathbf{\Tilde{s}}_{kn}$), where the $k$th source component vector at index $n$ is defined as:
\begin{equation}
    \mathbf{\Tilde{s}}_{kn} = \left[s_{k1n} \ldots s_{kDn}\right]^{\mathsf{T}},\quad k = 1,\ldots,M.
\end{equation}
The source component vector takes into account the sample dependence. For spherical super-Gaussian distributions, an auxiliary function that bounds from above the negative log-likelihood function is minimized. Auxiliary function based IVA (AuxIVA) \cite{ono2012auxIVA} solves the following problem:
    \begin{equation}
        \min_{\{\mathbf{w}_{kd}\}} \sum_{k=1}^M \mathbf{w}_{kd}^{\mathsf{H}}\mathbf{V}_{kd}\mathbf{w}_{kd} -2\log\lvert\mathbf{W}_{d}\rvert,
    \label{eqn:auxIVAprob}
    \end{equation}
    where $\mathbf{V}_{kd}$ is a weighing covariance matrix that depends on the auxiliary variables $r_{kn} = \|\mathbf{\Tilde{s}}_{kn}\|$, a spherical super-Gaussian contrast function $G(r_{kn})$ and the observation vectors $\mathbf{y}_{dn}$:
\begin{equation}
    \mathbf{V}_{kd} = \frac{1}{N}\sum_{n=1}^N \frac{G'(r_{kn})}{2r_{kn}}\mathbf{y}_{dn}\mathbf{y}_{dn}^{\mathsf{H}},
\end{equation}
    where $G'(r_{kn})$ is the derivate of $G(r_{kn})$.
    The problem in \eqref{eqn:auxIVAprob} can be solved using the MM4MM methodology. For notational convenience, we drop the subscript $d$ and restate the IVA problem as follows:
\begin{equation}
        \min_{\{\mathbf{w}_k\}} \sum_{k=1}^M \mathbf{w}_{k}^{\mathsf{H}}\mathbf{V}_{k}\mathbf{w}_{k} -2\log\lvert\mathbf{W}\rvert.
\label{eqn:minf_IVA}
\end{equation}
    For the non-convex term $-2\log\lvert\mathbf{W}\rvert$, we use max formulation in {Example} \ref{eglogWWHiva} to write $f(\mathbf{W})$ as:
\begin{equation}
\begin{split}
    f(\mathbf{W}) = \max_{\mathbf{Z}\succ 0} \Bigg\{{g}(\mathbf{W},\mathbf{Z})& =  \sum_{k=1}^M\mathbf{w}_k^{\mathsf{H}}\mathbf{V}_k\mathbf{w}_k +\log\lvert\mathbf{Z}\rvert -  \textrm{Tr}
    (\mathbf{W}\mathbf{W}^{\mathsf{H}}\mathbf{Z}) \Bigg\}
\end{split}
\label{eqn:maxg_IVA}
\end{equation}
    Combining \eqref{eqn:minf_IVA} and \eqref{eqn:maxg_IVA} leads to the following min-max problem:
\begin{equation}
\begin{split}
    \min_{\{\mathbf{w}_k\}}~\max_{\mathbf{Z}\succ 0}~ \Bigg\{{g}(\mathbf{W},\mathbf{Z}) =  \sum_{k=1}^M\mathbf{w}_k^{\mathsf{H}}\mathbf{V}_k\mathbf{w}_k + \log\lvert\mathbf{Z}\rvert - \sum_{k=1}^M\mathbf{w}_k^{\mathsf{H}}\mathbf{Z}\mathbf{w}_k
    \Bigg\}
\end{split}
\end{equation}
    The last term is concave in $\mathbf{w}_k$, and can be majorized as follows:
\begin{equation}
    -\mathbf{w}_k^{\mathsf{H}}\mathbf{Z}\mathbf{w}_k  \leq -2\mathrm{Re}((\mathbf{w}_k^t)^{\mathsf{H}}\mathbf{Z}\mathbf{w}_k) +(\mathbf{w}_k^t)^{\mathsf{H}}\mathbf{Z}\mathbf{w}_k^t \quad \forall\; k.
\label{eqn:majcvxIVA}
\end{equation}
    Using \eqref{eqn:majcvxIVA} we arrive at the following min-max problem with a convex-concave surrogate function $g_s(\mathbf{W},\mathbf{Z}\mid\mathbf{W}^t)$:
\begin{equation}
\begin{split}
    \min_{\{\mathbf{w}_k\}}~\max_{\mathbf{Z}\succ 0}~ \Bigg\{{g}_s&(\mathbf{W},\mathbf{Z}\mid\mathbf{W}^t) = \sum_{k=1}^M\mathbf{w}_k^{\mathsf{H}}\mathbf{V}_k\mathbf{w}_k + \log\lvert\mathbf{Z}\rvert \\&- \sum_{k=1}^M2\mathrm{Re}((\mathbf{w}_k^t)^{\mathsf{H}}\mathbf{Z}\mathbf{w}_k) +\sum_{k=1}^M(\mathbf{w}_k^t)^{\mathsf{H}}\mathbf{Z}\mathbf{w}_k^t
    \Bigg\}
\end{split}
\end{equation}
    We swap the max and min operators, and solve the inner minimization problem in $\{\mathbf{w}_k\}$, whose solution is:
\begin{equation}
    \mathbf{\Tilde{w}}_k(\mathbf{Z}\mid\mathbf{W}^t) = \mathbf{V}_k^{-1}\mathbf{Z}\mathbf{w}_k^t.
\label{eqn:wtil_IVA}
\end{equation}
    Next we substitute \eqref{eqn:wtil_IVA} in the objective function ${g}_s(\mathbf{W},\mathbf{Z}\mid\mathbf{W}^t)$, and solve the resulting convex problem in $\mathbf{Z}$:
\begin{equation}
\begin{split}
    \max_{\mathbf{Z}\succ 0} \Bigg\{h(\mathbf{Z}\mid\mathbf{W}^t) =  \log\lvert\mathbf{Z}\rvert - \sum_{k=1}^M(\mathbf{w}_k^t)^{\mathsf{H}}\mathbf{Z}\mathbf{V}_k^{-1}\mathbf{Z}\mathbf{w}_k^t + \sum_{k=1}^M(\mathbf{w}_k^t)^{\mathsf{H}}\mathbf{Z}\mathbf{w}_k^t\Bigg\},
\end{split}
\end{equation}
    whose solution gives the update for $\mathbf{Z}$:
\begin{equation}
    \mathbf{Z}^{t+1} = \arg ~\max_{\mathbf{Z}\succ 0} ~h(\mathbf{Z}\mid\mathbf{W}^t).
\end{equation}
    The corresponding update of $\mathbf{W}$ is obtained from \eqref{eqn:wtil_IVA}:
\begin{equation}
    \mathbf{w}_k^{t+1} = \mathbf{V}_k^{-1}\mathbf{Z}^{t+1}\mathbf{w}_k^t \quad \forall \; k
\end{equation}

\section{Dual-Function Beamforming Design}\label{DFRC}

A dual-function radar and communication (DFRC) system performs joint radar and communication functions while using the same hardware and spectral resources. The DFRC systems, thus, provide low cost solutions to the problem of spectral congestion \cite{hassanien2019dfrcAsSpecSol} and encompass applications such as autonomous vehicles, unmanned aerial vehicles and military equipment \cite{hassanien2016signalDFRC}. Recent DFRC methods use multiple-input-multiple-output (MIMO) systems, which allow for greater waveform diversity, better estimation performance, and especially the ability to synthesise multiple beams at the same time for communication users and radar targets. \cite{dfrc24Tang} has proposed a design algorithm of the transmit beamforming matrix that minimizes a tight upperbound on Cram\'er-Rao bound (CRB) of the estimated target angles, under a constraint on the signal-to-interference-plus-noise ratio (SINR) of the communication users. In the following we present a derivation of this algorithm within the MM4MM framework.

Consider a MIMO DFRC system that comprises a transmit array with $n_T$ antennas and a receive array (standard uniform linear array) with $n_R$ antennas. Let there be $K$ communication users and $P$ targets with directions of $\theta_1,\ldots,\theta_P$. If the data stream of length $L$ from the transmit array to the $k$th communication user is denoted by $s_k \in \mathbb{C}^{L \times 1}$, then the transmit waveform matrix $\mathbf{X}\in\mathbb{C}^{n_T \times L}$ is given by:
\begin{equation}
    \mathbf{X} = \mathbf{W}\mathbf{S},
\end{equation}
    where $\mathbf{W} = [\mathbf{w}_1, \ldots, \mathbf{w}_K] \in \mathbb{C}^{n_T \times K}$ is the beamforming matrix and $\mathbf{S} = [\mathbf{s}_1,\ldots,\mathbf{s}_K]^{\mathsf{T}}\in\mathbb{C}^{K \times L}$. If the channel response matrix (assumed to be known) is denoted $\mathbf{H} = [\mathbf{h}_1,\ldots,\mathbf{h}_K]^{\mathsf{H}}\in\mathbb{C}^{K \times n_T}$, the matrix $\mathbf{Y}_C \in \mathbb{C}^{K \times L}$ of signals received by the communication users can be written as:
\begin{equation}
\begin{split}
    \mathbf{Y}_C &= \mathbf{H}\mathbf{X} + \boldsymbol{\Sigma}_C,\\
\implies     \mathbf{Y}_C &= \mathbf{H}\mathbf{W}\mathbf{S} + \boldsymbol{\Sigma}_C,
\end{split}
\label{eqn:sigAtCommUsers}
\end{equation}
where $\boldsymbol{\Sigma}_C\in\mathbb{C}^{K \times L}$ is the noise matrix at the $K$ communication users. Let $\mathbf{A}_{T} = [\mathbf{a}_{1,T},\ldots,\mathbf{a}_{P,T}]\in\mathbb{C}^{n_T \times P}$ and $\mathbf{A}_{R} = [\mathbf{a}_{1,R},\ldots,\mathbf{a}_{P,R}]\in\mathbb{C}^{n_R \times P}$, where $\mathbf{a}_{p,T}$ and $\mathbf{a}_{p,R}$ are the transmit and receive array steering vectors at $\theta_p$. Defining $\mathbf{B}$ as a diagonal matrix whose $p^{\mathrm{th}}$ diagonal element $\beta_p$ is the amplitude of the $p^{\mathrm{th}}$ target, the signals received by the DFRC system (under the far-field assumption) are given by:
\begin{equation}
    \mathbf{Y}_{R} = \mathbf{A}_{R}\mathbf{B}\mathbf{A}_{T}^{\mathsf{T}}\mathbf{X} + \boldsymbol{\Sigma_R},
\end{equation}
where $\boldsymbol{\Sigma}_R \in \mathbb{C}^{n_R \times L}$ is the noise matrix at the receiver.
    Each data stream $\mathbf{s}_k$ is assumed to have an average power of 1 and be independent of the other data streams for all $k=1,\ldots,K$. This implies that
\begin{equation}
\label{eqn:avgPower1}
    \frac{1}{L}\mathbf{S}\mathbf{S}^{\mathsf{H}} \approx \mathbf{I}_{K}.
\end{equation}
    Thus, if the maximum transmit energy is $\hat{e}_T$, then
\begin{align}
    \textrm{Tr}(\mathbf{X}\mathbf{X}^{\mathsf{H}}) &= \textrm{Tr}(\mathbf{W}\mathbf{S}\mathbf{S}^{\mathsf{H}}\mathbf{W}^{\mathsf{H}})\\
 & \approx L\textrm{Tr}(\mathbf{W}\mathbf{W}^{\mathsf{H}})\leq \hat{e}_T\\
~~~\implies &\textrm{Tr} (\mathbf{W}\mathbf{W}^{\mathsf{H}}) \leq e_T,&
\label{eqn:maxTranEng}
\end{align}
    where $e_{T} = \hat{e}_T/L$.

The noise power at each of the $K$ communication receivers is assumed to be same, viz. $\sigma_C^2$. Thus, using \eqref{eqn:sigAtCommUsers} and the fact that each data stream has a unity average power (i.e., $\mathbb{E}\{\|\mathbf{s}_k\|^2\} = 1$), the SINR of the $k^{\mathrm{th}}$  user can be written as:
\begin{equation}
    \mathrm{SINR}_k = \frac{|\mathbf{h}_k^{\mathsf{H}}\mathbf{w}_k|^2}{\displaystyle\sum_{\substack{{j=1}\\j\neq k}}^K |\mathbf{h}_k^{\mathsf{H}}\mathbf{w}_j|^2 + \sigma_C^2} \quad \forall~ k = 1,\ldots,K.
\end{equation}
    If $\Hat{\Gamma}_k$ is the minimum SINR required to  guarantee the communication quality of service (QoS), then $\mathrm{SINR}_k \geq \Hat{\Gamma}_k ~\forall~k$, which implies that
\begin{align}
    \frac{|\mathbf{h}_k^{\mathsf{H}}\mathbf{w}_k|^2}{\displaystyle\sum_{\substack{{j=1}\\j\neq k}}^K |\mathbf{h}_k^{\mathsf{H}}\mathbf{w}_j|^2 + \sigma_C^2}\quad &\geq \Hat{\Gamma}_k, \\
    \implies ~~~|\mathbf{h}_k^{\mathsf{H}}\mathbf{w}_k|^2 - \Hat{\Gamma}_k\sum_{\substack{{j=1}\\j\neq k}}^K |\mathbf{h}_k^{\mathsf{H}}\mathbf{w}_j|^2~~&\geq   {\Gamma}_k,
\label{eqn:sinrIneq}
\end{align}
    where $\Gamma_k = \Hat{\Gamma}_k\sigma_C^2$. Defining $\mathbf{H}_k = \mathbf{h}_k\mathbf{h}_k^{\mathsf{H}}$ and $\boldsymbol{\Lambda}_k$ as the diagonal matrix of size $K \times K$ having the $k^{\mathrm{th}}$ diagonal element equal to $1$ and rest equal to $-\Hat{\Gamma}_k$, the inequality in \eqref{eqn:sinrIneq} can be written as:
\begin{align}
    &\textrm{Tr}(\boldsymbol{\Lambda}_k\mathbf{W}^{\mathsf{H}}\mathbf{H}_k\mathbf{W}) ~\geq~ \Gamma_k,\\
    \implies ~~&~~\mathbf{w}\mathbf{\Hat{T}}_k\mathbf{w} ~\geq~ \Gamma_k,
\label{eqn:ineqMinSINR}
\end{align}
    where $\mathbf{w} = \mathrm{vec}(\mathbf{W^*})$ and $\mathbf{\Hat{T}}_k = \boldsymbol{\Lambda}_k\otimes\mathbf{H}_k^{\mathsf{T}}$. Subtracting the minimum eigenvalue of $\mathbf{\Hat{T}}_k$, denoted $\lambda_{\mathrm{min}}(\mathbf{\Hat{T}}_k)$, from the diagonal elements of $\mathbf{\Hat{T}}_k$, we define the matrix $\mathbf{T}_k = \mathbf{\Hat{T}}_k - \lambda_{\mathrm{min}}(\mathbf{\Hat{T}}_k)\mathbf{I}_{n_TK} \left(\succeq \mathbf{0}\right)$ and rewrite the inequality \eqref{eqn:ineqMinSINR} as:
\begin{equation}
    \mathbf{w}^{\mathsf{H}}\mathbf{T}_k\mathbf{w} \geq \gamma_k,
\end{equation}
    where $\gamma_k = \Gamma_k - \lambda_{\mathrm{min}}(\mathbf{T}_k)e_T$.
    Under the assumption that the columns of the noise matrix $\boldsymbol{\Sigma}_R$ are independent and identically distributed, and that their distribution is circularly symmetric complex Gaussian with zero mean and covariance matrix $\sigma_R^2\mathbf{I}_{n_R}$, \cite{dfrc24Tang} has derived an upper bound $\mathbf{C}$ on the CRB matrix for the estimates of $\theta_p$ when the number of receivers $n_R$ is sufficiently large. The $p^{\mathrm{th}}$ diagonal element of $\mathbf{C}$ is:
\begin{equation}
    \mathbf{C}(p,p) \leq \frac{6\sigma_R^2}{|\beta_p|^2n_R^3\mathbf{a}_{p,T}^{\mathsf{H}}\mathbf{X}\mathbf{X}^{\mathsf{H}}\mathbf{a}_{p,T}}.
\end{equation}
    Then the transmit beamforming design problem is to minimize $\textrm{Tr}(\mathbf{C})$ while satisfying the SINR constraint for QoS communication:
\begin{equation}
\begin{split}
    \min_{\mathbf{W}}~&\left\{f(\mathbf{W}) = \sum_{p=1}^P \frac{1}{|\beta_p|^2\mathbf{a}_{p,T}^{\mathsf{T}}\mathbf{W}\mathbf{W}^{\mathsf{H}}\mathbf{a}_{p,T}^*}\right\}\\
    \mathrm{s.t.}~~ &\mathbf{w}^{\mathsf{H}}\mathbf{w} = e_T \\ &\mathbf{w}^{\mathsf{H}}\mathbf{T}_k\mathbf{w}\geq \gamma_k \quad \forall~ k,
\end{split}
\label{eqn:dfrc_desProb}
\end{equation}
    The objective function $f(\mathbf{W})$ can be re-written as:
\begin{equation}
\begin{split}
    f(\mathbf{W}) &= \sum_{p=1}^P \frac{1}{|\beta_p|^2\mathbf{a}_{p,T}^{\mathsf{T}}\mathbf{W}\mathbf{W}^{\mathsf{H}}\mathbf{a}_{p,T}^*}\\
     &= \sum_{p=1}^P \frac{1}{|\beta_p|^2\textrm{Tr}(\mathbf{W}^{\mathsf{H}}\mathbf{a}_{p,T}^*\mathbf{a}_{p,T}^{\mathsf{T}}\mathbf{W})}\\
     &= \sum_{p=1}^P \frac{1}{\mathbf{w}^{\mathsf{H}}\mathbf{A}_p\mathbf{w}},
\end{split}
\end{equation}
    where $\mathbf{A}_p = |\beta_p|^2(\mathbf{I}_K \otimes (\mathbf{a}_{p,T}\mathbf{a}_{p,T}^{\mathsf{H}}))$. The max formulation of $\displaystyle\frac{1}{\mathbf{w}^{\mathsf{H}}\mathbf{A}_p\mathbf{w}}$ presented in {Example} \ref{egDFRCmm} is:
\begin{equation}
     \frac{1}{\mathbf{w}^{\mathsf{H}}\mathbf{A}_p\mathbf{w}} = \max_{z_p\geq 0}~ -z_p\mathbf{w}^{\mathsf{H}}\mathbf{{A}}_p\mathbf{w} + 2\sqrt{z_p}\quad \forall~p.
\label{eqn:maxFfp}
\end{equation}
    Using \eqref{eqn:maxFfp} leads to the following constrained min-max problem:
\begin{equation}
\begin{split}
    \min_{\mathbf{w}}~\max_{\{z_p\} \geq 0}~&\sum_{p=1}^P -z_p\mathbf{w}^{\mathsf{H}}\mathbf{{A}}_p\mathbf{w} + 2\sqrt{z_p}\\
    \mathrm{s.t.}~~~&\mathbf{w}^{\mathsf{H}}\mathbf{{T}}_k\mathbf{w} \geq \gamma_k\quad\forall~k\\
&~\mathbf{w}^{\mathsf{H}}\mathbf{w} = e_{T}.
\end{split}
\end{equation}
    The Lagrangian can be used to handle the $K$ inequality constraints:
\begin{equation}
\mbox{\fontsize{8}{10}\selectfont$\displaystyle\begin{split}
    \min_{\mathbf{w}}\max_{\substack{\{z_p\} \geq 0\\\{\lambda_k\}\geq 0}}& \left\{g(\mathbf{w},\mathbf{z},\boldsymbol{\lambda}) = \displaystyle\sum_{p=1}^P \left[-z_p\mathbf{w}^{\mathsf{H}}\mathbf{{A}}_p\mathbf{w} + 2\sqrt{z_p}\right] +\displaystyle\sum_{k=1}^K \lambda_k\left(\gamma_k -\mathbf{w}^{\mathsf{H}}\mathbf{{T}}_k\mathbf{w}\right)\right\}\\
\mathrm{s.t.}~~&~\mathbf{w}^{\mathsf{H}}\mathbf{w} = e_{T},
\end{split}$}
\label{eqn:minmzxg}
\end{equation}
    where $\{\lambda_k\}$ are the Lagrange multipliers.
    Let
\begin{equation}
    \mathbf{B}(\mathbf{z},\boldsymbol{\lambda}) = \displaystyle\sum_{p=1}^P z_p\mathbf{{A}}_p + \sum_{k=1}^K \lambda_k\mathbf{{T}}_k (\succeq \mathbf{0})
\end{equation}
    and
\begin{equation}
    \hat{g}(\mathbf{z},\boldsymbol{\lambda}) = \displaystyle\sum_{p=1}^P 2\sqrt{z_p} + \sum_{k=1}^K \lambda_k\gamma_k.
\end{equation}
    Then the function $g(\mathbf{w},\mathbf{z},\boldsymbol{\lambda})$ can be written as:
\begin{equation}
    g(\mathbf{w},\mathbf{z},\boldsymbol{\lambda}) = -\mathbf{w}^{\mathsf{H}}\mathbf{B}\mathbf{w} + \hat{g}(\mathbf{z},\boldsymbol{\lambda}).
\label{eqn:augFuncdfrc}
\end{equation}
    The first term in \eqref{eqn:augFuncdfrc} is concave in $\mathbf{w}$ and can be majorized at $\mathbf{w}^t$ as follows:
\begin{equation}
\begin{split}
    -\mathbf{w}^{\mathsf{H}}\mathbf{B}\mathbf{w} \leq -2\mathrm{Re}((\mathbf{w}^t)^{\mathsf{H}}\mathbf{B}\mathbf{w}) + {(\mathbf{w}^t)^{\mathsf{H}}\mathbf{B}\mathbf{w}^t}
\end{split}
\label{eqn:majCcvDfrc}
\end{equation}
    Making use of \eqref{eqn:majCcvDfrc} leads to the following surrogate problem:
\begin{equation}
\begin{split}
    \min_{\mathbf{w}}\max_{\substack{\{z_p\}\geq 0\\\{\lambda_k\}\geq 0}}& \{g_s(\mathbf{w},\mathbf{z},\boldsymbol{\lambda}\mid\mathbf{w}^t) = -2\mathrm{Re}((\mathbf{w}^t)^{\mathsf{H}}\mathbf{B}\mathbf{w}) + {(\mathbf{w}^t)^{\mathsf{H}}\mathbf{B}\mathbf{w}^t} + \hat{g}(\mathbf{z},\boldsymbol{\lambda})\}\\
\mathrm{s.t.}~~&~\mathbf{w}^{\mathsf{H}}\mathbf{w} = e_{T}.
\end{split}
\end{equation}
    The function $g_s(\mathbf{w},\mathbf{z},\boldsymbol{\lambda}\mid\mathbf{w}^t)$ is linear in $\mathbf{w}$ and hence we can relax the constraint $\mathbf{w}^{\mathsf{H}}\mathbf{w} = e_{T}$:
\begin{equation}
\begin{split}
    \min_{\mathbf{w}}~\max_{\substack{\{z_p\} \geq 0\\\{\lambda_k\}\geq 0}}~& -2\mathrm{Re}((\mathbf{w}^t)^{\mathsf{H}}\mathbf{B}\mathbf{w}) + {(\mathbf{w}^t)^{\mathsf{H}}\mathbf{B}\mathbf{w}^t} + \hat{g}(\mathbf{z},\boldsymbol{\lambda})\\
\mathrm{s.t.}~~~&~\mathbf{w}^{\mathsf{H}}\mathbf{w} \leq e_{T}.
\end{split}
\end{equation}
    We can now invoke the minimax theorem to swap the min and max operators, which results in the following max-min problem:
\begin{equation}
\begin{split}
    \max_{\substack{\{z_p\}\geq 0\\\{\lambda_k\}\geq 0}}~\min_{\mathbf{w}}~& -2\mathrm{Re}((\mathbf{w}^t)^{\mathsf{H}}\mathbf{B}\mathbf{w}) + {(\mathbf{w}^t)^{\mathsf{H}}\mathbf{B}\mathbf{w}^t} + \hat{g}(\mathbf{z},\boldsymbol{\lambda})\\
\mathrm{s.t.}~~&~\mathbf{w}^{\mathsf{H}}\mathbf{w} \leq e_{T}.
\end{split}
\label{eqn:maxmingsdfrc}
\end{equation}
    The inner minimization problem has the following solution:
\begin{equation}
    \mathbf{\Tilde{w}}(\mathbf{z},\boldsymbol{\lambda}\mid\mathbf{w}^t) = \sqrt{e_T}\frac{\mathbf{B}\mathbf{w}^t}{\|\mathbf{B}\mathbf{w}^t\|}.
\label{eqn:updw}
\end{equation}
    Substituting $\mathbf{\Tilde{w}}(\mathbf{z},\boldsymbol{\lambda}\mid\mathbf{w}^t)$ in \eqref{eqn:maxmingsdfrc} we obtain the following maximization problem in the variables $\mathbf{z}$ and $\boldsymbol{\lambda}$:
\begin{equation}
\begin{split}
    \max_{\substack{\{z_p\}\geq 0\\\{\lambda_k\}\geq 0}} \Bigg\{h&(\mathbf{z},\boldsymbol{\lambda}\mid\mathbf{w}^t) = -2\sqrt{e_T}\|\mathbf{B}(\mathbf{z},\boldsymbol{\lambda})\mathbf{w}^t\| ~+ \\&{(\mathbf{w}^t)^{\mathsf{H}}\mathbf{B}(\mathbf{z},\boldsymbol{\lambda})\mathbf{w}^t} + \sum_{p=1}^P 2\sqrt{z_p} + \sum_{k=1}^K\lambda_k\gamma_k\Bigg\},
\end{split}
\label{eqn:dualh}
\end{equation}
    which is a convex problem that can be solved using any convex solver. The solution of \eqref{eqn:dualh} is the update $(\mathbf{z}^{t+1},\boldsymbol{\lambda}^{t+1})$. The update of $\mathbf{w}$ is then computed using \eqref{eqn:updw}:
\begin{equation}
    \mathbf{w}^{t+1} = \sqrt{e_T}\frac{\mathbf{B}(\mathbf{z}^{t+1},\boldsymbol{\lambda}^{t+1})\mathbf{w}^t}{\|\mathbf{B}(\mathbf{z}^{t+1},\boldsymbol{\lambda}^{t+1})\mathbf{w}^t\|}.
\end{equation}

\section{Optimal Experiment Design}\label{applNSahuOptExp}

Optimal experiment design has a long history and many applications, see e.g., \cite{mandal2015optimalDesignReview}. In this section we will focus on the optimal experiment design for the following linear regression model:
\begin{equation}
    {y}_j = \mathbf{r}_j^{\mathsf{T}}\boldsymbol{\beta} + {\nu}_j,
\end{equation}
    where ${y}_j$ is the $j^{\mathrm{th}}$ observation, $\mathbf{r}_j\in\mathbb{R}^n$ is the regression vector, $\boldsymbol{\beta}\in\mathbb{R}^n$ is the parameter vector (to be estimated), and ${\nu}_j$ is the noise in the $j^{\mathrm{th}}$ observation. We assume that $\{\nu_j\}$ are independent and identically distributed with zero mean and variance $\sigma^2$. The design space is $\left\{\mathbf{a}_i\in\mathbb{R}^n\right\}_{i=1}^M$ of size $M$. The problem is to find weights $p_i$ such that selecting $\mathbf{r}_j$ from the design space (that is, $\mathbf{a}_i$ in proportion of $p_i$) is optimal with respect to the design criterion.

    The E-optimal design minimizes the maximum eigenvalue of the covariance matrix of the estimated parameter vector:
\begin{equation}
\begin{split}
    \min_{\{p_i\}}\;\;\; &\lambda_{\mathrm{max}}[(\mathbf{A}\mathbf{P}\mathbf{A}^{\mathsf{T}})^{-1}]\\
    \mathrm{s.t.}\;\;\; &\textrm{Tr}(\mathbf{P})= 1, \mathbf{P}\succeq\mathbf{0},
\end{split}
\label{eqn:EdesignProb}
\end{equation}
    where $\lambda_{\mathrm{max}}(\cdot)$ denotes the maximum eigenvalue, $\mathbf{P}$ is the diagonal matrix with $\{p_i\}$ as its diagonal elements, and $\mathbf{A}\in\mathbb{R}^{n \times M}$ is a matrix with $\{\mathbf{a}_i\}$ as column vectors.
    The above problem is convex (more exactly it can be re-written as a semi-definite program) but using a convex solver for it can be computationally inefficient especially in high dimensional cases.
    A faster algorithm for solving \eqref{eqn:EdesignProb} was proposed in \cite{NSahuEoptimal}. Here, we derive this algorithm making use of the MM4MM methodology.

    Using scaled variables $x_i = \alpha p_i$ and $\mathbf{X} = \alpha\mathbf{P}$, the problem in \eqref{eqn:EdesignProb} can be re-written as:
\begin{equation}
\begin{split}
    \min_{\{x_i\},\alpha}\;\;&\alpha\\
    \mathrm{s.t.}\;\;&\sum_{i=1}^M x_i = \alpha, x_i \geq 0, \lambda_{\mathrm{max}}\left[(\mathbf{A}\mathbf{X}\mathbf{A}^{\mathsf{T}})^{-1}\right] \leq 1.
\end{split}
\end{equation}
    The above problem can also be expressed as follows:
\begin{equation}
\begin{split}
    \min_{\{x_i\}}\;\; &\bigg\{f(\mathbf{x}) = \sum_{i=1}^M x_i\bigg\}\\
    \mathrm{s.t.}\;\; &x_i\geq 0,\; (\mathbf{A}\mathbf{X}\mathbf{A}^{\mathsf{T}})^{-1} \preceq \mathbf{I},
\end{split}
\label{eqn:minfxunconOptExp}
\end{equation}
    where $\mathbf{x} = [x_1,\ldots,x_M]$. The Lagrangian associated with \eqref{eqn:minfxunconOptExp} (without the constraint $x_i>0$) is:
\begin{equation}
   g(\mathbf{x},\mathbf{\Bar{Z}}) = \sum_{i=1}^M x_i + \textrm{Tr}(\mathbf{\Bar{Z}}((\mathbf{A}\mathbf{X}\mathbf{A}^{\mathsf{T}})^{-1} - \mathbf{I})),
\end{equation}
    where $\mathbf{\Bar{Z}}$ is the Lagrangian multiplier matrix. The matrix $(\mathbf{A}\mathbf{X}\mathbf{A}^{\mathsf{T}})^{-1}$ can be majorized at $\mathbf{X}^t$ ($\mathbf{X}^t$ is a diagonal matrix with $\mathbf{x}^t$ as diagonal elements) as shown below (assuming $\mathbf{X}\succ \mathbf{0}$) \cite{mm}:
\begin{equation}
    (\mathbf{A}\mathbf{X}\mathbf{A}^{\mathsf{T}})^{-1} \preceq \mathbf{B}^t\mathbf{X}^{-1}(\mathbf{B}^t)^{\mathsf{T}},
\label{eqn:majCovM}
\end{equation}
    where $\mathbf{B}^t = (\mathbf{A}\mathbf{X}^t\mathbf{A}^{\mathsf{T}})^{-1}\mathbf{A}\mathbf{X}^t$ and the equality holds for $\mathbf{X} = \mathbf{X}^t$. It follows that:
\begin{equation}
    \textrm{Tr}(\mathbf{\Bar{Z}}(\mathbf{A}\mathbf{X}\mathbf{A}^{\mathsf{T}})^{-1}) \leq \textrm{Tr}(\mathbf{\Bar{Z}}\mathbf{B}^t\mathbf{X}^{-1}(\mathbf{B}^t)^{\mathsf{T}}).
\label{eqn:IneqEOpt}
\end{equation}
    Using \eqref{eqn:IneqEOpt} we get the following surrogate problem:
\begin{equation}
    \min_{\{x_i > 0\}}~\max_{\mathbf{\Bar{Z}}\succeq \mathbf{0}}~\left\{g_s(\mathbf{x},\mathbf{\Bar{Z}}\mid\mathbf{x}^t) = \sum_{i=1}^M x_i + \textrm{Tr}(\mathbf{\Bar{Z}}\mathbf{B}^t\mathbf{X}^{-1}(\mathbf{B}^t)^{\mathsf{T}}) - \textrm{Tr}({\mathbf{\Bar{Z}}})\right\},
\end{equation}
    Because $g_s(\mathbf{x},\mathbf{\Bar{Z}}\mid\mathbf{x}^t)$ is a convex-concave function, invoking the minimax theorem leads to the following max-min problem:
\begin{equation}
    \max_{\mathbf{\Bar{Z}}\succeq \mathbf{0}}\;\min_{\{x_i>0\}}\sum_{i=1}^M x_i + \textrm{Tr}(\mathbf{\Bar{Z}}\mathbf{B}^t\mathbf{X}^{-1}(\mathbf{B}^t)^{\mathsf{T}}) - \textrm{Tr}({\mathbf{\Bar{Z}}})
\label{eqn:maxminEOptED}
\end{equation}
    The solution of the inner minimization problem is:
\begin{equation}
    \Tilde{x}_i(\mathbf{\Bar{Z}}\mid\mathbf{x}^t) = \sqrt{(\mathbf{b}_i^t)^{\mathsf{T}}\mathbf{\Bar{Z}}\mathbf{b}_i^t},
\label{eqn:xitil_OptExp}
\end{equation}
    where $\mathbf{b}_i^t$ is the $i^{\mathrm{th}}$ column of the matrix $\mathbf{B}^t$. Substituting \eqref{eqn:xitil_OptExp} in \eqref{eqn:maxminEOptED} yields the following convex problem:
\begin{equation}
    \max_{\mathbf{\Bar{Z}}\succeq \mathbf{0}}\; h(\mathbf{\Bar{Z}}\mid\mathbf{x}^t) = 2\sum_{i=1}^M \sqrt{\left(\mathbf{b}_i^t\right)^{\mathsf{T}}\mathbf{\Bar{Z}}\mathbf{b}_i^t} - \textrm{Tr}(\mathbf{\Bar{Z}}),
\label{eqn:maxhzEOptED}
\end{equation}
    which can be solved using a convex solver or in a computationally more efficient way using the MM technique. To use the MM we let $\mathbf{Z}$ denote a square root of $\mathbf{\Bar{Z}}$ (i.e., $\mathbf{\Bar{Z}} = \mathbf{Z}\mathbf{Z}^{\mathsf{T}}$). Then it follows from the Cauchy-Schwartz inequality that:
\begin{equation}
    \|\mathbf{Z}^{\mathsf{T}}\mathbf{b}_i^t\| \geq \frac{\left(\mathbf{b}_i^t\right)^{\mathsf{T}}\mathbf{Z}\mathbf{Z}^k\mathbf{b}_i^t}{\|\left(\mathbf{Z}^k\right)^{\mathsf{T}}\mathbf{b}_i^t\|},
\end{equation}
    where the equality holds at $\mathbf{Z} = \mathbf{Z}^k$ and $\mathbf{Z}^k$ is the point at the current iteration. The above inequality can be used to minorize the objective function in \eqref{eqn:maxhzEOptED}. The maximizer of the so obtained surrogate function is:
\begin{equation}
    \mathbf{Z}^{k+1} = \sum_{i=1}^M \frac{\mathbf{b}_i^t\left(\mathbf{b}_i^t\right)^{\mathsf{T}}\mathbf{Z}^k}{\|\left(\mathbf{Z}^k\right)^{\mathsf{T}}\mathbf{b}_i^t\|}.
\end{equation}
    The variable $\mathbf{Z}$ is iteratively updated till the value of the function $h(\mathbf{\Bar{Z}}\mid \mathbf{x}^t)$ converges and the maximizer $\mathbf{Z}^{t+1}$ of this function is used in \eqref{eqn:xitil_OptExp} to update $\left\{p_i\right\}$:
\begin{equation}
    p_i^{t+1} = \frac{1}{\alpha} \sqrt{(\mathbf{b}_i^t)^{\mathsf{T}}\mathbf{Z}^{t+1}\left(\mathbf{Z}^{t+1}\right)^{\mathsf{T}}\mathbf{b}_i^t}
\end{equation}

\section{Quantum State Discrimination}\label{applQSD}

Determining the quantum states is a key problem for applications such as quantum communication, information processing, and cryptography \cite{bae2015qsdApplcs}. In quantum communication, the quantum states are transmitted through a noisy quantum channel and are detected at the receiver. Distinguishing the quantum states at the receiver's end requires designing a measurement scheme that minimizes the detection error. Let $\boldsymbol{\rho}_i\in\mathbb{C}^{n\times n}$ be the $M$ $n \times n$ positive semi-definite density matrices that are such that $\displaystyle\sum_{i=1}^M \textrm{Tr}(\boldsymbol{\rho}_i) = 1$. The problem of quantum state discrimination (QSD) is to design $M$ positive semi-definite Hermitian measurement operators $\boldsymbol{\Pi}_i\in\mathbb{C}^{n\times n}$, with $\displaystyle\sum_{i=1}^M \boldsymbol{\Pi}_i = \mathbf{I}$, such that the probability of correctly detecting the quantum state is maximized \cite{eldar2003qsdSDP,Wang2008qsdSDP2}:
\begin{equation}
\begin{split}
    \min_{\{\boldsymbol{\Pi}_i\}}\; &-\sum_{i=1}^M\;p_i\textrm{Tr}\left(\boldsymbol{\rho}_i\boldsymbol{\Pi}_i\right) \\
    \mathrm{s.t.}\; &\sum_{i=1}^M \boldsymbol{\Pi}_i = \mathbf{I}, \boldsymbol{\Pi}_i \succeq 0 \quad \forall \;i=1,\ldots,M
\end{split}
\label{eqn:qsdProb}
\end{equation}
    where $\textrm{Tr}\left(\boldsymbol{\rho}_i\boldsymbol{\Pi}_i\right)$ is the probability of correctly detecting $\boldsymbol{\rho}_i$ and $p_i$ is the prior probability of $i^{\mathrm{th}}$ quantum state.

    The above problem is convex but the computational cost of solving it as a semi-definite program (SDP) can be rather high for large values of $M$. In this section we derive a computationally efficient algorithm for \eqref{eqn:qsdProb} using the MM4MM methodology. Letting $\mathbf{X}_i$ denote a square root of $\boldsymbol{\Pi}_i$ ($\boldsymbol{\Pi}_i = \mathbf{X}_i\mathbf{X}_i^{\mathsf{H}}$), and relaxing the semi-orthogonality constraint in \eqref{eqn:qsdProb}, we can re-write the problem as:
\begin{equation}
\begin{split}
    \min_{\{\mathbf{X}_i\}}\;& \left\{f(\mathbf{X}) =  -\sum_{i=1}^M \textrm{Tr}\left(\mathbf{X}_i^{\mathsf{H}}p_i\boldsymbol{\rho}_i\mathbf{X}_i\right)\right\}\\
    \mathrm{s.t.}\;& \sum_{i=1}^M \mathbf{X}_i\mathbf{X}_i^{\mathsf{H}} \preccurlyeq \mathbf{I},
\end{split}
\label{eqn:minfxi_qsdProb}
\end{equation}
    which is possible because as shown below the solution to \eqref{eqn:minfxi_qsdProb} satisfies the equality constraint in \eqref{eqn:qsdProb}.
    Using the Lagrangian we write \eqref{eqn:minfxi_qsdProb} in the following equivalent form:
\begin{equation}
\begin{split}
    \min_{\{\mathbf{X}_i\}}\;\max_{\mathbf{Z}\succeq 0} \Bigg\{{g}(\mathbf{X},\mathbf{Z}) =\sum_{i=1}^M-\textrm{Tr}(\mathbf{X}_i^{\mathsf{H}}p_i\boldsymbol{\rho}_i\mathbf{X}_i) \\+ \textrm{Tr}\left(\mathbf{Z}\left(\sum_{i=1}^M\mathbf{X}_i\mathbf{X}_i^{\mathsf{H}} - \mathbf{I}\right)\right)\Bigg\}.
\end{split}
\end{equation}
    The first term in ${g}(\mathbf{X}, \mathbf{Z})$ is concave in $\mathbf{X}_i$ and thus it can be majorized as follows:
\begin{equation}
    -\sum_{i=1}^M \textrm{Tr}(\mathbf{X}_i^{\mathsf{H}}p_i\boldsymbol{\rho}_i\mathbf{X}_i) \leq -2\mathrm{Re}\left(\sum_{i=1}^M\textrm{Tr}(\mathbf{X}_i^{\mathsf{H}}p_i\boldsymbol{\rho}_i\mathbf{X}_i^t)\right) +\; \textrm{const.}
\end{equation}
    This observation leads to the following surrogate problem:
\begin{equation}
\begin{split}
    \min_{\{\mathbf{X}_i\}} \max_{\mathbf{Z}\geq 0} \Bigg\{g_s(\mathbf{X},\mathbf{Z}\mid\mathbf{X}^t) =  \sum_{i=1}^M -2\mathrm{Re}(\textrm{Tr}(\mathbf{X}_i^{\mathsf{H}}p_i\boldsymbol{\rho}_i\mathbf{X}_i^t))\\ + \sum_{i=1}^M\textrm{Tr}(\mathbf{X}_i^{\mathsf{H}}\mathbf{Z}\mathbf{X}_i) - \textrm{Tr}(\mathbf{Z})\Bigg\}.
\end{split}
\label{eqn:minmaxSurr_qsd}
\end{equation}
    Because the function $g_s(\mathbf{X},\mathbf{Z}\mid\mathbf{X}^t)$ is convex-concave, we can invoke the minimax theorem to swap the min and max operators in \eqref{eqn:minmaxSurr_qsd}. The solution of the inner minimization of the so obtained problem is:
\begin{equation}
    \mathbf{\Tilde{X}}_i(\mathbf{Z}\mid\mathbf{X}^t) = p_i\mathbf{Z}^{-1}\boldsymbol{\rho}_i\mathbf{X}_i^t
\label{eqn:xtil_qsd}
\end{equation}
    Substituting it in $g_s(\mathbf{X},\mathbf{Z}\mid\mathbf{X}^t)$ leaves the following problem in $\mathbf{Z}$ to be solved:
\begin{equation}
\begin{split}
    \min_{\mathbf{Z}\geq 0} \; \Bigg\{&-g_s(\{\mathbf{\Tilde{X}}_i(\mathbf{Z}\mid\mathbf{X}^t)\},\mathbf{Z}\mid\mathbf{X}^t) = \\&\sum_{i=1}^M p_i^2\textrm{Tr}((\mathbf{X}_i^t)^{\mathsf{H}}\boldsymbol{\rho}_i^{\mathsf{H}}\mathbf{Z}^{-1}\boldsymbol{\rho}_i\mathbf{X}_i^t) + \textrm{Tr}(\mathbf{Z})
    \Bigg\}
\end{split}
\label{eqn:minZsurr_qsd}
\end{equation}
    The minimizer of \eqref{eqn:minZsurr_qsd} is the update of $\mathbf{Z}$:
\begin{equation}
    \mathbf{Z}^{t+1} = \left(\sum_{i=1}^M p_i^2\boldsymbol{\rho}_i\mathbf{X}_i^t(\mathbf{X}_i^t)^{\mathsf{H}}\boldsymbol{\rho}_i^{\mathsf{H}}\right)^{1/2},
\end{equation}
    The update of $\mathbf{X}$ is obtained from \eqref{eqn:xtil_qsd}:
\begin{equation}
    \mathbf{X}_i^{t+1} = p_i(\mathbf{Z}^{t+1})^{-1}\boldsymbol{\rho}_i\mathbf{X}_i^t,
\end{equation}
    which satisfies the equality constraint in \eqref{eqn:qsdProb}, i.e., $\displaystyle\sum_{i=1}^M \mathbf{X}_i^{t+1}(\mathbf{X}_i^{t+1})^{\mathsf{H}}\break =  \mathbf{I}$.

\section{Fair PCA}\label{appFairPCA}

Principal component analysis (PCA) is a dimensionality reduction technique that has found a wide-range of applications in finance, biomedicine and signal processing \cite{abdi2010pcaApplbook}. Given a mean centered data matrix $\mathbf{Y}\in\mathbb{R}^{n\times M}$, the goal is to find $r$ ($r\leq n$) orthogonal principal components $\{\mathbf{x}_i\}_{i=1}^r\in\mathbb{R}^n$ such that the variance of the data along the $\left\{\mathbf{x}_i\right\}$ is maximized, that is,
\begin{equation}
\begin{split}
    \max_{\{\mathbf{x}_i\}} \;\; &\sum_{i=1}^r \mathbf{x}_i^{\mathsf{T}}\mathbf{C}\mathbf{x}_i\\
    \mathrm{s.t.}\;\; &\|\mathbf{x}_i\| = 1; \mathbf{x}_i^{\mathsf{T}}\mathbf{x}_j = 0 \quad \forall\;\; i\neq j=1,\ldots,r
\end{split}
\end{equation}
    where $\mathbf{C} = \mathbf{Y}\mathbf{Y}^{\mathsf{T}}$ is the sample covariance matrix. With $\mathbf{X} = [\mathbf{x}_1,\ldots,\mathbf{x}_r]$, the PCA problem can be written as:
\begin{equation}
\begin{split}
    \max_{\mathbf{X}}\;\; &\textrm{Tr}(\mathbf{X}^{\mathsf{T}}\mathbf{C}\mathbf{X})\\
    \mathrm{s.t.}\;\;&\mathbf{X}^{\mathsf{T}}\mathbf{X} = \mathbf{I}
\end{split}
\label{eqn:maxAvgVarPCA}
\end{equation}
    The solution of \eqref{eqn:maxAvgVarPCA}, i.e., the  eigenvectors of $\mathbf{C}$ corresponding to the $r$ largest eigenvalues, may provide a good fit for some data samples but not for others, especially in the case that the data samples come from different classes. Assume that there are $k=1,\ldots,K$ classes from which the $M$ samples in $\mathbf{Y}$ are drawn. Let $\mathbf{Y}_k$ denote the data matrix of the $k^{\mathrm{th}}$ class and let $\mathbf{C}_k$ be the sample covariance matrix of $\mathbf{Y}_k$. To ensure fairness for all classes, the fair PCA problem can be formulated as follows:
\begin{equation}
\begin{split}
    \max_{\mathbf{X}}~&f(\mathbf{X}) =\bigg\{ \min_{k\in[1,K]}\;f_k(\mathbf{X}) = \textrm{Tr}(\mathbf{X}^{\mathsf{T}}\mathbf{C}_k\mathbf{X})\bigg\}\\
    \mathrm{s.t.}\;\;\;&\mathbf{X}^{\mathsf{T}}\mathbf{X} = \mathbf{I}
\end{split}
\label{eqn:fPCAprob}
\end{equation}
\cite{PBabuFPCA} has introduced an algorithm for fair PCA. In this section we derive this algorithm within the MM4MM framework.
\par
    Using the result of {Example} \ref{egMaxElt} we can re-write \eqref{eqn:fPCAprob} in the following form:
\begin{equation}
\begin{split}
    \max_{\mathbf{X}} \;\min_{\mathbf{z}}\; \left\{{g}(\mathbf{X},\mathbf{z}) = \sum_{k=1}^K z_kf_k(\mathbf{X})\right\}\\
    \mathrm{s.t.}\;\;\;z_k\geq 0, \mathbf{z}^{\mathsf{T}}\mathbf{1} = 1, \mathbf{X}^{\mathsf{T}}\mathbf{X} = \mathbf{I},
\end{split}
\label{eqn:maxminfpcaProb}
\end{equation}
    where $\mathbf{z} = [z_1,\ldots,z_K]^{\mathsf{T}}$ is the auxiliary variable. The problem \eqref{eqn:maxminfpcaProb} is not convex in $\mathbf{X}$. Let us relax the non-convex constraint $\mathbf{X}^{\mathsf{T}}\mathbf{X} = \mathbf{I}$ as shown below:
\begin{align}
    \mathbf{X}^{\mathsf{T}}\mathbf{X} \preccurlyeq \mathbf{I},\\
    \implies \begin{bmatrix}
        \mathbf{I}_r& \mathbf{X}^{\mathsf{T}}\\
        \mathbf{X} & \mathbf{I}_n
    \end{bmatrix} \succcurlyeq {0}
\label{eqn:relConst_FPCA}
\end{align}
    We will show later that the relaxed constraint is tight (i.e., its solution satisfies the equality constraint in \eqref{eqn:maxminfpcaProb}).

    Next we note that minorizing the convex function $f_k(\mathbf{X}) =\break \textrm{Tr}(\mathbf{X}^{\mathsf{T}}\mathbf{C}_k\mathbf{X})$ at a feasible point $\mathbf{X}^t$, we get:
\begin{equation}
    \textrm{Tr}(\mathbf{X}^{\mathsf{T}}\mathbf{C}_k\mathbf{X})\geq 2\textrm{Tr}((\mathbf{X}^t)^{\mathsf{T}}\mathbf{C}_k\mathbf{X}) - \textrm{Tr}((\mathbf{X}^t)^{\mathsf{T}}\mathbf{C}_k\mathbf{X}^t).
\label{eqn:lb_fpca}
\end{equation}
    Using \eqref{eqn:relConst_FPCA} and \eqref{eqn:lb_fpca} we arrive at the following surrogate problem with a concave-convex objective function:
\begin{equation}
\begin{split}
    \max_{\mathbf{X}}\;\min_{\mathbf{z}}\;&\left\{g_s(\mathbf{X},\mathbf{z}\mid\mathbf{X}^t) = 2\textrm{Tr}(\left(\mathbf{A}(\mathbf{z})\right)^{\mathsf{T}}\mathbf{X}) + \mathbf{z}^{\mathsf{T}}\mathbf{d}\right\}\\
    \mathrm{s.t.}\;\; &z_k\geq 0,\; \mathbf{z}^{\mathsf{T}}\mathbf{1} = 1, \begin{bmatrix}
        \mathbf{I}_r& \mathbf{X}^{\mathsf{T}}\\
        \mathbf{X} & \mathbf{I}_n
    \end{bmatrix} \succcurlyeq {0},
\end{split}
\label{eqn:maxmminFPCA}
\end{equation}
    where
\begin{equation}
    \mathbf{A}(\mathbf{z}) = \sum_{k=1}^K z_k\mathbf{C}_k\mathbf{X}^t,
\end{equation}
    and $\mathbf{d}$ is a vector with the following $k^{\mathrm{th}}$ element:
\begin{equation}
    d_k = - \textrm{Tr}((\mathbf{X}^t)^{\mathsf{T}}\mathbf{C}_k\mathbf{X}^t).
\end{equation}
    As $g_s(\mathbf{X},\mathbf{z}\mid\mathbf{X}^t)$ is concave-convex, we can use minimax theorem to swap the max and min operators in \eqref{eqn:maxmminFPCA}:
\begin{equation}
\begin{split}
    \min_{\mathbf{z}}\;\max_{\mathbf{X}}\;\;&2\textrm{Tr}(\mathbf{A}^{\mathsf{T}}\mathbf{X}) + \mathbf{z}^{\mathsf{T}}\mathbf{d}\\
    \mathrm{s.t.}~~~ &z_k\geq 0,\; \mathbf{z}^{\mathsf{T}}\mathbf{1} = 1, \begin{bmatrix}
        \mathbf{I}_r& \mathbf{X}^{\mathsf{T}}\\
        \mathbf{X} & \mathbf{I}_n
    \end{bmatrix} \succcurlyeq {0}.
\end{split}
\label{eqn:minmaxFPCA}
\end{equation}
    If $\left\{\sigma_i(\mathbf{A})\right\}$ and $\left\{\sigma_i(\mathbf{X})\right\}$ denote the non-zero singular values of the matrices $\mathbf{A}$ and $\mathbf{X}$, then using von-Neumann inequality \cite{marshall1979inequalities}, we have that:
\begin{equation}
\begin{split}
    \textrm{Tr}(\mathbf{A}^{\mathsf{T}}\mathbf{X}) &\leq \sum_{i=1}^r \sigma_i(\mathbf{A})\sigma_i(\mathbf{X})\\
     &\leq \sum_{i=1}^r\sigma_i(\mathbf{A}) = \textrm{Tr}((\mathbf{A}^{\mathsf{T}}\mathbf{A})^{\frac{1}{2}})
\end{split}
\label{eqn:ineqNeuFPCA}
\end{equation}
    The second inequality above is implied by the fact that $\sigma_i(\mathbf{X})\leq 1$. It follows from \eqref{eqn:ineqNeuFPCA} that the maximizer $\mathbf{X}$ is given by:
\begin{equation}
    \mathbf{\Tilde{X}}(\mathbf{z}\mid\mathbf{X}^t) = \mathbf{A}(\mathbf{A}^{\mathsf{T}}\mathbf{A})^{-\frac{1}{2}}.
\label{eqn:xtil_FPCA}
\end{equation}
    Observe that $\mathbf{\Tilde{X}}(\mathbf{z}\mid\mathbf{X}^t)$ satisfies the equality constraint in \eqref{eqn:maxminfpcaProb}, as required. Substituting $\mathbf{\Tilde{X}}(\mathbf{z}\mid\mathbf{X}^t)$ in \eqref{eqn:minmaxFPCA} leads to the following problem, whose solution is the update of $\mathbf{z}$:
\begin{equation}
\begin{split}
    \mathbf{z}^{t+1} = \arg \;&\min_{\mathbf{z}}\; 2\sum_{i=1}^r\sigma_i(\mathbf{A}) + \mathbf{z}^{\mathsf{T}}\mathbf{d}\\
    &\mathrm{s.t.}\; z_k\geq0, \;\mathbf{z}^{\mathsf{T}}\mathbf{1} = 1.
\end{split}
\label{eqn:zUpdFPCA}
\end{equation}
    The above nuclear norm minimization problem is convex and it can be solved by any of a number of convex solvers \cite{recht2010NucNormSDP}. The update $\mathbf{X}^{t+1}$ is simply $\mathbf{\Tilde{X}}(\mathbf{z}\mid\mathbf{X}^t)$ in \eqref{eqn:xtil_FPCA} with $\mathbf{z}=\mathbf{z}^{t+1}$.

    When there is prior information that the \textit{fair principal components are sparse}, solving the following $\ell_1$ norm penalized fair PCA problem becomes of interest:
\begin{equation}
\begin{split}
    \max_{\mathbf{X}}\;&\Tilde{f}(\mathbf{X}) = \left\{\min_{k\in[1,K]}\; f_k(\mathbf{X}) - \lambda\|\mathbf{X}\|_1\right\}\\
    \mathrm{s.t.}\;&\; \mathbf{X}^{\mathsf{T}}\mathbf{X} = \mathbf{I},
\end{split}
\label{eqn:maxfPCA}
\end{equation}
    where $\lambda$ is a regularization parameter that controls sparsity. For the first term in \eqref{eqn:maxfPCA} we use the max formulation in \eqref{eqn:maxminfpcaProb}. For the $\ell_1$ norm we extend the representation in \eqref{eq:dualL1norm} to matrices:
\begin{equation}
    \|\mathbf{X}\|_1 = \max_{|w_{ij}|\leq1} \textrm{Tr}(\mathbf{W}^{\mathsf{T}}\mathbf{X}),
\end{equation}
    where $w_{ij}$ are the elements of the auxiliary variable $\mathbf{W}\in \mathbb{R}^{n\times r}$. This implies that $-\lambda\|\mathbf{X}\|_1$ can be represented as:
\begin{equation}
    -\lambda\|\mathbf{X}\|_1  = \min_{|w_{ij}|\leq1} \lambda\textrm{Tr}(\mathbf{W}^{\mathsf{T}}\mathbf{X}).
\end{equation}
    and the function $\Tilde{f}(\mathbf{X})$ as:
\begin{equation}
\begin{split}
    \Tilde{f}(\mathbf{X}) = \min_{\mathbf{z},\mathbf{W}}&\Bigg\{{{g}}(\mathbf{X},\mathbf{z},\mathbf{W}) = \displaystyle\sum_{k=1}^K z_kf_k(\mathbf{X}) + \lambda\textrm{Tr}(\mathbf{W}^{\mathsf{T}}\mathbf{X})\Bigg\}\\
    \mathrm{s.t.}\; &z_k\geq0, \;\mathbf{z}^{\mathsf{T}}\mathbf{1} = 1, |w_{ij}|\leq 1 \quad \forall\;k,i,j
\end{split}
\end{equation}
    Using the surrogate function in \eqref{eqn:lb_fpca} and the relaxed constraint in \eqref{eqn:relConst_FPCA} once again, the max-min problem becomes:
\begin{equation}
\begin{split}
    \displaystyle\max_{\mathbf{X}}\;\displaystyle\min_{\mathbf{z},\mathbf{W}}\;&\big\{{g_s}(\mathbf{X},\mathbf{z},\mathbf{W}\mid\mathbf{X}^t) =\\ &2\textrm{Tr}(\mathbf{A}^{\mathsf{T}}\mathbf{X}) + \mathbf{z}^{\mathsf{T}}\mathbf{d} + \lambda \textrm{Tr}(\mathbf{W}^{\mathsf{T}}\mathbf{X})\big\}\\
    \mathrm{s.t.}~&~z_k\geq0, \;\mathbf{z}^{\mathsf{T}}\mathbf{1} = 1, |w_{ij}|\leq 1, \;\begin{bmatrix}
        \mathbf{I}_r& \mathbf{X}^{\mathsf{T}}\\
        \mathbf{X} & \mathbf{I}_n
    \end{bmatrix} \succcurlyeq {0}.
\end{split}
\label{eqn:maxminfspca}
\end{equation}
    Invoking the minimax theorem and solving the inner maximization problem in $\mathbf{X}$ gives:
\begin{equation}
    \mathbf{\Tilde{X}}(\mathbf{z},\mathbf{W}\mid\mathbf{X}^t) = \mathbf{V}(\mathbf{V}^{\mathsf{T}}\mathbf{V})^{-\frac{1}{2}},
\label{eqn:xtil_FSPCA}
\end{equation}
    where
\begin{equation}
    \mathbf{V}(\mathbf{z},\mathbf{W}) = \mathbf{A}(\mathbf{z}) + \frac{\lambda}{2}\mathbf{W}.
\end{equation}
    Observe that the above $\mathbf{\Tilde{X}}(\mathbf{z},\mathbf{W}\mid\mathbf{X}^t)$ satisfies the equality constraint in \eqref{eqn:maxminfpcaProb} as it should. Substituting $\mathbf{\Tilde{X}}(\mathbf{z},\mathbf{W}\mid\mathbf{X}^t)$ in \eqref{eqn:maxminfspca} leads to the following convex problem whose solutions are the updates of $\mathbf{z}$ and $\mathbf{W}$:
\begin{equation}
\begin{split}
    \mathbf{z}^{t+1},\mathbf{W}^{t+1} = \arg\;\min_{\mathbf{z},\mathbf{W}}&\left\{h(\mathbf{z},\mathbf{W}\mid\mathbf{X}^t) = 2\textrm{Tr}((\mathbf{V}^{\mathsf{T}}\mathbf{V})^{\frac{1}{2}}) + \mathbf{z}^{\mathsf{T}}\mathbf{d}\right\}\\
    \mathrm{s.t.}\;&z_k\geq0, \;\mathbf{z}^{\mathsf{T}}\mathbf{1} = 1, |w_{ij}|\leq 1.
\end{split}
\end{equation}
    Finally the update $\mathbf{X}^{t+1}$ is obtained using $\mathbf{z} = \mathbf{z}^{t+1}$ and $\mathbf{W} = \mathbf{W}^{t+1}$ in \eqref{eqn:xtil_FSPCA}.




%

\chapter{Conclusions}\label{sec:conclusion}

This monograph introduced a general framework for reformulating a minimization problem as a min-max problem and using MM to develop an MM4MM algorithm that can be used to find a solution of the original possibly non-convex problem. As a pre-requisite for developing the MM4MM algorithm, we have presented the theory that underlines the max formulations of several non-convex functions. The discussed examples of max formulations include some non-convex functions that appear in signal processing applications and pave the way for devising such formulations for other non-convex functions in a host of application areas. Next we presented the main steps of the MM4MM methodology. Ten algorithms developed using MM4MM for ten important signal processing optimization problems were described in detail.

We believe that the MM4MM framework can be used to develop efficient algorithms for optimization problems that appear in many other applications besides those discussed in this monograph. The alternating direction method of multipliers (ADMM) \cite{boydadmm} is another optimization methodology, which has been used in the last years for solving several signal processing optimization problems. We finish this monograph with a brief comparison of MM4MM and ADMM:
\begin{itemize}
\item For a given minimization problem, deriving either an ADMM or MM4MM based algorithm requires some ingenuity. For instance, in the case of MM4MM, one has to come up with a suitable choice of the augmented function $g(\mathbf{x},\mathbf{z})$. Similarly ADMM requires the judicious choice of certain auxiliary variables which is done via variable splitting. Nonetheless, there are many guidelines for making these choices (see section V for MM4MM and \cite{boydadmm} for ADMM) that can aid in the derivation of the respective algorithms, therefore this is not considered to be a serious impediment of either methodology.
\item ADMM requires the selection of a penalty parameter that dictates the speed of convergence of the algorithm. This selection is not trivial and the need for making it can be seen as a drawback of ADMM. MM4MM, on the other hand, is hyper-parameter free.
\item Because MM4MM uses the MM technique, it inherits its property of monotonic convergence to a stationary point under mild conditions. In contrast to this, the iterates of ADMM do not monotonically decrease the cost function and moreover for non-convex problems an ADMM algorithm with an imprecise choice of the penalty parameter might not converge to a stationary point of the objective function.
\end{itemize}

\chapter*{Appendix}

\chaptermark{Appendix}

\makeatletter
\addcontentsline{toc}{chapter}{Appendix}
\makeatother

\def\theequation{A.\arabic{equation}}
\setcounter{equation}{0}

\def\thetheorem{A.\arabic{theorem}}
\setcounter{theorem}{0}

\begin{lemma}
\label{fact_dcvx}
    Let
\begin{equation}
    d(\mathbf{x}) = \max_{\mathbf{z}\in \mathbb{Z}} g(\mathbf{x},\mathbf{z}), \;\; \mathbf{x}\in \mathbb{X}
\label{eqn:dx}
\end{equation}
    If $g(\mathbf{x},\mathbf{z})$ is strictly convex in $\mathbf{x}$ for $\forall \;\mathbf{z} \in \mathbb{Z}$ (but not necessarily concave in $\mathbf{z}$, nor differentiable) then $d(\mathbf{x})$ is a strictly convex function.
\end{lemma}
\renewcommand\qedsymbol{$\blacksquare$}
\begin{proof}
    For any $\lambda \in \left(0, 1\right)$ and $\mathbf{x}_1 \neq \mathbf{x}_2 \in \mathbb{X}$ we have:
\begin{equation}
\begin{split}
    g(\lambda \mathbf{x}_1 + (1-\lambda)\mathbf{x}_2, \mathbf{z}) &< \lambda g(\mathbf{x}_1, \mathbf{z}) + (1-\lambda) g(\mathbf{x}_2, \mathbf{z}) \\&\leq \lambda d(\mathbf{x}_1) + (1-\lambda)d(\mathbf{x}_2),
\end{split}
\label{eqn:convIneqg}
\end{equation}
    where the first inequality follows from the fact that $g(\mathbf{x},\mathbf{z})$ is strictly convex in $\mathbf{x}$, and the second from the definition of $d(\mathbf{x})$. The right hand side in \eqref{eqn:convIneqg} does not depend on $\mathbf{z}$. Consequently \eqref{eqn:convIneqg} implies:
\begin{equation}
\begin{split}
    \lambda d(\mathbf{x}_1) + (1 - \lambda)d(\mathbf{x}_2) &> \max_{\mathbf{z} \in \mathbb{Z}} \; g(\lambda \mathbf{x}_1 + (1-\lambda)\mathbf{x}_2, z)\\
    & = d(\lambda \mathbf{x}_1 + (1-\lambda) \mathbf{x}_2)
\end{split}
\end{equation}
    which proves the strict convexity of $d(\mathbf{x})$.
\end{proof}
    Next assume that $g(\mathbf{x},\mathbf{z})$ is strictly convex in $\mathbf{x}\in \mathbb{X}$ and strictly concave in $\mathbf{z}\in \mathbb{Z}$, $\mathbb{X}$ and $\mathbb{Z}$ are convex sets and $g(\mathbf{x},\mathbf{z})$ is a differentiable function. We also assume that the following functions are defined and differentiable on $\mathbb{X}$ and, respectively, $\mathbb{Z}$:
\begin{equation}
    \Tilde{\mathbf{x}}(\mathbf{z}) = \arg \min_{\mathbf{x} \in \mathbb{X}} \; g(\mathbf{x},\mathbf{z}) \quad  \forall\; \mathbf{z} \in \mathbb{Z}
\label{eqn:xTil}
\end{equation}
\begin{equation}
    \Tilde{\mathbf{z}}(\mathbf{x}) = \arg \max_{\mathbf{z} \in \mathbb{Z}} \; g(\mathbf{x},\mathbf{z}) \quad  \forall\; \mathbf{x} \in \mathbb{X}.
\label{eqn:zTil}
\end{equation}
    The above assumptions are stronger than those under which the result presented below is known to hold (see e.g., \cite{MSionMinimax}), but they are satisfied in all applications of this monograph and considerably simplify the proof of Lemma~\ref{fact_minimax}.

\begin{lemma}
\label{fact_minimax}
    Under the stated assumptions,
\begin{equation}
    \min_{\mathbf{x} \in \mathbb{X}}\; \max_{\mathbf{z}\in\mathbb{Z}}\; g(\mathbf{x},\mathbf{z})  = \max_{\mathbf{z}\in\mathbb{Z}} \;\min_{\mathbf{x}\in\mathbb{X}} \;g(\mathbf{x},\mathbf{z}).
\label{eqn:minmaxmaxmin}
\end{equation}
    Furthermore the minimax/maximin optimal pair $(\mathbf{x}^*,\mathbf{z}^*)$, which is given by
\begin{equation}
    \mathbf{x}^* = \arg \min_{\mathbf{x} \in \mathbb{X}}\; \underbrace{g(\mathbf{x},\Tilde{\mathbf{z}}(\mathbf{x}))}_{d(\mathbf{x})}
\end{equation}
\begin{equation}
    \mathbf{z}^* = \arg \max_{\mathbf{z} \in \mathbb{Z}}\; \underbrace{g(\Tilde{\mathbf{x}}(\mathbf{z}),\mathbf{z})}_{h(\mathbf{z})},
\label{eqn:zopt_defn}
\end{equation}
    is unique (as $d(\mathbf{x})$ and $h(\mathbf{z})$ are strictly convex and, respectively, concave functions. See Lemma~\ref{fact_dcvx}) and is the saddle point of $g(\mathbf{x},\mathbf{z})$:
\begin{equation}
    g(\mathbf{x}^*,\mathbf{z}) \leq g(\mathbf{x}^*,\mathbf{z}^*) \leq g(\mathbf{x},\mathbf{z}^*)\quad  \forall\; \mathbf{x} \in \mathbb{X}\quad
\emph{and}\quad  \forall\; \mathbf{z} \in \mathbb{Z}.
\end{equation}
\end{lemma}

\begin{proof}
    From the inequality
\begin{equation}
    \max_{\mathbf{z} \in\mathbb{Z}} g(\mathbf{x},\mathbf{z}) \geq g(\mathbf{x},\mathbf{z}) \quad  \forall\;\mathbf{x} \in \mathbb{X}\quad  \textrm{and}\quad  \forall\;\mathbf{z} \in \mathbb{Z},
\end{equation}
    we have that
\begin{equation}
    \min_{\mathbf{x} \in \mathbb{X}}\; \max_{\mathbf{z} \in \mathbb{Z}}\; g(\mathbf{x},\mathbf{z}) \geq \min_{\mathbf{x} \in \mathbb{X}} g(\mathbf{x},\mathbf{z}) \quad  \forall\;\mathbf{z} \in \mathbb{Z}
\end{equation}
    which, in turn, implies:
\begin{equation}
    \min_{\mathbf{x} \in \mathbb{X}}\; \max_{\mathbf{z} \in \mathbb{Z}}\; g(\mathbf{x},\mathbf{z}) \geq \max_{\mathbf{z} \in \mathbb{Z}} \; \min_{\mathbf{x} \in \mathbb{X}} \; g(\mathbf{x},\mathbf{z}).
\end{equation}
    To prove \eqref{eqn:minmaxmaxmin} we will show that, under the assumptions made, we also have:
\begin{equation}
    \min_{\mathbf{x} \in \mathbb{X}}\; \max_{\mathbf{z} \in \mathbb{Z}}\; g(\mathbf{x},\mathbf{z}) \leq \max_{\mathbf{z} \in \mathbb{Z}}\; \min_{\mathbf{x} \in \mathbb{X}}\; g(\mathbf{x},\mathbf{z}).
\label{eqn:minmaxleq}
\end{equation}
    A straightforward use of the chain rule for derivatives yields the following equality:
\begin{equation}
\begin{split}
    \diff{g(\mathbf{x},\Tilde{\mathbf{z}}(\mathbf{x}))}{\mathbf{x}} &= \frac{\partial g(\mathbf{x},\mathbf{z})}{\partial \mathbf{x}}\bigg\rvert_{\mathbf{z} = \Tilde{\mathbf{z}}(\mathbf{x})} + \underbrace{\frac{\partial g(\mathbf{x},\mathbf{z})}{\partial \mathbf{z}}\bigg\rvert_{\mathbf{z} = \Tilde{\mathbf{z}}(\mathbf{x})}}_{0}\diff{\Tilde{\mathbf{z}}(\mathbf{x})}{\mathbf{x}}\\
    &= \frac{\partial g(\mathbf{x},\mathbf{z})}{\partial \mathbf{x}}\bigg\rvert_{\mathbf{z} = \Tilde{\mathbf{z}}(\mathbf{x})}
\end{split}
\end{equation}
    and this implies that
\begin{equation}
    \frac{\partial g(\mathbf{x},\mathbf{z}^*)}{\partial \mathbf{x}}\bigg\rvert_{\mathbf{x} = \mathbf{x}^*} = 0
\label{eqn:partder_opt}
\end{equation}
    where
\begin{equation}
    \mathbf{z}^* = \Tilde{\mathbf{z}}(\mathbf{x}^*)
\label{eqn:zopt}
\end{equation}
    (note that using the same symbol $\mathbf{z}^*$ in \eqref{eqn:zopt_defn} and \eqref{eqn:zopt} is not coincidental, see below). Because $g(\mathbf{x},\mathbf{z})$ is strictly convex in $\mathbf{x}$, it follows from \eqref{eqn:partder_opt} that $\mathbf{x}^*$ is the unique minimizer of $g(\mathbf{x},\mathbf{z}^*)$:
\begin{equation}
    g(\mathbf{x}^*,\mathbf{z}^*) \leq g(\mathbf{x},\mathbf{z}^*) \quad  \forall \; \mathbf{x} \in \mathbb{X}.
\label{eqn:leqg}
\end{equation}
    Next we note that $\mathbf{z}^*$ is the unique maximizer of $g(\mathbf{x}^*,\mathbf{z})$, see the definition of $\Tilde{\mathbf{z}}(\mathbf{x})$ in \eqref{eqn:zTil}:
\begin{equation}
    g(\mathbf{x}^*,\mathbf{z}^*) \geq g(\mathbf{x}^*,\mathbf{z})\quad  \forall \; \mathbf{z} \in \mathbb{Z}
\label{eqn:geqg}
\end{equation}
    It follows from \eqref{eqn:leqg} and \eqref{eqn:geqg} that:
\begin{equation}
    g(\mathbf{x}^*,\mathbf{z}^*) = \min_{\mathbf{x} \in \mathbb{X}}\; g(\mathbf{x},\mathbf{z}^*) \leq \max_{\mathbf{z} \in \mathbb{Z}}\; \min_{\mathbf{x} \in \mathbb{X}}\; g(\mathbf{x},\mathbf{z})
\end{equation}
and
\begin{equation}
    g(\mathbf{x}^*,\mathbf{z}^*) = \max_{\mathbf{z}\in \mathbb{Z}} \; g(\mathbf{x}^*,\mathbf{z}) \geq \min_{\mathbf{x} \in \mathbb{X}}\; \max_{\mathbf{z} \in \mathbb{Z}}\; g(\mathbf{x},\mathbf{z})
\end{equation}
    With this observation the proof of \eqref{eqn:minmaxleq} is concluded, and \eqref{eqn:minmaxmaxmin} follows as a consequence:
\begin{equation}
    \min_{\mathbf{x} \in \mathbb{X}}\; \max_{\mathbf{z} \in \mathbb{Z}} \; g(\mathbf{x},\mathbf{z}) = g(\mathbf{x}^*,\mathbf{z}^*) = \max_{\mathbf{z} \in \mathbb{Z}}\; \min_{\mathbf{x} \in \mathbb{X}}\; g(\mathbf{x},\mathbf{z})
\end{equation}
(the fact that the pair $(\mathbf{x}^*,\mathbf{z}^*)$ is the saddle point of $g(\mathbf{x},\mathbf{z})$ follows from \eqref{eqn:leqg} and \eqref{eqn:geqg}).
\end{proof}

\setcounter{biburlnumpenalty}{200}
\setcounter{biburlucpenalty}{200}
\setcounter{biburllcpenalty}{200}

\printbibliography

\end{document}